\documentclass[cernpreprint,texlive=2016,txfonts,UKenglish]{latex/atlasdoc}
\pdfoutput=1
\usepackage[biblatex=true]{latex/atlaspackage}
\usepackage{latex/atlasphysics}

\newcommand{\AtlasCoordFootnote}{
ATLAS uses a right-handed coordinate system with its origin at the
nominal interaction point (IP) in the centre of the detector and the
$z$-axis along the beam pipe. The $x$-axis points from the IP to the
centre of the LHC ring, and the $y$-axis points upwards. Cylindrical
coordinates $(r,\phi)$ are used in the transverse plane, $\phi$ being
the azimuthal angle around the $z$-axis. The pseudorapidity is defined
in terms of the polar angle $\theta$ as $\eta = -\ln
\tan(\theta/2)$. Angular distance is measured in units of $\Delta R
\equiv \sqrt{(\Delta\eta)^{2} + (\Delta\phi)^{2}}$. The rapidity is
defined as $y=0.5\ln[(E+p_z)/(E-p_z)]$, where $E$ is the energy and
$p_z$ is the $z$-component of the momentum, and transverse energy
is defined as $E_{\mathrm{T}}=E\sin\theta$.}

\graphicspath{{logos/}}

\usepackage{latex/atlasbiblatex}

\def\fb1{fb$^{-1}$}
\def\pb1{pb$^{-1}$}

\def\qq{q\bar q}

\def\kt{k_t}

\def\as{\alpha_s}
\def\oalphas2{{\cal O}(\alpha\as^2)}

\def\etg{E_{\mathrm{T}}^{\gamma}}
\def\ptjet{p_{\mathrm{T}}^{\mathrm{jet}}}
\def\rapjet{y^{\mathrm{jet}}}
\def\etag{\eta^{\gamma}}

\def\mgjn{m^{\gamma-{\mathrm{jet1}}}}
\def\ctgjn{\cos\theta^*}

\def\drgj{\Delta R^{\gamma-{\mathrm{jet1}}}}
\def\drjsj{\Delta R^{{\mathrm{jet1}}-{\mathrm{jet2}}}}
\def\drgsj{\Delta R^{\gamma-{\mathrm{jet2}}}}

\def\pt{p_{\mathrm{T}}}

\def\etisop{E_{\mathrm{T,part}}^{\mathrm{iso}}}
\def\etisod{E_{\mathrm{T,det}}^{\mathrm{iso}}}

\def\etisocut{$4.8\ \mathrm{\GeV}+4.2\cdot 10^{-3}\cdot\etg$}

\def\ptjetl{p_{\mathrm{T}}^{\mathrm{jet1}}}
\def\rapjetl{y^{\mathrm{jet1}}}
\def\etajetl{\eta^{\mathrm{jet1}}}

\def\ptjetsl{p_{\mathrm{T}}^{\mathrm{jet2}}}

\def\ptjetssl{p_{\mathrm{T}}^{\mathrm{jet3}}}

\def\delphijjlsl{\Delta\phi^{\mathrm{jet1-jet2}}}
\def\delphijjlssl{\Delta\phi^{\mathrm{jet1-jet3}}}
\def\delphijjslssl{\Delta\phi^{\mathrm{jet2-jet3}}}

\def\delphigsl{\Delta\phi^{\gamma-{\mathrm{jet2}}}}
\def\delphigssl{\Delta\phi^{\gamma-{\mathrm{jet3}}}}

\def\betaj{\beta^{\mathrm{jet1}}}
\def\betag{\beta^{\gamma}}

\def\sher{{\textsc{Sherpa}}}
\def\pyt{{\textsc{Pythia}}}
\def\jetp{{\textsc{Jetphox}}}
\def\blh{{\textsc{Blackhat}}}

\def\figdir{figures/}

\AtlasTitle{High-{\boldmath $E_{\mathrm{T}}$} isolated-photon plus jets
production in {\boldmath $pp$} collisions at {\boldmath $\sqrt
s=8$}~\TeV\ with the ATLAS detector}

\author{The ATLAS Collaboration}

\AtlasJournalRef{Nucl. Phys. B 918 (2017) 257}
\AtlasDOI{doi:10.1016/j.nuclphysb.2017.03.006} 

\AtlasAbstract{
The dynamics of isolated-photon plus one-, two- and three-jet
production in $pp$ collisions at a centre-of-mass energy of $8$~\TeV\
are studied with the ATLAS detector at the LHC using a data set with
an integrated luminosity of $20.2$~\fb1. Measurements of
isolated-photon plus jets cross sections are presented as functions of
the photon and jet transverse momenta. The cross sections as functions
of the azimuthal angle between the photon and the jets, the azimuthal
angle between the jets, the photon--jet invariant mass and the
scattering angle in the photon--jet centre-of-mass system are
presented. The pattern of QCD radiation around the photon and the
leading jet is investigated by measuring jet production in an annular
region centred on each object; enhancements are observed around the
leading jet with respect to the photon in the directions towards the
beams. The experimental measurements are compared  to several
different theoretical calculations, and overall a good description of
the data is found.
}

\usepackage[notindex,nottoc,notlof,notlot]{tocbibind}

\begin{document}

\maketitle

\tableofcontents

\section{Introduction}
\label{intro}

The production of prompt photons in association with jets in
proton--proton collisions, $pp\rightarrow\gamma+{\mathrm{jets}}+{\mathrm{X}}$,
provides a testing ground for perturbative QCD (pQCD) with a hard
colourless probe less affected by hadronisation effects than jet
production. The measurements of the angular correlations between the
photon and the jets can be used to probe the dynamics of the
hard-scattering process. Since the dominant production mechanism in
$pp$ collisions at the Large Hadron Collider (LHC) proceeds via the
$qg\rightarrow q\gamma$ process, measurements of prompt-photon plus
jet production are useful in constraining the gluon density in the
proton~\cite{np:b860:311,epl:101:61002}. These measurements can also
be used to tune the Monte Carlo (MC) models and to test $t$-channel
quark exchange~\cite{np:b875:483}.

At leading order (LO) in pQCD, the reaction 
$pp\rightarrow\gamma+{\mathrm{jet}}+{\mathrm{X}}$ is understood to proceed via
two separate production mechanisms: direct photons (D), which
originate from the hard process, and fragmentation photons (F), which
arise from the fragmentation of a coloured, high transverse momentum
($\pt$) parton~\cite{pr:d76:034003,pr:d79:114024}. The direct and
fragmentation contributions are only well defined at LO; at higher
orders such distinction is no longer possible.  Measurements of
prompt-photon production in a final state with accompanying hadrons
require isolation of photons to avoid the large contribution from
neutral-hadron decays into photons. The production of inclusive
isolated photons in $pp$ collisions was studied by the 
ATLAS~\cite{pr:d83:052005,pl:b706:150,pr:d89:052004,incpap} and
CMS~\cite{prl:106:082001,pr:d84:052011} collaborations. The
cross section for isolated photons in association with jets as a
function of the photon transverse energy\footnote{\AtlasCoordFootnote}
($\etg$) in different regions of rapidity of the highest-$\pt$ jet was
measured by ATLAS~\cite{pr:d85:092014}. The production of isolated
photons in association with jets was also measured by 
CMS~\cite{pr:d88:112009,jhep:1406:009,jhep:1510:128}.

\begin{figure}[h]
\vfill
\setlength{\unitlength}{1.0cm}
\begin{picture} (18.0,4.9)
\put (4.0,0.0){\includegraphics[width=3cm]{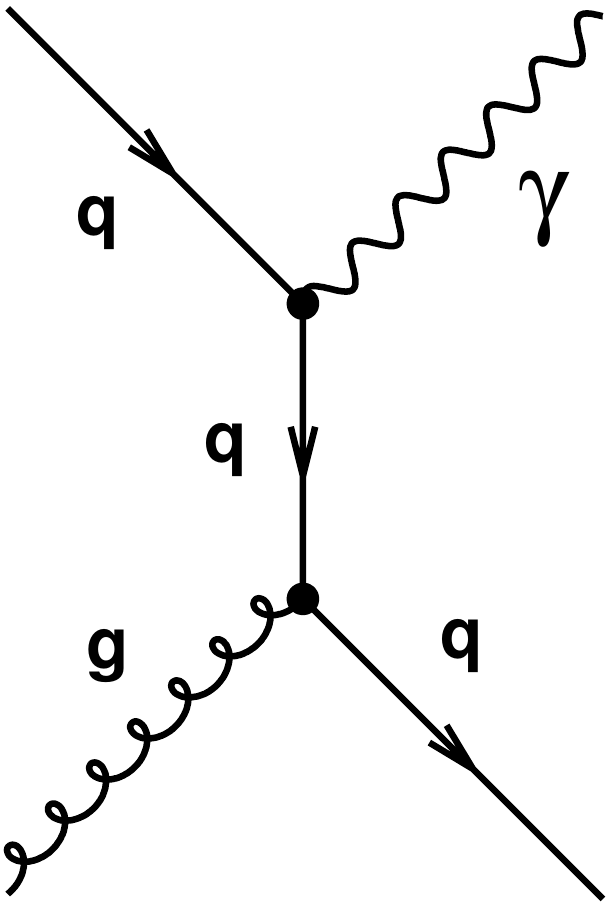}}
\put (8.0,0.0){\includegraphics[width=3.7cm]{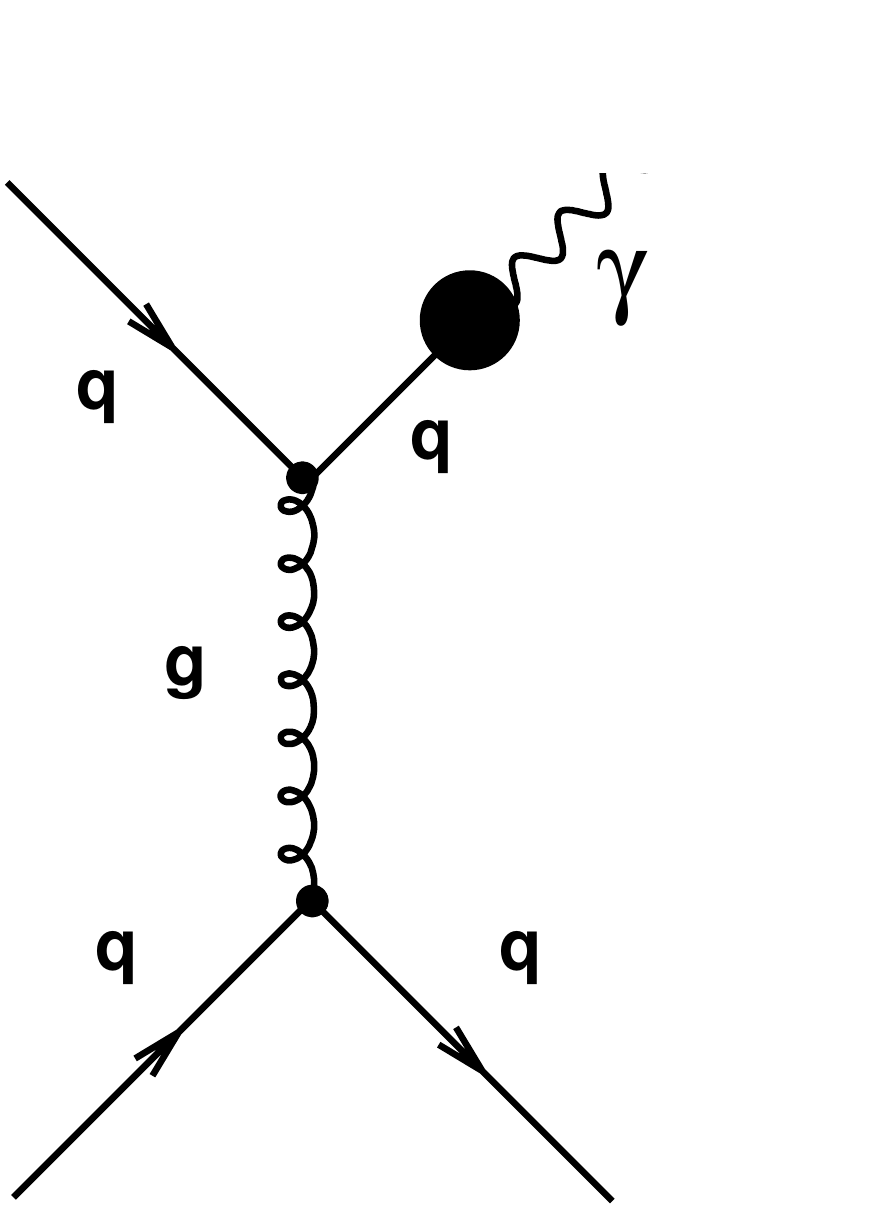}}
\put (5.0,-0.5){{\textbf{\small (a)}}}
\put (9.0,-0.5){{\textbf{\small (b)}}}
\end{picture}
\vspace{0.1cm}
\caption
{Examples of diagrams for (a) $\gamma+{\mathrm{jet}}$ production through
  direct-photon processes and (b) $\gamma+{\mathrm{jet}}$ production
  through fragmentation processes.} 
\label{fig01}
\end{figure}

The dynamics of the underlying processes in $2\rightarrow 2$ hard
scattering can be investigated using the variable $\theta^*$, where
$\cos\theta^*\equiv\tanh(\Delta y/2)$ and $\Delta y$ is the difference
between the rapidities of the two final-state particles. The variable
$\theta^*$ coincides with the scattering angle in the centre-of-mass
frame for collinear scattering of massless particles, and its
distribution is sensitive to the spin of the exchanged particle. For
processes dominated by $t$-channel gluon exchange, such as dijet
production in $pp$ collisions, the cross section behaves as
$(1-|\cos\theta^*|)^{-2}$ when $|\cos\theta^*|\rightarrow 1$. In
contrast, processes dominated by $t$-channel quark exchange, such as
$W/Z+{\mathrm{jet}}$ production, are expected to have an asymptotic
$(1-|\cos\theta^*|)^{-1}$ behaviour. This prediction from QCD can be
tested in photon plus jet production in high-energy hadron--hadron
collisions. The direct-photon contribution, as shown in
Figure~\ref{fig01}(a), is expected to exhibit a
$(1-|\cos\theta^*|)^{-1}$ dependence when $|\cos\theta^*|\rightarrow 1$,  
whereas that of fragmentation processes, as shown in
Figure~\ref{fig01}(b), is predicted to be the same as in dijet
production, namely $(1-|\cos\theta^*|)^{-2}$. For both processes,
there are also $s$-channel contributions which are, however,
non-singular when $|\cos\theta^*|\rightarrow 1$. At higher orders,
direct processes such as $qq\rightarrow qq\gamma$ are dominated by
$t$-channel gluon exchange and contribute to the distribution in
$|\cos\theta^*|$ with a component similar to that of
fragmentation. However, a measurement of the cross section for
prompt-photon plus jet production as a function of $|\cos\theta^*|$ is
still sensitive to the relative contributions of the direct and
fragmentation components and allows a test of the dominance of the
$t$-channel quark exchange, such as that shown in
Figure~\ref{fig01}(a).

Colour connection between the partons in the initial and final states
modifies the pattern of QCD radiation around the final-state
partons. Colour-coherence effects were studied at the
Tevatron~\cite{pr:d50:5562,pl:b414:419} using dijet events by
comparing the measurements with predictions with and without such
effects. Photon plus two-jet events are optimal for investigating jet
production around the photon and the highest-$\pt$ jet: the partons
are colour-connected while the photon is colourless.

The results presented in this paper include measurements of cross
sections for isolated-photon plus \mbox{one-,} two- and three-jet final
states as functions of $\etg$ and the transverse momentum of the
leading jet (jet1, $\ptjetl$), the second-highest-$\pt$ jet (jet2,
$\ptjetsl$) and the third-highest-$\pt$ jet (jet3,
$\ptjetssl$)~\cite{pr:d79:114024,pl:b269:445,pr:d84:114002,pr:d87:034026}.
The analysis is performed using a dataset with an integrated
luminosity of $20.2$~\fb1\ of $pp$ collisions at $\sqrt s=8$~\TeV. The
dynamics of the photon plus one-jet system are studied by measuring
the photon--jet invariant mass ($\mgjn$) and
$\ctgjn$~\cite{pr:d79:114024}. In addition, the azimuthal angles
between the photon and each jet ($\delphigsl$, $\delphigssl$) and
between the jets ($\delphijjlsl$, $\delphijjlssl$, $\delphijjslssl$)
are measured for photon plus two- and three-jet
events~\cite{pr:d84:114002,pr:d87:034026}. The production of jet2
around the photon and jet1 is measured separately to investigate the
differences between the two configurations. The scale evolution of the
photon plus one-jet system is studied by measuring the cross sections
as functions of $\ctgjn$ in different regions of $\mgjn$. For photon
plus two- and three-jet events, the scale evolution is investigated by
measuring the angular correlations in different regions of $\etg$.

The predictions from the event generators \pyt~\cite{cpc:178:852} and
\sher~\cite{jhep:0902:007} are compared with the measurements. The
next-to-leading-order (NLO) QCD predictions from \jetp~\cite{jhep:0205:028,pr:d73:094007} are compared with the
photon plus one-jet measurements, whereas those from \blh~\cite{pr:d78:036003,prl:109:042001} are compared with the
photon plus two-jet and photon plus three-jet measurements.

\section{The ATLAS detector}
\label{detector}
The ATLAS detector~\cite{PERF-2007-01} at the LHC covers nearly the
entire solid angle around the collision point. It consists of an inner
tracking detector surrounded by a thin superconducting solenoid,
electromagnetic and hadronic calorimeters, and a muon spectrometer
incorporating three large superconducting toroidal magnets. 

The inner-detector system (ID) is immersed in a \SI{2}{\tesla} axial
magnetic field and provides charged-particle tracking in the range
$|\eta| < 2.5$. A high-granularity silicon pixel detector covers the
interaction region and typically provides three measurements per
track. It is followed by a silicon microstrip tracker, which provides
eight two-dimensional measurement points per track. These silicon
detectors are complemented by a transition radiation tracker, which
enables radially extended track reconstruction up to $|\eta| =
2.0$. The transition radiation tracker also provides electron
identification information based on the fraction of the typically 30
total hits which are above a higher energy-deposit threshold
corresponding to transition radiation.

The calorimeter system covers the pseudorapidity range $|\eta| <
4.9$. Within the region $|\eta|< 3.2$, electromagnetic calorimetry is
provided by barrel and endcap high-granularity lead/liquid-argon (LAr)
electromagnetic calorimeters, with an additional thin LAr presampler
covering $|\eta| < 1.8$ to correct for energy loss in material
upstream of the calorimeters; for $|\eta|<2.5$ the LAr calorimeter is
divided into three layers in depth, which are finely segmented in
$\eta$ and $\phi$. Hadronic calorimetry is provided by a
steel/scintillator-tile calorimeter, segmented into three barrel
structures within $|\eta| < 1.7$, and two copper/LAr hadronic endcap
calorimeters, which cover $1.5 < |\eta| < 3.2$. The solid angle
coverage is completed with forward copper/LAr and tungsten/LAr
calorimeter modules optimised for electromagnetic and hadronic
measurements, respectively.

The muon spectrometer (MS) comprises separate trigger and
high-precision tracking chambers measuring the deflection of muons in
a magnetic field generated by superconducting air-core toroids. The
tracking chamber system covers the region $|\eta| < 2.7$ with three
layers of monitored drift tubes, complemented by cathode-strip
chambers in the forward region. The muon trigger system covers the
range $|\eta| < 2.4$ with resistive-plate chambers in the barrel and
thin-gap chambers in the endcap regions.

A three-level trigger system is used to select interesting
events~\cite{PERF-2011-02}. The level-1 trigger is implemented in
hardware and uses a subset of detector information to reduce the event
rate to at most \SI{75}{\kHz}. This is followed by two software-based
trigger levels which together reduce the event rate to about
\SI{400}{\Hz}.

\section{Data selection}
\label{datsel}
The data used in this analysis were collected during the
proton--proton collision running period of 2012, when the LHC operated
at a centre-of-mass energy of $\sqrt s=8$~\TeV. Only events taken in
stable beam conditions and passing detector and data-quality
requirements are considered. Events were recorded using a
single-photon trigger, with a nominal transverse energy threshold of
$120$~\GeV; this trigger is used offline to select events in which the
photon transverse energy, after reconstruction and calibration, is
greater than $130$~\GeV. For isolated photons with $\etg>130$~\GeV\
and pseudorapidity $|\etag|< 2.37$ the trigger efficiency is higher
than $99.8\%$. The integrated luminosity of the collected sample is
$20.2\pm 0.4$~\fb1~\cite{1608.03953}.

The sample of isolated-photon plus jets events is selected using
offline criteria similar to those reported in previous
publications~\cite{np:b875:483,incpap}. Events are required to have a
reconstructed primary vertex consistent with the average beam-spot
position, with at least two associated charged-particle tracks with
$\pt>400$~\MeV. If more than one such vertex is present in the event,
the one with the highest sum of the $\pt^2$ of the associated tracks
is selected as the primary vertex.

During the 2012 data-taking period there were on average $19$
proton--proton interactions per bunch crossing. The methods used to
mitigate the effects of the additional $pp$ interactions (pile-up) on
the photon isolation and jet reconstruction are described below.

\subsection{Photon selection}
\label{photsel}
The selection of photon candidates is based on energy clusters
reconstructed in the electromagnetic calorimeter with transverse
energies exceeding $2.5$~\GeV. The clusters matched to
charged-particle tracks, based on the distance in ($\eta$, $\phi$)
between the cluster barycentre and the track impact point extrapolated
to the second layer of the LAr calorimeter, are classified as electron
candidates. Those clusters without matching tracks are classified as
unconverted photon candidates, and clusters matched to pairs of tracks
originating from reconstructed conversion vertices in the inner
detector or to single tracks with no hit in the innermost layer of the
pixel detector are classified as converted photon
candidates~\cite{1606.01813}. From MC simulations, $96\%$ of prompt
photons with $\etg>25$~\GeV\ are expected to be reconstructed as
photon candidates, while the remaining $4\%$ are incorrectly
reconstructed as electrons but not as photons. The efficiency to
reconstruct photon conversions decreases at high $\etg$ ($>
150$~\GeV), where it becomes more difficult to separate the two tracks
from the conversions. Such conversions with very nearby tracks are
often not recovered as single-track conversions because of the tighter
selections applied to single-track conversion candidates. The overall
photon reconstruction efficiency is thus reduced to about $90\%$ for
$\etg\sim 1$~\TeV~\cite{1606.01813}.

The energy measurement is performed using calorimeter and tracking
information. A dedicated energy calibration~\cite{epj:c74:3071} is
applied separately for converted and unconverted photon candidates to
account for upstream energy loss and both lateral and longitudinal
leakage.

The direction of the photon is determined from the barycentre of the
energy cluster in the electromagnetic calorimeter and the position of
the primary vertex. Events with at least one photon candidate with
calibrated $\etg>130$~\GeV\ and $|\etag|<2.37$ are selected;
candidates in the region $1.37<|\etag|<1.56$, which includes the
transition region between the barrel and endcap calorimeters, are not
considered. The same shower-shape and isolation requirements as
described in previous
publications~\cite{pr:d83:052005,pl:b706:150,incpap,pr:d85:092014,1606.01813}
are applied to the candidates; these requirements are referred to as
``tight'' identification criteria. The photon identification
efficiency for $\etg>130$~\GeV\ varies in the range $(94$--$100)\%$
depending on $\etag$ and whether the candidate is classified as an
unconverted or converted photon~\cite{1606.01813}.

The photon candidate is required to be isolated based on the amount of
transverse energy in a cone of size $\Delta R=0.4$ around the
photon. The isolation transverse energy is computed from
three-dimensional topological clusters of calorimeter cells (see
Section 3.3)~\cite{1603.02934} and is denoted by $\etisod$. The
measured value of $\etisod$ is corrected for leakage of the photon's
energy into the isolation cone and the estimated contributions from
the underlying event and pile-up. The latter correction is performed
using the jet-area method~\cite{jhep:0804:005} to estimate the ambient
transverse energy density on an event-by-event basis; this estimate is
used to subtract the joint contribution of the underlying event and
pile-up to $\etisod$ and amounts to $1.5$--$2$~\GeV\ in the 2012
data-taking period. After these corrections, $\etisod$ is required to
be lower than  \etisocut~[\GeV]~\cite{incpap}; the requirement is
$\etg$-dependent so that in simulation the fraction of identified
photons which are isolated stays high as $\etg$ increases. The
isolation requirement significantly reduces the main background, which
consists of multi-jet events where one jet typically contains a
$\pi^0$ or $\eta$ meson that carries most of the jet energy and is
misidentified as a prompt photon.

A small fraction of the events contain more than one photon candidate
satisfying the selection criteria. In such events, the highest-$\etg$
photon is considered for further study.

\subsection{Jet selection}
\label{jetsel}
Jets are reconstructed using the anti-$\kt$
algorithm~\cite{jhep:0804:063} with radius parameter $R=0.6$. The
inputs to the jet reconstruction are three-dimensional topological
clusters of calorimeter cells. This method first clusters
topologically connected calorimeter cells and classifies these
clusters as either electromagnetic or hadronic. The classification
uses a local cluster weighting (LCW) calibration scheme based on
cell-energy density and longitudinal depth within the
calorimeter~\cite{epj:c75:17}. Based on this classification, energy
corrections derived from single-pion MC simulations are
applied. Dedicated corrections are derived for the effects of the
non-compensating response of the calorimeter, signal losses due to
noise-suppression threshold effects, and energy lost in
non-instrumented regions. The jet four-momenta are computed from the
sum of the topological cluster four-momenta, treating each as a
four-vector with zero mass. These jets are referred to as
detector-level jets. The direction of the jet is then corrected such
that the jet originates from the selected primary vertex of the
event. Prior to the final calibration, the contribution from the
underlying event and pile-up is subtracted on a jet-by-jet basis using
the jet-area method. An additional jet-energy calibration is derived
from MC simulations as a correction relating the calorimeter response
to the true jet energy. To determine these corrections, the jet
reconstruction procedure applied to the topological clusters is also
applied to the generated stable particles, which are defined as those
with a lifetime $\tau$ longer than $10$~ps, including muons and
neutrinos; these jets are referred to as particle-level jets. In
addition, sequential jet corrections, derived from MC simulated events
and using global properties of the jet such as tracking information,
calorimeter energy deposits and muon spectrometer information, are
applied~\cite{ATLAS-CONF-2015-002}. Finally, the detector-level jets
are further calibrated with additional correction factors derived in
situ from a combination of $\gamma+{\mathrm{jet}}$, $Z+{\mathrm{jet}}$ and dijet
balance methods~\cite{epj:c75:17,ATLAS-CONF-2015-037}. 

Jets reconstructed from calorimeter signals not originating from a
$pp$ collision are rejected by applying jet-quality
criteria~\cite{epj:c75:17,epj:c73:2304}. These criteria suppress
spurious jets from electronic noise in the calorimeter, cosmic rays
and beam-related backgrounds. Remaining jets are required to have
calibrated transverse momenta greater than $50$~\GeV\ and rapidity
$|\rapjet|<4.4$. Jets overlapping with the candidate photon are not
considered if the jet axis lies within a cone of size $\Delta R=1.0$
around the photon candidate; this requirement prevents any overlap
between the photon isolation cone ($\Delta R=0.4$) and the jet cone
($\Delta R=0.6$).

\subsection{Event categorisation}
\label{catego}
To investigate the production of jets in association with a photon,
six samples are selected; the requirements are listed in
Table~\ref{tabzero}: 
\begin{itemize}
\item ``Photon plus one-jet sample'' (P1J): it is used to study the
  major features of an inclusive sample of events with a photon and at
  least one jet. In this sample, jet1 is required to have
  $\ptjetl>100$~\GeV; asymmetric requirements on $\etg$ and $\ptjetl$
  are applied to reduce the infrared sensitivity of the NLO QCD
  calculations~\cite{np:b507:315}.
\item ``Photon plus one-jet $\mgjn$ and $\ctgjn$ sample'' (P1JM): for
  the measurements of the cross sections as functions of $\mgjn$ and
  $|\ctgjn|$ additional constraints are needed to remove biases due to
  the rapidity and transverse momentum requirements on the photon and
  jet1~\cite{np:b875:483}. To perform unbiased measurements, the
  requirements $|\etag+\rapjetl|<2.37$, $|\ctgjn|<0.83$ and
  $\mgjn>467$~\GeV\ are applied.\footnote{The maximal (minimal) value
    of $|\ctgjn|$ ($\mgjn$) for which the measurement is unbiased
    corresponds to $\tanh(2.37/2)$ ($2\cdot\etg/\sin\theta^*$) with
    $\etg=130$~\GeV\ and $\cos\theta^*=0.83$.} These selections define
  a kinematic region where the acceptance is independent of the
  variables being studied.
\item ``Photon plus two-jet sample'' (P2J): it is used to study the
  major features of an inclusive sample of events with a photon and at
  least two jets and the azimuthal correlations between the photon and
  jet2 as well as between jet1 and jet2. Due to the resolution in
  $\pt$ the highest- and next-to-highest-$\pt$ particle-level jets can
  end up being reconstructed as jet2 and jet1, respectively. To
  suppress such migrations, asymmetric requirements are applied:
  $\ptjetl>100$~\GeV\ and $\ptjetsl>65$~\GeV.
\item ``Photon plus three-jet sample'' (P3J): it is used to
  investigate the major characteristics of an inclusive sample of
  events with a photon and at least three jets; in addition,
  measurements of the azimuthal correlations between the photon and
  jet3, jet1 and jet3, as well as between jet2 and jet3 are
  performed. Asymmetric requirements are applied to suppress the
  migrations in $\pt$ between the three highest-$\pt$ jets:
  $\ptjetl>100$~\GeV, $\ptjetsl>65$~\GeV\ and $\ptjetssl>50$~\GeV.
\end{itemize}

\begin{figure}[p]
\setlength{\unitlength}{1.0cm}
\begin{picture} (18.0,18.0)
\put (0.0,0.5){\centerline{\includegraphics[bb = 0 0 567 567,width=11cm,trim=0cm 0cm 0cm 0cm,clip=true]{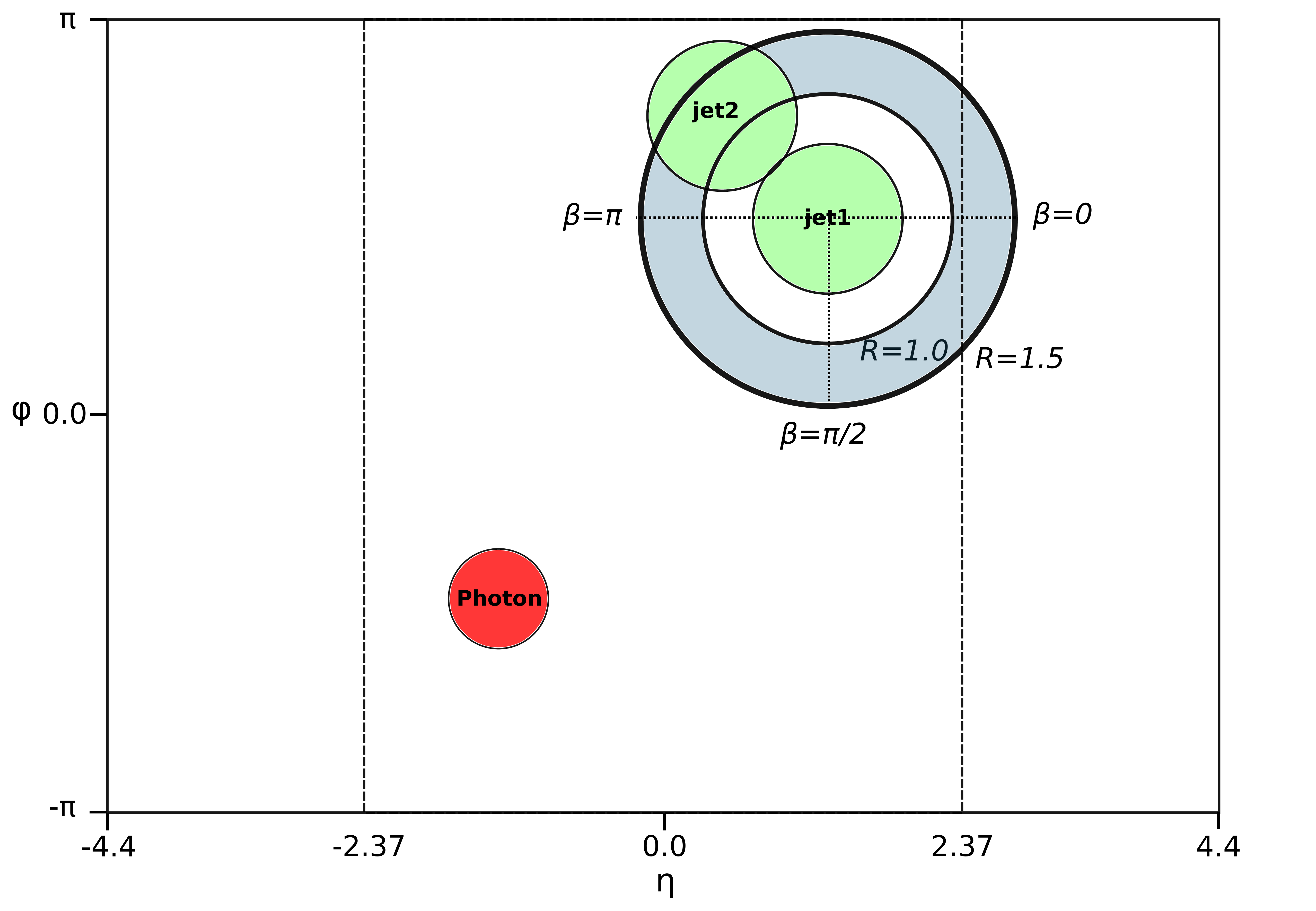}}}
\put (0.0,9.0){\centerline{\includegraphics[bb = 0 0 567 567,width=11cm,trim=0cm 0cm 0cm 0cm,clip=true]{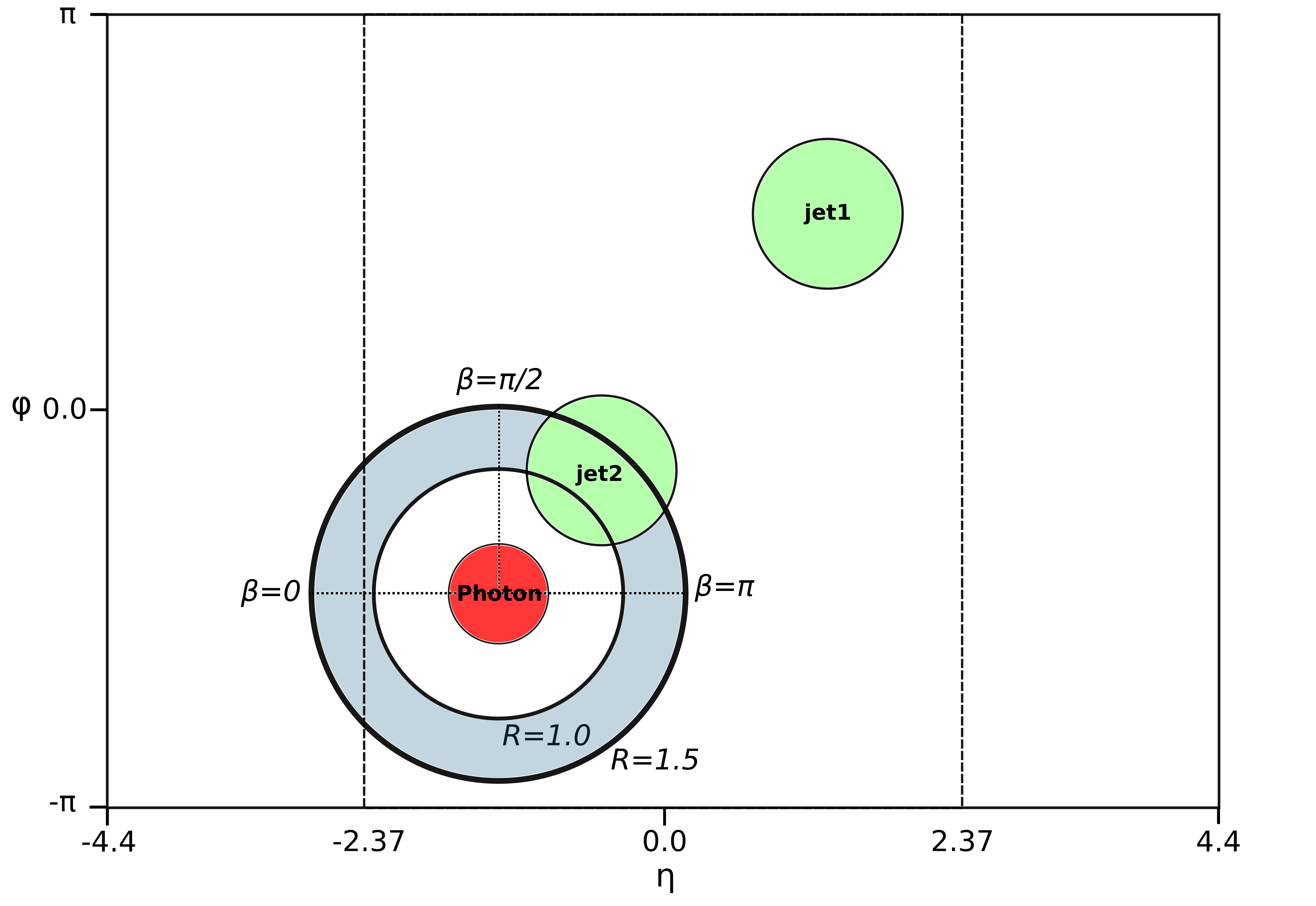}}}
\put (7.8,8.5){{\textbf{\small (a)}}}
\put (7.8,0.0){{\textbf{\small (b)}}}
\end{picture}
\caption
{
  Schematic diagrams that show the definitions of (a) $\betag$ and (b)
  $\betaj$.
}
\label{figbeta}
\end{figure}

To compare the pattern of QCD radiation around the photon and jet1,
two additional samples of photon plus two-jet events are selected. The
phase-space regions are defined to avoid biases due to different $\pt$
and $\eta$ requirements on the final-state objects as well as to have
no overlap between the two samples. The following requirements are
common to the two samples:
\begin{itemize}
\item The jets must satisfy $\ptjetl>130$~\GeV, $|\etajetl|<2.37$ and
  $\ptjetsl>50$~\GeV. The first two requirements are imposed to be the
  same as for the photon so as to compare additional jet production in
  similar regions of phase space. The third requirement is chosen to
  select jets with the lowest $\pt$ threshold, while suppressing the
  contribution from the underlying event and pile-up.
\item The angular distance between the photon and jet1, $\drgj$, is
  restricted to $\drgj>3$ to avoid any overlap between the two samples
  and any bias within the regions that are used to study additional
  jet production.
\end{itemize}
The requirements specific to each of the two samples are listed below:
\begin{itemize}
\item ``Photon plus two-jet $\betag$ selection'' (P2JBP): it is used
  to measure the production of jet2 around the photon. The cross
  section is measured as a function of the observable
  $\betag$~\cite{pr:d50:5562,pl:b414:419}, which is defined
  as\footnote{In the definitions of $\betag$ and $\betaj$, the
    arctangent function with two arguments is used to keep track of
    the proper quadrant.}

  \begin{equation}
  \betag=\tan^{-1} \frac{|\phi^{\mathrm{jet2}}-\phi^{\gamma}|}{\mbox{sign}(\eta^{\gamma})\cdot(\eta^{\mathrm{jet2}} - \eta^{\gamma})}.
  \end{equation}
  The phase space is restricted to $1\!<\!\drgsj\!<\!1.5$; the lower
  requirement avoids the overlap with the photon isolation cone while
  the upper requirement is the largest value which makes this sample
  and the next one non-overlapping. In addition, $\ptjetsl<\etg$ is
  imposed for comparison with the other sample.
\item ``Photon plus two-jet $\betaj$ selection'' (P2JBJ): it is used
  to measure the production of jet2 around jet1 using the observable
  $\betaj$, defined as

  \begin{equation}
  \betaj=\tan^{-1} \frac{|\phi^{\mathrm{jet2}}-\phi^{\mathrm{jet1}}|}{\mbox{sign}(\eta^{\mathrm{jet1}})\cdot(\eta^{\mathrm{jet2}} - \eta^{\mathrm{jet1}})}.
  \end{equation}
  To compare on equal footing with the measurement of the previous
  sample, the restriction\\ 
 \mbox{$1\!<\!\drjsj\!<\!1.5$} is applied.
\end{itemize}
Schematic diagrams for the definitions of $\betag$ and $\betaj$ are
shown in Figure~\ref{figbeta}. The variable $\betag$ ($\betaj$) is
defined in such a way that $\betag=0$ or $\pi$ ($\betaj=0$ or $\pi$)
corresponds to a plane in space containing jet2, the beam axis and the
photon (jet1); $\beta=0$ ($\pi$) always points to the beam which is
closer to (farther from) the photon or jet1 in the $\eta$--$\phi$
plane.

The number of selected events in data for each of the six samples is
included in Table~\ref{tabzero}. The overlap between the different
samples is as follows: (a) P3J is contained within P2J, which in turn
is a subset of P1J; (b) P1JM is contained within P1J; (c) P2JBP and
P2JBJ have no overlap. 

\begin{table}[h]
\setlength{\unitlength}{1.0cm}
\begin{picture} (18.0,11.0)
\put (-0.1,5.5){\scalebox{1.0}{
\begin{tabular}{|l|c|c|c|c|c|c|}
\hline
 & \multicolumn{6}{c|}{Sample} \\ \cline{2-7}
 & P1J & P1JM & P2J & P2JBP & P2JBJ & P3J \\ \hline
Common &  \multicolumn{6}{c|}{ $\etg>130$ \GeV\ and $|\etag|<2.37$, excluding $1.37<|\etag|<1.56$ } \\
requirements &  \multicolumn{6}{c|}{ $|y^{\mathrm{jet}}|<4.4$ and $\Delta R^{\gamma-{\mathrm{jet}}}>1$ } \\ \hline
$\ptjetl$ [\GeV] & $> 100$ & $> 100$ & $> 100$ & $> 130$ & $> 130$ & $> 100$ \\ 
$\ptjetsl$ [\GeV] & $-$ & $-$ & $> 65$ & $> 50$ & $> 50$ & $> 65$ \\ 
$\ptjetssl$ [\GeV] & $-$ & $-$ & $-$ & $-$ & $-$ & $> 50$ \\ 
$|\etag+\rapjetl|$ & $-$ & $<2.37$ & $-$ & $-$ & $-$ & $-$ \\ 
$|\ctgjn|$ & $-$ & $<0.83$ & $-$ & $-$ & $-$ & $-$ \\ 
$\mgjn$ [\GeV] & $-$ & $>467$ & $-$ & $-$ & $-$ & $-$ \\
$\drgj$  & $-$ & $-$ & $-$ & $> 3$ & $> 3$ & $-$ \\
$\drgsj$ & $-$ & $-$ & $-$ & $1\!<\!\drgsj\!<\!1.5$ & $-$ & $-$ \\
$\drjsj$ & $-$ & $-$ & $-$ & $-$ & $1\!<\!\drjsj\!<\!1.5$ & $-$ \\
$|\etajetl|$ & $-$ & $-$ & $-$ & $<2.37$ & $<2.37$ & $-$ \\
$\ptjetsl, \etg$ & $-$ & $-$ & $-$ & $\ptjetsl<\etg$ & $-$ & $-$ \\\hline\hline
Number of &  &  & &  &  &  \\ 
Events & $2\,451\,236$ & $344\,572$ & $567\,796$ & $40\,537$ & $37\,429$ & $164\,062$ \\ \hline\hline
Normalisation  & & & & & & \\
factor & & & & & & \\
\sher\ (\pyt) & $1.0\ (1.1)$ & $1.0\ (1.2)$ & $1.1\ (1.2)$ & $1.0\ (1.2)$ & $1.0\ (1.2)$ & $1.1\ (1.1)$ \\ \hline
\end{tabular}
}}
\end{picture}
\caption{
  Characteristics of the six samples of $\gamma+{\mathrm{jet(s)}}$ events:
  kinematic requirements, number of selected events in data and
  normalisation factors applied to the MC predictions.
}
\label{tabzero}
\end{table}

\section{Monte Carlo simulations}
\label{mc}

Samples of MC events were generated to study the characteristics of
signal events. The MC samples were also used to determine the response
of the detector and the correction factors necessary to obtain the
particle-level cross sections. In addition, these samples were used to
estimate hadronisation corrections to the NLO QCD calculations.

The MC programs \pyt~8.165~\cite{cpc:178:852} and
\sher~1.4.0~\cite{jhep:0902:007} were used to generate the simulated
events (see Table~\ref{tab:mc}). In both generators, the partonic
processes were simulated using LO matrix elements, with the inclusion
of initial- and final-state parton showers. Fragmentation into hadrons
was performed using the Lund  string model~\cite{prep:97:31} in the
case of  \pyt, and a modified version of the cluster
model~\cite{epj:c36:381} in the case of  \sher, for which it is the
default treatment. The LO CTEQ6L1~\cite{jhep:0207:012} (NLO
CT10~\cite{pr:d82:074024}) parton distribution functions (PDF) were
used to parameterise the proton structure in \pyt\ (\sher). Both
samples included a simulation of the underlying event. The event
generator parameters were set according to the ``AU2
CTEQ6L1''~\cite{ATL-PHYS-PUB-2012-003} tune for \pyt\ and the ``CT10''
tune for \sher.  All samples of generated events were passed through
the  {\textsc{Geant4}}-based~\cite{nim:a506:250} ATLAS detector- and
trigger-simulation programs~\cite{epj:c70:823}. They were
reconstructed and analysed by the same program chain as the data.

The \pyt\ simulation of the signal included LO photon plus jet events
from both direct processes (the hard subprocesses $qg\rightarrow
q\gamma$ and $\qq\rightarrow g\gamma$, called ``hard component'') and
photon bremsstrahlung in QCD dijet events (called ``brem
component''). The \sher\ samples were generated with LO matrix
elements for photon plus jet final states with up to three additional
partons, supplemented with parton showers. While the brem component
was modelled in \pyt\ by final-state QED radiation arising from
calculations of all $2\rightarrow 2$ QCD processes, it was accounted
for in \sher\ through the matrix elements of $2\rightarrow n$
processes with $n\geq 3$; in the evaluation of the matrix elements the
photon was required to be farther than $\Delta R=0.3$ from any
parton.

All samples were simulated taking into account the effects of the
pile-up appropriate for 2012 data. The additional interactions were
modelled by overlaying simulated hits from events with exactly one
high momentum-transfer (signal) collision per bunch crossing with hits
from minimum-bias events that were produced with the \pyt\ 8.160
program~\cite{cpc:178:852} using the A2M
tune~\cite{ATL-PHYS-PUB-2012-003} and the MSTW2008
LO~\cite{epj:c64:653} PDF set.

Dedicated \pyt\ and \sher\ samples of events were generated at
particle and parton levels, switching off the mechanisms that account
for the underlying event to correct the NLO calculations for
hadronisation and underlying-event effects.

The particle-level isolation variable on the photon was built from the
transverse energy of all stable particles, except for muons and
neutrinos, in a cone of size $\Delta R=0.4$ around the photon
direction after the contribution from the underlying event was
subtracted; in this case, the same underlying-event subtraction
procedure used on data was applied at the particle level. The
isolation transverse energy at particle level is denoted by
$\etisop$. The particle-level requirement on $\etisop$ was determined
using the \pyt\ and \sher\ MC samples, by comparing the calorimeter
isolation transverse energy with the particle-level isolation on an
event-by-event basis. The effect of the experimental isolation
requirement used in the data is close to a particle-level requirement
of $\etisop<10$~\GeV\ over the measured $\etg$ range. The measured
cross sections refer to particle-level jets and photons that are
isolated by requiring $\etisop<10$~\GeV.

The MC predictions at particle level are normalised to the measured
integrated cross sections. The normalisation factors are applied
globally for each sample defined in Section~\ref{catego} and are
listed in Table~\ref{tabzero}.

\begin{table}[h]
\begin{center}
\begin{tabular}{|l|l|l|l|}
\hline
Name & Matrix elements & PDF & Tune \\
\hline
\pyt~8.165 & $2\rightarrow 2$ & LO CTEQ6L1 & AU2 CTEQ6L1 \\
\sher~1.4.0 & $2\rightarrow n$, $n=2,...,5$ & NLO CT10 & CT10 \\
\hline
\end{tabular}
\end{center}
\caption{\label{tab:mc}
The generators used for correcting the data are listed, together with
their matrix elements, the PDF and the tunes.
}
\end{table}

\section{Signal extraction}

\subsection{Backgrounds}
\label{bacsub}
A non-negligible background contribution from jets remains in the
selected sample, even after the application of the tight
identification and isolation requirements on the photon. The
background subtraction uses a data-driven method based on
signal-suppressed control regions. The background contamination in the
selected sample is estimated using the same two-dimensional sideband
technique as in the previous
analyses~\cite{pr:d83:052005,pl:b706:150,pr:d85:092014,incpap,np:b875:483}
and then subtracted bin-by-bin from the observed yield. In the
two-dimensional plane formed by $\etisod$ and the photon
identification variable, four regions are defined: 
\begin{itemize}
\item{$A$}: the ``signal'' region, containing tight isolated photon
  candidates.
\item{$B$}: the ``non-isolated'' background control region, containing
  tight non-isolated photon candidates. A candidate photon is
  considered to be non-isolated if $\etisod>(4.8+2)\ {\mathrm
    {\GeV}}+4.2\cdot 10^{-3}\cdot\etg$~[\GeV]; the threshold is
  $2$~\GeV\ higher than the isolation requirement for the signal
  region.
\item{$C$}: the ``non-tight'' background control region, containing
  isolated non-tight photon candidates. A candidate photon is labelled
  as ``non-tight'' if it fails at least one among four of the tight
  requirements on the shower-shape variables computed from the energy
  deposits in the first layer of the electromagnetic calorimeter, but
  satisfies the tight requirement on the total lateral shower
  width~\cite{1606.01813} in the first layer and all the other tight
  identification criteria in other layers.
\item{$D$}: the background control region containing non-isolated
  non-tight photon candidates. 
\end{itemize}

The signal yield in region $A$, $N_A^{\mathrm{sig}}$, is estimated from the
numbers of events in regions $A$, $B$, $C$ and $D$ and takes into
account the expected number of signal events in the three background
control regions via signal leakage fractions, which are extracted from
MC simulations of the signal. The only hypothesis is that the
isolation and identification variables are uncorrelated in background
events, thus  
$R^{\mathrm{bg}}=(N_A^{\mathrm{bg}}\cdot N_D^{\mathrm{bg}})/(N_B^{\mathrm{bg}}\cdot N_C^{\mathrm{bg}})=1$,
where $N_K^{\mathrm{bg}}$ with $K=A,B,C,D$ is the number of background
events in each region. This assumption is verified~\cite{incpap} both
in simulated background samples and in data in a background-dominated
region. Deviations from unity are taken as systematic uncertainties
(see Section~\ref{syst}). In addition, a systematic uncertainty is
assigned to the modelling of the signal leakage fractions. Since the
simulation does not accurately describe the electromagnetic shower
profiles, a correction factor for each simulated shape variable is
applied to better match the data~\cite{pr:d83:052005,pl:b706:150}. 

There is an additional background from electrons misidentified as
photons, mainly produced in $Z\rightarrow e^+e^-$ and 
$W\rightarrow e\nu$ processes. Such misidentified electrons are
largely suppressed by the photon selection. The remaining electron
background is estimated using MC simulations and found to be negligible
in the phase-space region of the analysis presented here.

\subsection{Signal yield}
\label{fit}
The signal purity, defined as $N_A^{\mathrm{sig}}/N_A$, is typically above
$0.9$ and is similar whether \pyt\ or \sher\ is used to extract the
signal leakage fractions. The signal purity increases as $\etg$,
$\ptjetl$ and $\mgjn$ increase and decreases as $|\ctgjn|$ increases.

For most of the distributions studied, the shapes of the hard and brem
components in the signal MC simulated by \pyt\ are somewhat
different. Therefore, in each case, the shape of the total MC
distribution depends on the relative fraction of the two
contributions. To improve the description of the data by the \pyt\ MC
samples, a fit~\cite{np:b875:483} to each data distribution is
performed with the weight of the hard contribution, $\alpha$, as the
free parameter; the weight of the brem contribution is given by
$1-\alpha$. The fitted values of $\alpha$ are in the range
$0.26$--$0.86$. After these fits, a good description of the data is
obtained from the \pyt\ MC simulations for all the observables, except
for the distributions in the azimuthal angle between the jets. The
simulations of \sher\ give a good description of the data, except for
the tail of the distributions in $\etg$.

The integrated efficiency, including the effects of trigger,
reconstruction, particle identification and event selection, is
evaluated from the simulated signal samples described in
Section~\ref{mc}. The integrated efficiency is computed as
$\varepsilon=N^{\mathrm{det,part}}/N^{\mathrm{part}}$, where  $N^{\mathrm{det,part}}$
is the number of MC events that pass all the selection requirements at
both the detector and particle levels and $N^{\mathrm{part}}$ is the number
of MC events that pass the selection requirements at the particle
level. The integrated efficiency using \sher\ (\pyt) is found to be
$81.3\%$ ($81.5\%$) for the photon plus one-jet, $74.6\%$ ($75.3\%$)
for the photon plus two-jet and $70.2\%$ ($70.6\%$) for the photon
plus three-jet sample. 

The bin-to-bin efficiency is computed as 
$\varepsilon_i=N^{\mathrm{det,part}}_i/N^{\mathrm{part}}_i$, where 
$N^{\mathrm{det,part}}_i$ is the number of MC events that pass all the
selection requirements at both the detector and particle levels and
are generated and reconstructed in bin $i$, and $N^{\mathrm{part}}_i$ is
the number of MC events that pass the selection requirements at the
particle level and are generated in bin $i$. The bin-to-bin
efficiencies are typically above $50\%$, except for the $\ptjet$
observables ($\gtrsim 40\%$) due to the resolution in these steeply
falling distributions, and are similar for \sher\ and \pyt.

The bin-to-bin reconstruction purity is computed as 
$\kappa_i=N^{\mathrm{det,part}}_i/N^{\mathrm{det}}_i$, where $N^{\mathrm{det}}_i$ is
the number of MC events that pass the selection requirements at the
detector level and are reconstructed in bin $i$. The bin-to-bin
reconstruction purities are typically above $55\%$, except for
$\ptjet$ ($\gtrsim 40\%$) for the same reason as the bin-to-bin
efficiency, and are similar for \sher\ and \pyt.

\section{Cross-section measurement procedure}
\label{cor}

The cross sections, after background subtraction, are corrected to the
particle level using a bin-by-bin correction procedure. The bin-by-bin
correction factors are determined using the MC samples; these
correction factors take into account the efficiency of the selection
criteria and the jet and photon reconstruction, as well as migration
effects. The \sher\ samples are used to compute the nominal correction
factors to the cross sections and the \pyt\ samples are used to
estimate systematic uncertainties due to the modelling of the parton
shower, hadronisation and signal (see Section~\ref{syst}).

The cross sections are corrected to the particle level via the formula
\begin{equation}
\frac{{\mathrm{d}}\sigma}{{\mathrm{d}}A}(i)=\frac{N_A^{\mathrm{sig}}(i)\ C^{\mathrm{MC}}(i)}{{\cal L}\
  \Delta A(i)},
\end{equation}
where $({\mathrm{d}}\sigma/{\mathrm{d}}A)(i)$ is the cross section as a function
of observable $A$, $N_A^{\mathrm{sig}}(i)$ is the signal yield in bin $i$,
$C^{\mathrm{MC}}(i)$ is the correction factor in bin $i$, ${\cal L}$ is the
integrated luminosity and $\Delta A(i)$ is the width of bin $i$. The
correction factors are computed as

\begin{equation}
C^{\mathrm{MC}}(i)=\frac{N^{{\mathrm{S}}{\textsc{herpa}}}_{\mathrm{part}}(i)}{N^{{\mathrm{S}}{\textsc{herpa}}}_{\mathrm{det}}(i)},
\label{eqtwosher}
\end{equation}
where $N^{{\mathrm{S}}{\textsc{herpa}}}_{\mathrm{det\ (part)}}(i)$ is the number
of events in the \sher\ samples at detector (particle) level in bin $i$.

For the systematic uncertainties estimated with the \pyt\ samples, the
acceptance correction factors are computed as 

\begin{equation}
C^{\mathrm{MC}}(i)=\frac{\alpha\
    N^{{\mathrm{P}}{\textsc{ythia}},{\mathrm{H}}}_{\mathrm{part}}(i)+(1\!-\!\alpha)\ N^{{\mathrm{P}}{\textsc{ythia}},{\mathrm{B}}}_{\mathrm{part}}(i)}{\alpha\ N^{{\mathrm{P}}{\textsc{ythia}},{\mathrm{H}}}_{\mathrm{det}}(i)+ (1\!-\!\alpha)\
    N^{{\mathrm{P}}{\textsc{ythia}},{\mathrm{B}}}_{\mathrm{det}}(i)},
\label{eqtwop}
\end{equation}
where $\alpha$ is the value obtained from the fit to the data
distribution of each observable and 
$N^{{\mathrm{P}}{\textsc{ythia}}}_{\mathrm{det\ (part)}}(i)$ is the number of
events in the \pyt\ samples at detector (particle) level in bin
$i$. The indices H and B correspond to the hard and brem \pyt\
components, respectively. The correction factors from \pyt\ and \sher\
are very similar and differ from unity by typically $\lesssim
20\%$. The average correction factor for each distribution is listed
in Table~\ref{tabcmc}. The results of the bin-by-bin unfolding
procedure are checked with a Bayesian unfolding
method~\cite{nim:a362:487}, giving consistent results.

\begin{table}[h]
\setlength{\unitlength}{1.0cm}
\begin{picture} (18.0,3.0)
\put (-1.0,1.5){\scalebox{0.90}{
\begin{tabular}{|l|l|l|l|l|l|}\hline
Sample & \multicolumn{5}{c|}{Distribution: $C^{\mathrm{MC}}$ using \sher\ (\pyt)} \\ \cline{2-6}
\hline\hline
P1J  & $\etg$: $1.18$ ($1.17$)  & $\ptjetl$: $1.20$ ($1.17$) & & & \\
P1JM & $\mgjn$: $1.21$ ($1.18$) & $|\ctgjn|$: $1.16$ ($1.14$)  & & & \\
P2J  & $\etg$: $1.17$ ($1.15$) & $\ptjetsl$: $1.22$ ($1.15$) &  $\delphigsl$: $1.13$ ($1.11$) & $\delphijjlsl$: $1.13$ ($1.11$) & \\
P3J & $\etg$: $1.15$ ($1.13$) & $\ptjetssl$: $1.19$ ($1.18$) & $\delphigssl$: $1.11$ ($1.09$) & $\delphijjlssl$: $1.11$ ($1.09$) & $\delphijjslssl$: $1.12$ ($1.10$)\\ \hline
\end{tabular}
}}
\end{picture}
\caption{\label{tabcmc}
Overview of the average correction factor $C^{\mathrm{MC}}$ for each
distribution using the \sher\ and \pyt\ samples.
}
\end{table}

\section{Systematic uncertainties}
\label{syst}
Several sources of systematic uncertainty are investigated. These
sources include the photon energy scale and resolution, the jet energy
scale and resolution, the parton-shower and hadronisation model
dependence, the photon identification efficiency, the choice of
background control regions, the signal modelling and the
identification and isolation correlation in the background. Each
source is discussed below. An overview of the systematic uncertainties
in the cross sections is given in Table~\ref{tabsyst}.

\subsection{Photon energy scale and resolution}
Differences between the photon energy scale and resolution in data and
the simulations lead to systematic uncertainties. A total of 20
individual components~\cite{epj:c74:3071} influencing the energy
measurement of the photon are identified and varied within their
uncertainties to assess the overall uncertainty in the energy
measurement. These uncertainties are propagated through the analysis
separately to maintain the full information about the
correlations. The total relative photon energy-scale uncertainty is in
the range $(0.3$--$0.9)\%$ for $|\etag|<1.37$, $(1.3$--$2.4)\%$ for
$1.56<|\etag|<1.81$ and $(0.3$--$0.7)\%$ for $1.81<|\etag|<2.37$
depending on the photon transverse energy and whether the candidate is
classified as an unconverted or converted photon.

Similarly to the energy scale uncertainty, the energy resolution is
also influenced by different contributions (seven components), which
are also propagated through the analysis separately to maintain the
full information about the correlations. 

The systematic uncertainties in the measured cross sections due to the
effects mentioned above are estimated by varying each individual
source of uncertainty separately in the MC simulations and then added
in quadrature. The largest contribution arises from the uncertainty in
the gain of the second layer of the electromagnetic calorimeter. The
photon energy scale contributes an uncertainty in the cross section
measured as a function of $\etg$ of $\pm 1\%\ (\pm 4\%)$ at low (high)
$\etg$, and typically less than $\pm 2\%$ when measured with the jet
observables. The photon energy resolution contributes an uncertainty
in the measured cross sections of less than $\pm 1\%$ for all
observables.

\subsection{Jet energy scale and resolution}
The jet energy scale (JES) uncertainty contains a full treatment of
bin-to-bin correlations for systematic uncertainties. A total of 67
individual components~\cite{ATLAS-CONF-2015-037} influencing the
energy measurement of the jets are identified and varied within their
uncertainties to assess the overall uncertainty in the jet energy
measurement. These parameters are propagated through the analysis
separately to maintain the full information about the
correlations. The total relative jet energy-scale uncertainty is
$\lesssim\pm 3\%$ in the phase-space region of the measurements.

The jet energy resolution (JER) uncertainty accounts for the
differences between data and simulated events. The impact of the JER
uncertainty is estimated by smearing the MC simulated distributions
and comparing the smeared and non-smeared results.

The systematic uncertainties in the measured cross sections due to the
effects mentioned above are estimated by varying each individual
source of uncertainty separately in the MC simulations and then added
in quadrature. The major contributions arise from uncertainties in (a)
the electron and photon energy scale, which affect the in situ
corrections obtained from $\gamma+{\mathrm{jet}}$ and $Z+{\mathrm{jet}}$ events,
(b) the modelling of the ambient transverse energy used in the
subtraction of the underlying event and pile-up, and (c) the modelling
of the quark and gluon composition of the jets. The resulting
uncertainty due to the JES is the dominant effect on the measured
cross sections, except for those as functions of $\etg$. As an
example, the effect on the measured cross section as a function of
$\ptjetl$ is below $\pm 6\%$ for $\ptjetl<600$~\GeV\ and grows to
$\approx\pm 15\%$ for $\ptjetl\sim 1$~\TeV. The JER contributes an
uncertainty in the measured cross sections which is smaller than $\pm
1\%$ for the photon plus one-jet observables; for the photon plus
two-jet  and photon plus three-jet observables it is below $\pm 4\%$.

\subsection{Parton-shower and hadronisation model dependence} 
The difference between the signal purities and the correction factors
estimated in \sher\ and \pyt\ is taken as an estimate of the
systematic uncertainty due to the parton-shower and hadronisation
models. The resulting uncertainty in the measured cross sections is
below  $\pm 3\%$ for the photon plus one-jet measurements, except for
$\ptjetl$ (for which the uncertainty increases to $\pm 9\%$ for
$\ptjetl\sim 1$~\TeV), below $\pm 6\%$ for the photon plus two-jet
measurements, except for $\ptjetsl$ (for which the uncertainty
increases to $\pm 13\%$ for $\ptjetsl\sim 1$~\TeV), and below $\pm
7\%$ for the photon plus three-jet measurements.

\subsection{Photon identification efficiency}
Scale factors are applied to the MC events to match the ``tight''
identification efficiency between data and
simulation~\cite{1606.01813}. The uncertainty in the photon
identification is estimated by propagating the uncertainty in these
scale factors through the analysis. These effects result in an
uncertainty in the measured cross sections which is smaller than $\pm
0.4\%$ for all observables.

\subsection{Choice of background control regions} 
The estimation of the background contamination in the signal region is
affected by the choice of background control regions. The latter are
defined by the lower limit on $\etisod$ in regions $B$ and $D$ and the
choice of inverted photon identification variables used in the
selection of non-tight photons. To study the dependence on the
specific choices these definitions are varied over a wide range. The
lower limit on $\etisod$ in regions $B$ and $D$ is  varied by $\pm
1$~\GeV, which is larger than any difference between data and
simulations and still provides enough events to perform the
data-driven subtraction. Likewise, the choice of inverted photon
identification variables is varied. The analysis is repeated using
different sets of variables: tighter (looser) identification criteria
are defined by applying tight requirements to an extended (restricted)
set of shower-shape variables in the first calorimeter
layer~\cite{incpap}. The effects of these variations on the measured
cross sections are typically smaller than $\pm 1\%$ for all
observables.

\subsection{Signal modelling}

The simulation of the signal from the \pyt\ MC samples is used to
estimate the systematic uncertainties arising from the modelling of
the hard and bremsstrahlung components, which affect the signal
leakage fractions in the two-dimensional sideband method for
background subtraction and the bin-by-bin correction factors.

To estimate the effect of the signal modelling on the signal leakage
fractions, the \pyt\ components are first mixed according to the
default value given by the MC cross section to determine the signal
yield. The uncertainty related to the simulation of the hard and brem
components in the signal leakage fractions is estimated by performing
the background subtraction using the admixture derived from the
fit. For this estimation, the bin-by-bin correction factors are
computed using Eq.~(\ref{eqtwop}).

To estimate the effect of the signal modelling on the bin-by-bin
correction factors, the components in \pyt\ are mixed according to
Eq.~(\ref{eqtwop}) but using $\alpha\pm\Delta\alpha$, where
$\Delta\alpha$ is the error from the fit (see Section~\ref{fit}).

These effects result in an uncertainty in the measured cross sections
which is typically smaller than $\pm 1\%$ for all observables.

\subsection{Identification and isolation correlation in the
    background} 
The isolation and identification photon variables used to define the
plane in the two-dimensional sideband method to subtract the
background (see Section~\ref{bacsub}) are assumed to be uncorrelated
for background events ($R^{\mathrm{bg}}=1$). Any correlation between these
variables would affect the estimation of the purity of the signal and
lead to systematic uncertainties in the background-subtraction
procedure.  It was shown that $R^{\mathrm{bg}}$ is consistent with unity
within $\pm 10\%$~\cite{incpap}. Therefore, $\pm 10\%$ is taken as the
uncertainty in $R^{\mathrm{bg}}$ related to the identification and
isolation correlation in the background. These effects result in an
uncertainty in the measured cross sections which is typically smaller
than $\pm 1\%$ for all observables.

\subsection{Total systematic uncertainty}
The total systematic uncertainty is computed by adding in quadrature
the sources of uncertainty listed in the previous sections and the
statistical uncertainty of the MC samples. The uncertainty in the
integrated luminosity is $\pm 1.9\%$~\cite{1608.03953}; this
uncertainty is fully correlated in all bins of all the measured cross
sections and is added in quadrature to the other uncertainties. 

\begin{table}[h]
\setlength{\unitlength}{1.0cm}
\begin{picture} (18.0,12.0)
\put (0.0,6.0){\scalebox{0.90}{
\begin{tabular}{|l|c|c|c|c|c|c|c|}\hline
 & \multicolumn{7}{c|}{Variable} \\ \cline{2-8}
Source of & \multicolumn{3}{c|}{Photon plus one-jet} &  \multicolumn{2}{|c|}{Photon plus two-jet} & \multicolumn{2}{|c|}{Photon plus three-jet} \\ \cline{2-8}
uncertainty & $\etg$ & $\ptjetl$ & $|\ctgjn|$ & $\ptjetsl$ & $\Delta\phi$ & $\ptjetssl$ & $\Delta\phi$ \\ \hline\hline
Photon energy & & & & & & & \\
scale and     & $(1$--$4)\%$ & $(0$--$3.5)\%$ & $(1$--$1.4)\%$ & $(0$--$2.5)\%$ & $(0$--$2.4)\%$ & $(0$--$1.9)\%$ & $(0$--$1.7)\%$ \\
resolution    & & & & & & & \\ \hline
Jet energy    & & & & & & & \\
scale     & $(0$--$1.7)\%$ &  $(2.4$--$15)\%$ & $(1.8$--$2.3)\%$ & $(3.6$--$10)\%$ & $(1.8$--$9)\%$ & $(5.5$--$14)\%$ & $(4.5$--$11)\%$ \\
              & & & & & & & \\ \hline
Jet energy    & & & & & & & \\
resolution    & $(0$--$0.3)\%$ &  $(0.1$--$1.0)\%$ & $(0.1$--$0.4)\%$ & $(0.1$--$1.5)\%$ & $(0.2$--$2.0)\%$ & $(1.1$--$4.0)\%$ & $(0.1$--$2.5)\%$ \\
              & & & & & & & \\ \hline
Parton shower     & & & & & & & \\
and hadronisation & $(0$--$0.8)\%$ & $(1.1$--$9)\%$ & $(0.6$--$1.3)\%$ & $(1$--$13)\%$ & $(0.8$--$4.6)\%$ & $(2.3$--$5.6)\%$ & $(2.1$--$7)\%$ \\
models         & & & & & & & \\ \hline
Photon         & & & & & & & \\
identification & $(0$--$0.4)\%$ & $(0$--$0.4)\%$ & $(0$--$0.4)\%$ & $(0$--$0.4)\%$ & $(0$--$0.4)\%$ & $(0$--$0.4)\%$ & $(0$--$0.4)\%$ \\
               & & & & & & & \\ \hline
Background    & & & & & & & \\
control       & $(0$--$1)\%$ & $(0$--$1.1)\%$ & $(0$--$0.6)\%$ & $(0$--$1.2)\%$ & $(0$--$0.5)\%$ & $(0$--$1.9)\%$ & $(0$--$1)\%$ \\
regions       & & & & & & & \\ \hline
Signal        & & & & & & & \\
modelling     & $(0$--$0.1)\%$ & $(0$--$0.14)\%$ & $(0$--$0.4)\%$ & $(0$--$0.6)\%$ & $(0$--$0.7)\%$ & $(0$--$0.5)\%$ & $(0$--$1.2)\%$ \\
              & & & & & & & \\ \hline
Correlation   & & & & & & & \\
in background & $(0$--$0.8)\%$ & $(0$--$0.7)\%$ & $(0$--$0.9)\%$ & $(0$--$0.6)\%$ & $(0$--$0.6)\%$ & $(0$--$0.6)\%$ & $(0$--$0.5)\%$ \\
              & & & & & & & \\ \hline
\end{tabular}
}}
\end{picture}
\caption{\label{tabsyst}
Overview of the relative systematic uncertainties in the cross sections.
}
\end{table}

\section{Fixed-order QCD calculations}
\label{nlo}
The measurements are compared to the highest fixed-order pQCD
prediction available for each final state. The details of the
calculations are given below.

\subsection{Calculations for photon plus one-jet final state}
The LO and NLO QCD calculations used in the photon plus one-jet
analysis presented here are performed using the program \jetp~1.3.2~\cite{jhep:0205:028,pr:d73:094007}. This program
includes a full NLO calculation of both the direct and fragmentation
QCD contributions to the cross section for the
$pp\rightarrow\gamma+{\mathrm{jet}}+{\mathrm{X}}$ reaction.

The calculation assumes five massless quark flavours. The
renormalisation ($\mu_{\mathrm{R}}$), factorisation ($\mu_{\mathrm{F}}$) and
fragmentation ($\mu_{\mathrm{f}}$) scales are chosen to be $\mu_{\mathrm{R}}=\mu_{\mathrm{F}}=\mu_{\mathrm{f}}=\etg$. The calculations are performed
using the CT10 parameterisations of the proton PDF and the NLO BFG set
II photon fragmentation function~\cite{epj:c2:529}. The strong
coupling constant is calculated at two loops with $\alpha_{\mathrm{s}}(m_Z)=0.118$.

The calculations are performed using a parton-level isolation
criterion which requires the total transverse energy from the partons
inside a cone of $\Delta R=0.4$ around the photon direction, called
cone isolation henceforth, to be below $10$~\GeV. The anti-$\kt$
algorithm with radius parameter $R=0.6$ is applied to the partons in
the events generated by this program to compute the cross-section
predictions.

\subsection{Calculations for photon plus two-jet
    and photon plus three-jet final states}

NLO QCD calculations are performed separately for photon plus two-jet
and photon plus three-jet final states using the program \blh+\sher~\cite{pr:d78:036003,prl:109:042001}. This program
includes a full NLO QCD calculation of only the direct contribution to
the cross section for the $pp\rightarrow\gamma+2\ {\mathrm{jets}}+{\mathrm{X}}$
and $pp\rightarrow\gamma+3\ {\mathrm{jets}}+{\mathrm{X}}$ reactions. Therefore,
the highest-order calculation used in this paper corresponds to that
of photon plus three-jet production and it is up to ${\cal
  O}(\alpha_{\mathrm{em}}\alpha_{\mathrm{s}}^4)$. The $\mu_{\mathrm{R}}$ and
$\mu_{\mathrm{F}}$ scales are chosen to be $\mu_{\mathrm{R}}=\mu_{\mathrm{F}}=\etg$. The settings for the number of flavours, $\alpha_{\mathrm{s}}(m_Z)$ and proton PDF are the same as for \jetp. The
calculations are performed using a parton-level isolation on the
photon based on the Frixione method~\cite{pl:b429:369}, called
Frixione isolation henceforth. As with \jetp, the anti-$\kt$
algorithm with radius parameter $R=0.6$ is applied to the final-state
partons.

\subsection{Hadronisation and underlying-event corrections to the
  NLO QCD calculations}
Since the measurements refer to jets of hadrons and include
underlying-event (UE) effects, whereas the NLO QCD calculations refer
to jets of partons without such effects, the cross-section predictions
are corrected to include UE effects at particle level using the MC
models. The correction factor, $C_{\mathrm{NLO}}$, is defined as the ratio
of the cross section for jets of hadrons with UE to that for jets of
partons. The correction factors for the photon plus one-jet
predictions are estimated using the \pyt\ samples (using cone
isolation) and those for the photon plus two/three-jet predictions are
estimated using the \sher\ samples; in the latter case, the cone
(Frixione) isolation  is used at the particle (parton) level to match
the measurements (predictions). The MC samples of \pyt\ (\sher) are
suited for estimating the correction factors for \jetp\ (\blh) since these NLO QCD calculations include (do not include)
the fragmentation contribution. These factors are close to unity for
the photon plus one-jet observables, except for $\ptjetl\gtrsim
500$~\GeV, where they can differ by up to $30\%$ from unity due to the
dominance of the bremsstrahlung component in that region. For photon
plus two-jet (three-jet) observables the average correction factor is
$1.10$ ($1.14$).

\subsection{Theoretical uncertainties}
The following sources of uncertainty in the theoretical predictions
are considered: 
\begin{itemize}
\item The uncertainty due to the scales is estimated by repeating the
  calculations using values of $\mu_{\mathrm{R}}$ and $\mu_{\mathrm{F}}$ scaled
  by factors $0.5$ and $2$. The two scales are varied individually. In
  the case of photon plus one-jet calculations, the $\mu_{\mathrm{f}}$
  scale is also varied.
\item The uncertainty due to the proton PDF is estimated by repeating
  the calculations using the 52 additional sets from the CT10 error
  analysis and taking the sum in quadrature of all the uncertainty
  components. The scaling factor of $1/1.645$ is applied to convert
  the $90\%$ confidence-level (CL) interval as provided in
  Ref.~\cite{pr:d82:074024} to a $68\%$ CL interval.
\item The uncertainty due to the value of $\alpha_{\mathrm{s}}(m_Z)$ is
  estimated by repeating the calculations using two additional sets of
  proton PDFs, for which different values of $\alpha_{\mathrm{s}}(m_Z)$ are
  assumed in the fits, namely $\alpha_{\mathrm{s}}(m_Z)=0.116$ and
  $0.120$. In addition, the same scaling factor mentioned above is
  also applied to obtain the uncertainty for the $68\%$ CL interval. 
\item The uncertainty on the hadronisation and underlying-event
  corrections is negligible compared to the other uncertainties on the
  theoretical predictions~\cite{np:b875:483}.
\end{itemize}

The dominant theoretical uncertainty is that arising from the scale
variations. The total theoretical uncertainty is obtained by adding in
quadrature the individual uncertainties listed above.

\section{Results}
\label{result}

\subsection{Fiducial regions and integrated cross sections}
\label{fiducial}
The measurements presented here refer to isolated prompt photons with
$\etisop<10$~\GeV\ (see Section~\ref{mc}) and jets of hadrons (see
Section~\ref{jetsel}). The details of the phase-space regions are
given in Table~\ref{tabzero}. The integrated cross sections for the
photon plus one-jet, photon plus two-jet and photon plus three-jet
final states are shown in Table~\ref{tabint}. The measured and
predicted integrated cross sections are consistent within the
experimental and theoretical uncertainties.

\begin{table}
\begin{center}
\begin{tabular}{|l||r @{ $\pm$ } l|r @{}l |r @{}l |r @{}l |}\hline
& \multicolumn{2}{c|}{} &  \multicolumn{2}{c|}{} &   \multicolumn{2}{c|}{} &  \multicolumn{2}{c|}{} \\
 & \multicolumn{2}{c|}{Measured}  & \multicolumn{2}{c|}{NLO QCD} &   \multicolumn{2}{c|}{\pyt} & 
             \multicolumn{2}{c|}{\sher} \\
Final state &  \multicolumn{2}{c|}{cross}&  \multicolumn{2}{c|}{prediction} &   \multicolumn{2}{c|}{prediction} &
            \multicolumn{2}{c|}{prediction} \\
           &  \multicolumn{2}{c|}{section} &  \multicolumn{2}{c|}{\jetp /} &   \multicolumn{2}{c|}{[pb]} &  \multicolumn{2}{c|}{[pb]} \\ 
           &  \multicolumn{2}{c|}{[pb]} &   \multicolumn{2}{c|}{\blh} &   \multicolumn{2}{c|}{} &  \multicolumn{2}{c|}{}\\ 
&  \multicolumn{2}{c|}{} &  \multicolumn{2}{c|}{[pb]} &   \multicolumn{2}{c|}{} &  \multicolumn{2}{c|}{}\\
\hline\hline
&  \multicolumn{2}{c|}{} &  \multicolumn{2}{c|}{} &   \multicolumn{2}{c|}{} &  \multicolumn{2}{c|}{}\\
Photon plus one-jet & $134$ & $4$ & $128$ & $_{-9}^{+11}$ (J) & $120$ & & $132$ & \\ 
&  \multicolumn{2}{c|}{} &  \multicolumn{2}{c|}{} &   \multicolumn{2}{c|}{} &  \multicolumn{2}{c|}{}\\
\hline
&  \multicolumn{2}{c|}{} &  \multicolumn{2}{c|}{} &   \multicolumn{2}{c|}{} &  \multicolumn{2}{c|}{}\\
Photon plus two-jet & $30.4$ & $1.8$ & $29.2$ & $_{-2.7}^{+2.8}$ (B) & $26$ & $.4$ & $27$ & $.4$ \\ 
&  \multicolumn{2}{c|}{} &  \multicolumn{2}{c|}{} &   \multicolumn{2}{c|}{} &  \multicolumn{2}{c|}{}\\
\hline
&  \multicolumn{2}{c|}{} &  \multicolumn{2}{c|}{} &   \multicolumn{2}{c|}{} &  \multicolumn{2}{c|}{}\\
Photon plus three-jet & $8.7$ & $0.8$ & $9.5$ & $_{-1.2}^{+0.9}$ (B) & $8$& $.2$ & $7$ & $.9$ \\ 
&  \multicolumn{2}{c|}{}&  \multicolumn{2}{c|}{}&   \multicolumn{2}{c|}{} &  \multicolumn{2}{c|}{}\\
\hline
\end{tabular}
\end{center}
\caption{
Measured and predicted integrated cross sections.
}
\label{tabint}
\end{table}

\subsection{Cross sections for isolated-photon plus
  one-jet production}

\begin{figure}[p]
\setlength{\unitlength}{1.0cm}
\begin{picture} (18.0,15.0)
\put (0.0,8.0){\includegraphics[width=9cm,height=9cm]{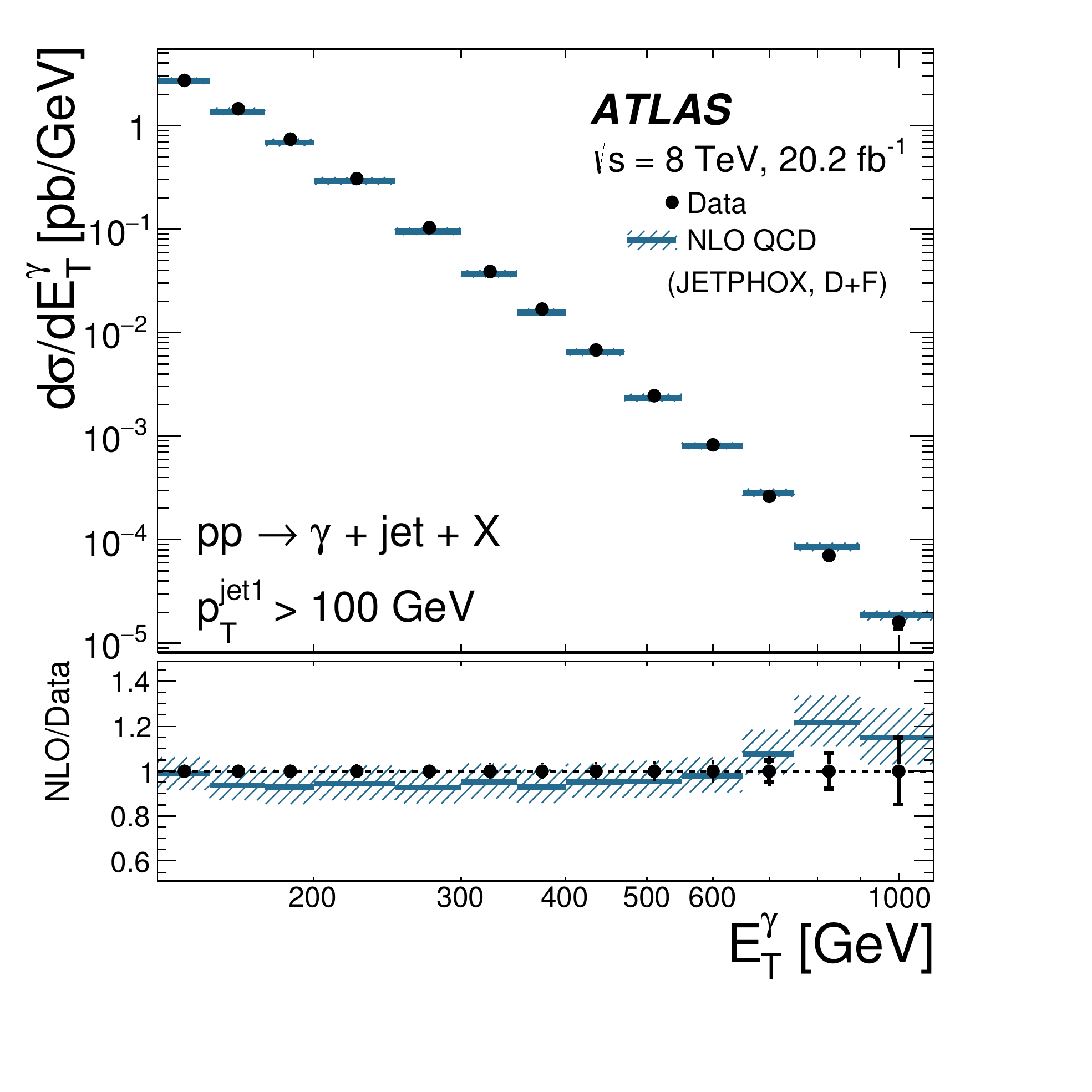}}
\put (8.0,8.0){\includegraphics[width=9cm,height=9cm]{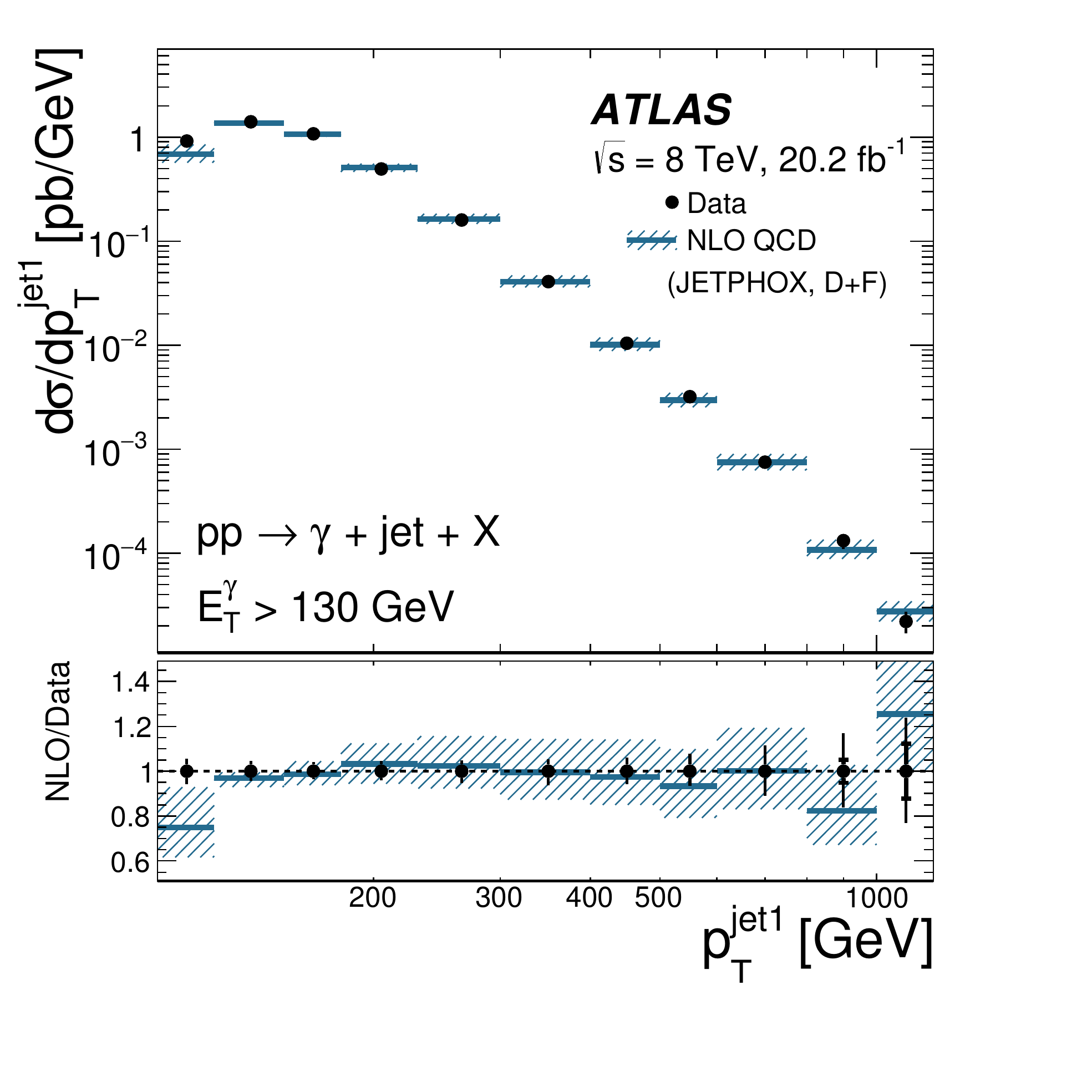}}
\put (0.0,0.0){\includegraphics[width=9cm,height=9cm]{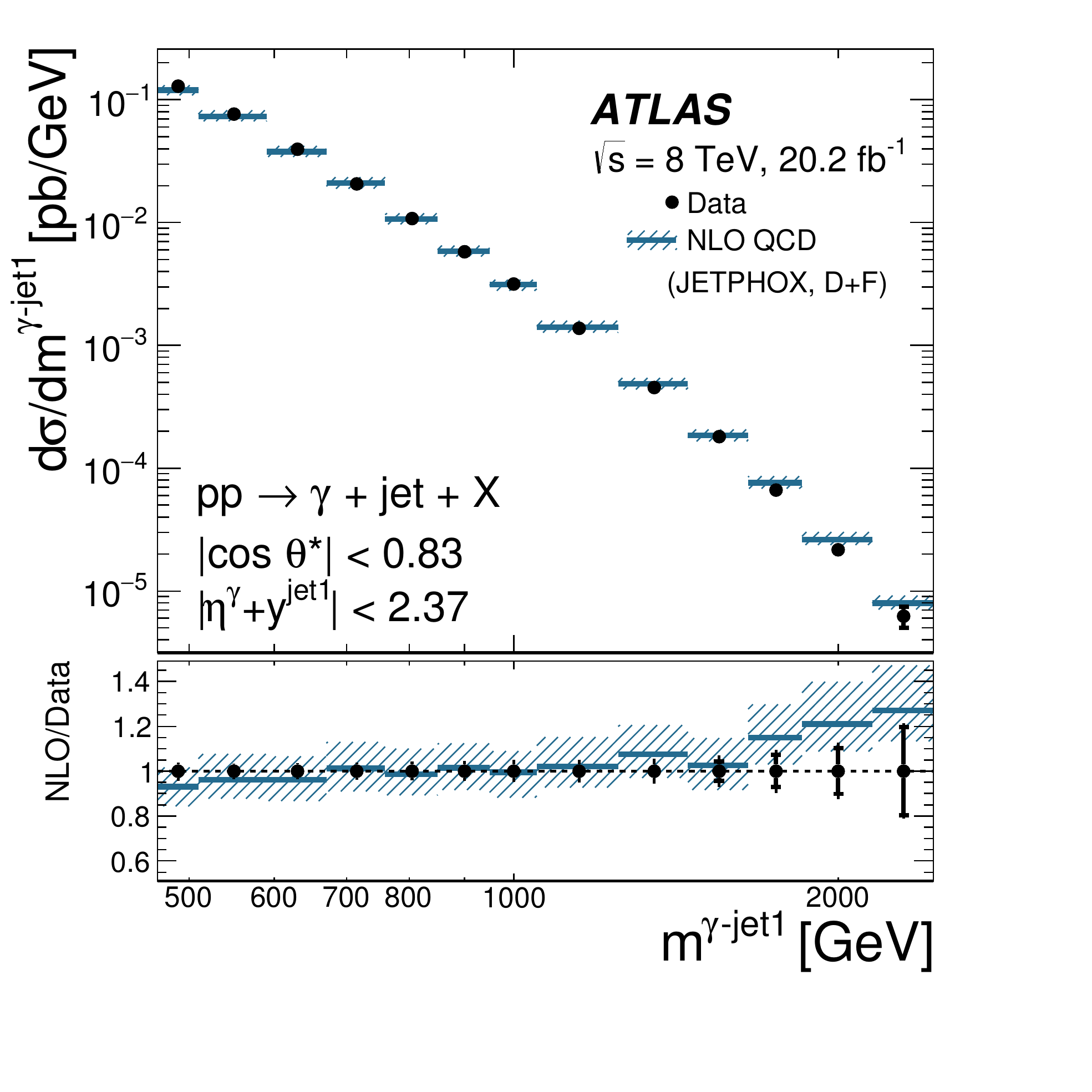}}
\put (8.0,0.0){\includegraphics[width=9cm,height=9cm]{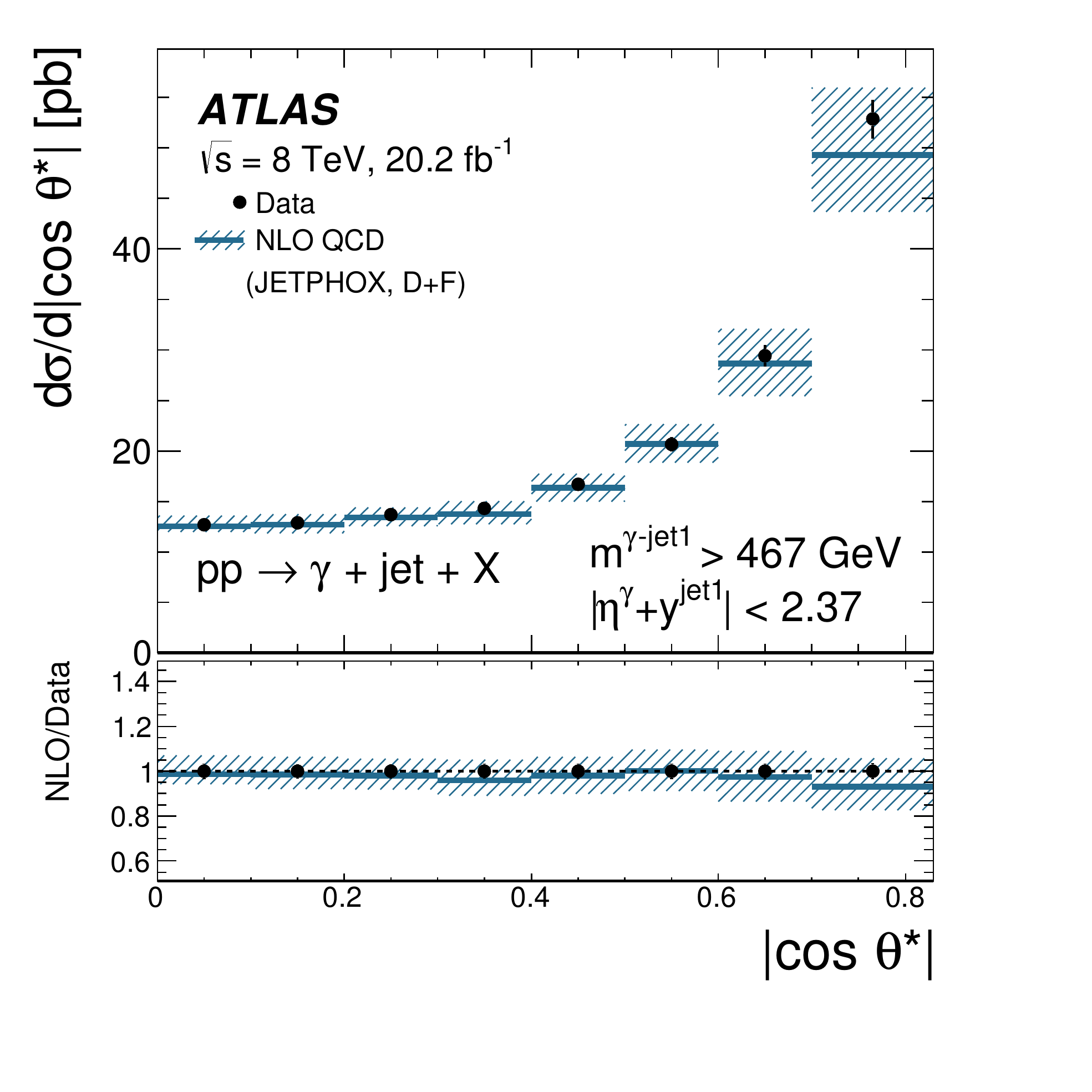}}
\put (3.7,9.0){{\textbf{\small (a)}}}
\put (11.7,9.0){{\textbf{\small (b)}}}
\put (3.7,1.0){{\textbf{\small (c)}}}
\put (11.7,1.0){{\textbf{\small (d)}}}
\end{picture}
\caption
{
  Measured cross sections for isolated-photon plus one-jet production
  (dots) as functions of (a) $\etg$, (b) $\ptjetl$, (c) $\mgjn$ and
  (d) $|\ctgjn|$. The NLO QCD predictions from \jetp\ corrected
  for hadronisation and underlying-event effects and using the CT10
  PDF set (solid lines) are also shown. These predictions include
  direct and fragmentation contributions (D+F). The bottom part of
  each figure shows the ratio of the NLO QCD prediction to the
  measured cross section. The inner (outer) error bars represent the
  statistical uncertainties (the statistical and systematic
  uncertainties added in quadrature) and the shaded band represents
  the theoretical uncertainty. For most of the points, the inner error
  bars are smaller than the marker size and, thus, not visible. 
  The cross sections in (c) and (d) include additional requirements on
  $|\etag+\rapjetl|$, $|\ctgjn|$ and $\mgjn$ (see Table~\ref{tabzero}).
}
\label{fig175}
\end{figure}

\begin{figure}[p]
\setlength{\unitlength}{1.0cm}
\begin{picture} (18.0,15.0)
\put (0.0,8.0){\includegraphics[width=9cm,height=9cm]{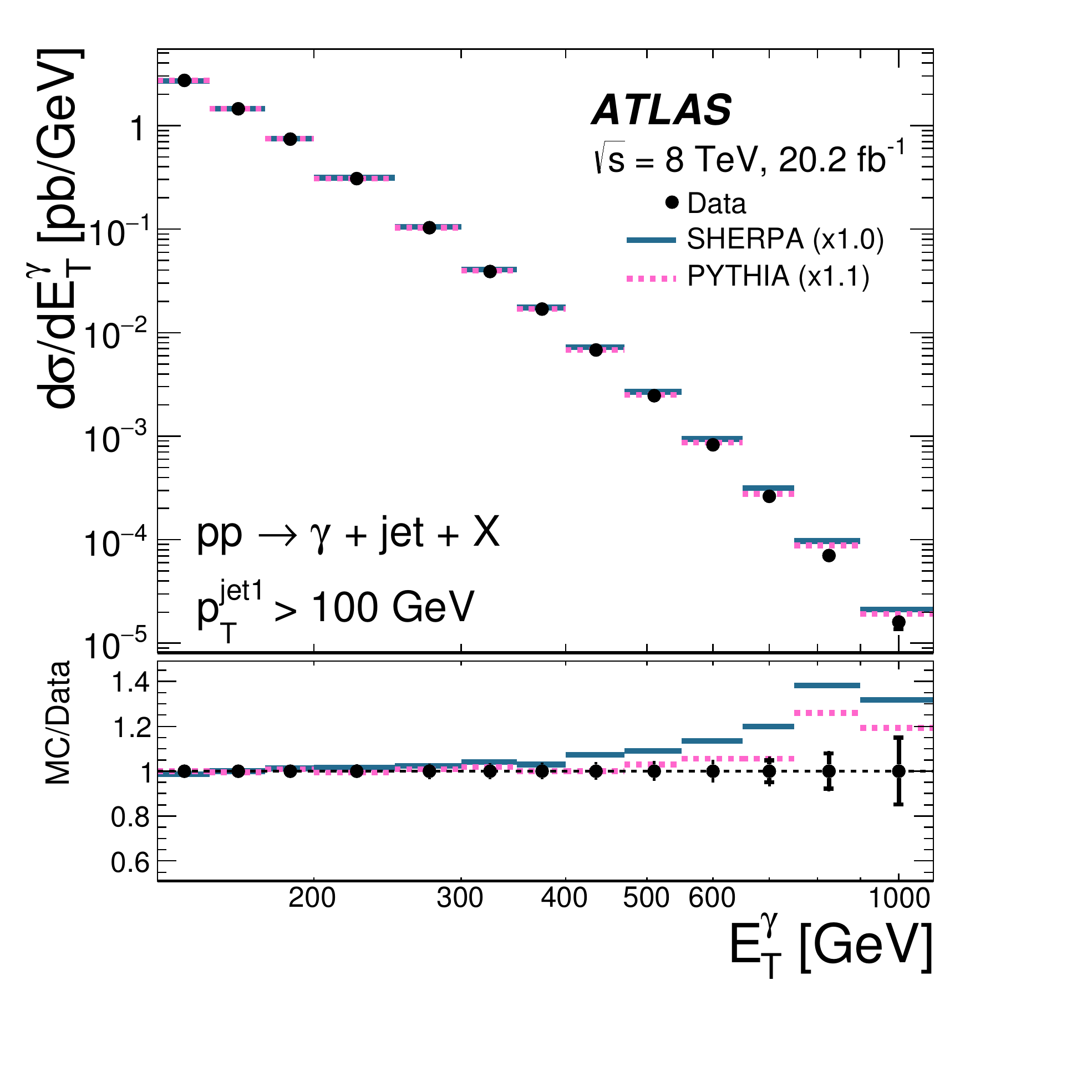}}
\put (8.0,8.0){\includegraphics[width=9cm,height=9cm]{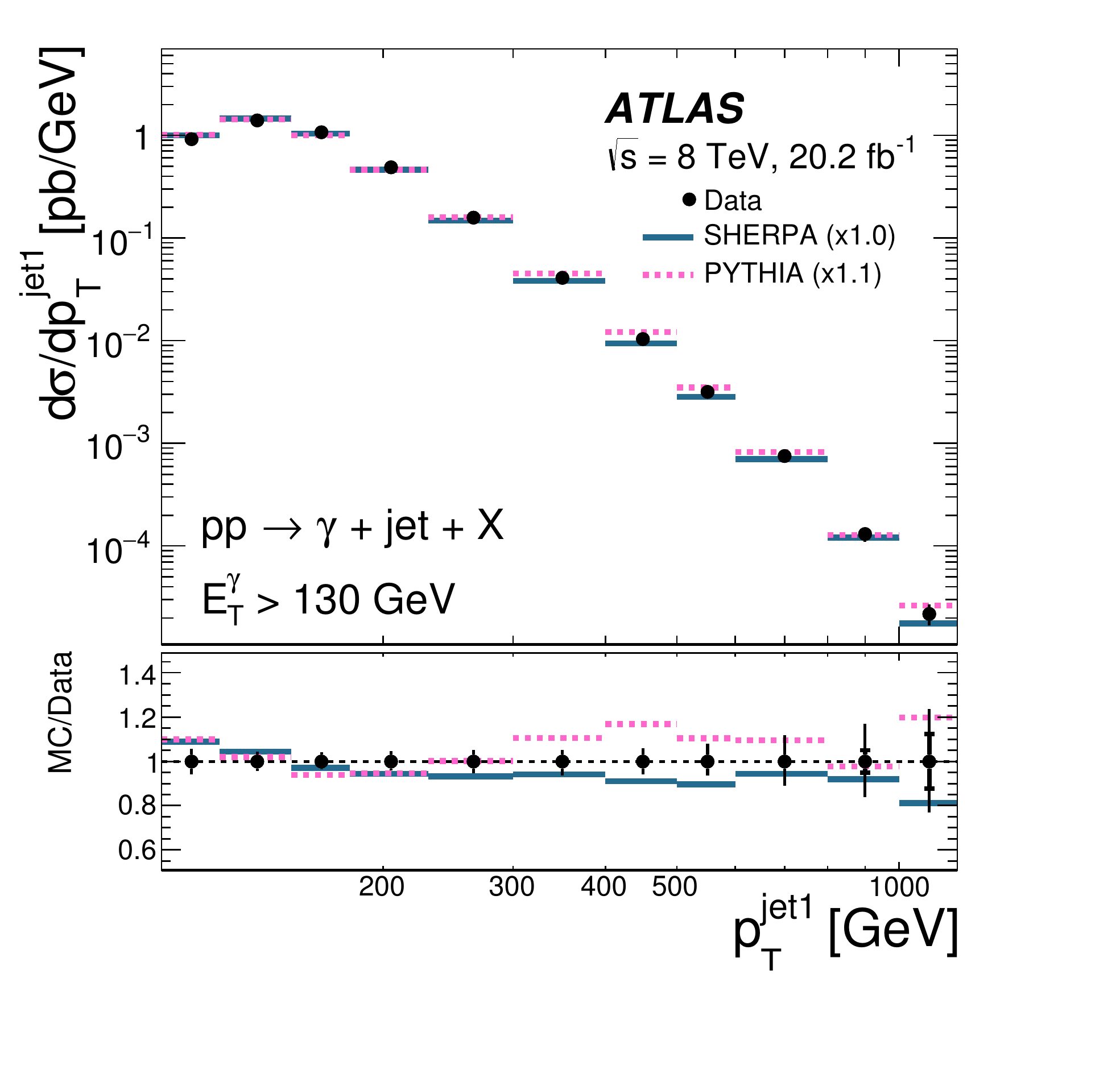}}
\put (0.0,0.0){\includegraphics[width=9cm,height=9cm]{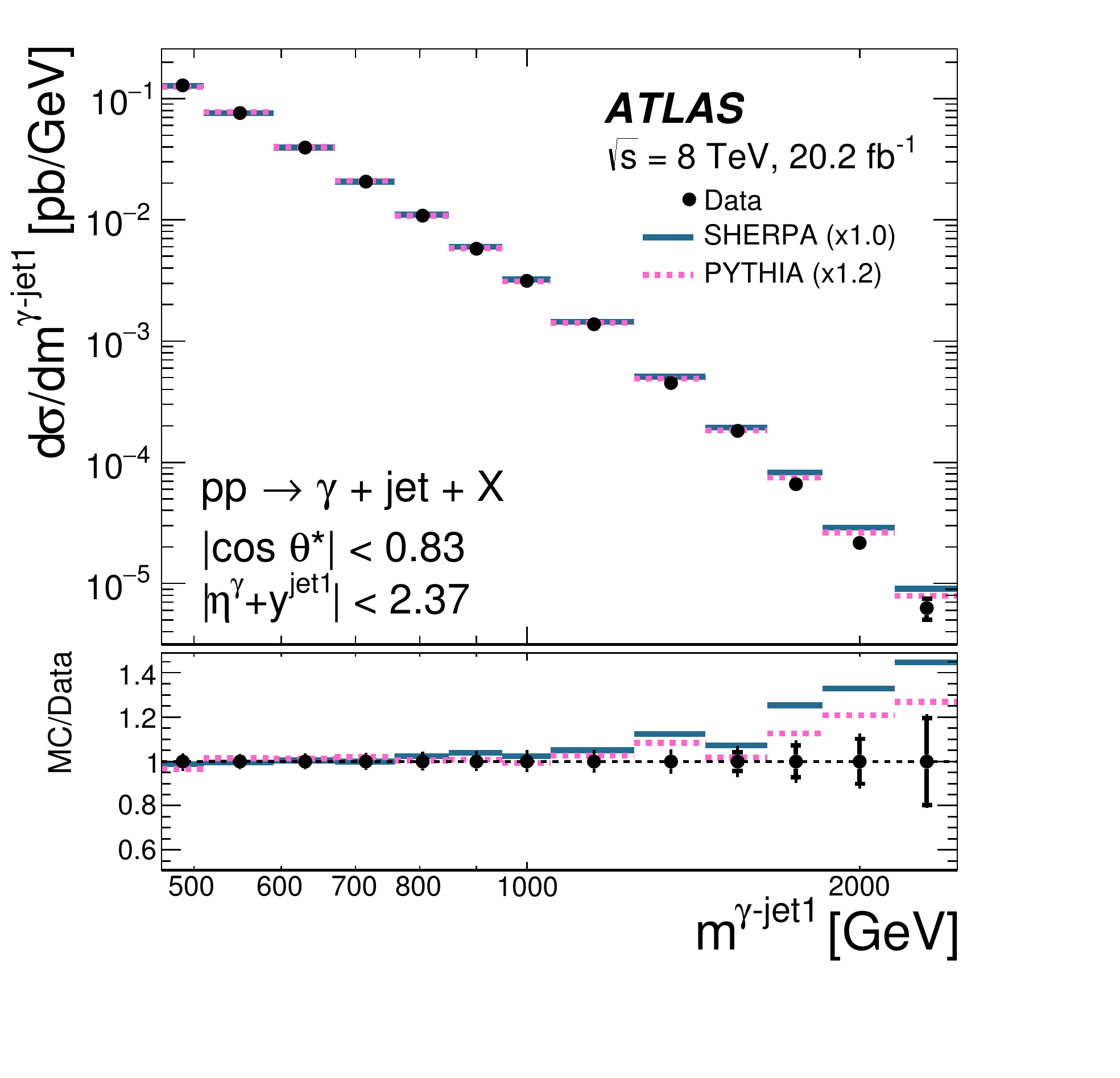}}
\put (8.0,0.0){\includegraphics[width=9cm,height=9cm]{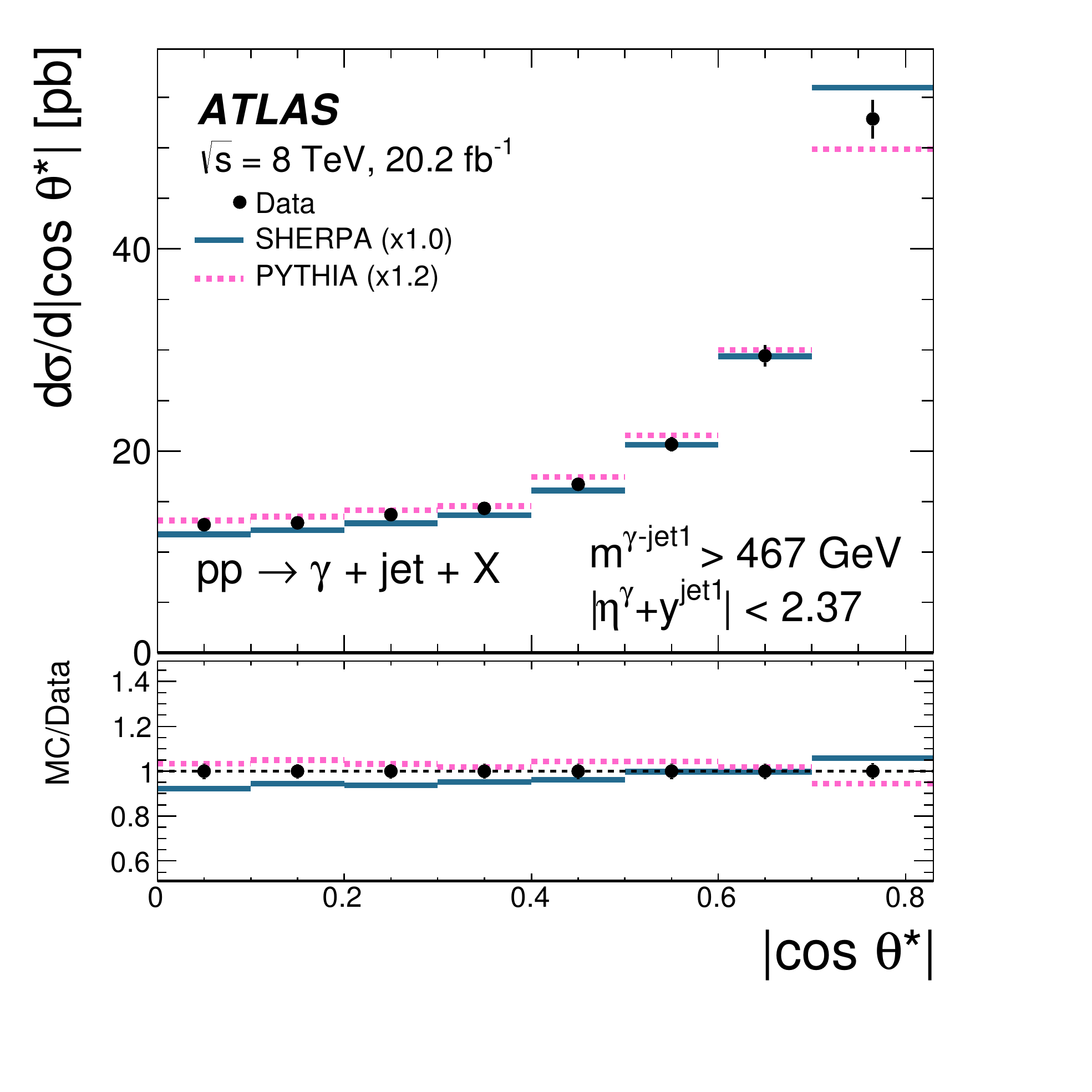}}
\put (3.7,9.0){{\textbf{\small (a)}}}
\put (11.7,9.0){{\textbf{\small (b)}}}
\put (3.7,1.0){{\textbf{\small (c)}}}
\put (11.7,1.0){{\textbf{\small (d)}}}
\end{picture}
\caption
{
  Measured cross sections for isolated-photon plus one-jet production
  (dots) as functions of (a) $\etg$, (b) $\ptjetl$, (c) $\mgjn$ and
  (d) $|\ctgjn|$, presented in Figure~\ref{fig175}. For comparison,
  the predictions from \sher\  (solid lines) and \pyt\ (dashed lines)
  normalised to the integrated measured cross sections (using the
  factors indicated in parentheses) are also shown. The bottom part of
  each figure shows the ratios of the MC predictions to the measured
  cross section. The inner (outer) error bars represent the
  statistical uncertainties (the statistical and systematic
  uncertainties added in quadrature). For most of the points, the
  inner error bars are smaller than the marker size and, thus, not
  visible.
  The cross sections in (c) and (d) include additional requirements on
  $|\etag+\rapjetl|$, $|\ctgjn|$ and $\mgjn$ (see Table~\ref{tabzero}).
}
\label{fig175b}
\end{figure}

The measured cross-section ${\mathrm{d}}\sigma/{\mathrm{d}}\etg$, shown in
Figure~\ref{fig175}(a), decreases by five orders of magnitude as
$\etg$ increases over the measured range. Values of $\etg$ up to
$1.1$~\TeV\ are measured. The experimental uncertainty is below $5\%$
for $\etg\lesssim 650$~\GeV, dominated by the photon energy scale
uncertainty, and grows to $15\%$ at $\etg\sim 1$~\TeV, dominated by
the statistical uncertainty in this region. The NLO QCD prediction
from \jetp\ is compared with the measurement in
Figure~\ref{fig175}(a). The NLO QCD prediction gives a good
description of the data within the experimental and theoretical
uncertainties. The theoretical uncertainty varies from $\approx 7\%$
for $\etg\sim 130$~\GeV\ to $\approx 10\%$ for $\etg\sim 1$~\TeV; it
is dominated by the contribution arising from scale uncertainties, in
particular from the variation of $\mu_{\mathrm{R}}$ ($7\%\ (5\%)$ at low
(high) $\etg$), although for $\etg\gtrsim 750$~\GeV\ the uncertainty
from the PDF grows to be of the same order and dominates for higher
$\etg$ values ($\approx 8\%$ for $\etg\sim 1$~\TeV). The predictions
from \sher\ and \pyt\ are compared with the measurements in
Figure~\ref{fig175b}(a). Both predictions give an adequate description
of the shape of the data distribution within the experimental and
theoretical uncertainties; the theoretical uncertainties are
necessarily at least as large as for the NLO QCD calculations.

The measured cross-section ${\mathrm{d}}\sigma/{\mathrm{d}}\ptjetl$, shown in
Figure~\ref{fig175}(b), decreases by five orders of magnitude from
$\ptjetl\sim 120$~\GeV\ to the highest transverse momentum available,
$\ptjetl\approx 1.2$~\TeV; for $\ptjetl<120$~\GeV\ the cross section
decreases due to the kinematic analysis requirements. The total
experimental uncertainty is below $6\%$ for $\ptjetl<500$~\GeV\ and
grows to $\approx 25\%$ for $\ptjetl\sim 1.1$~\TeV. It is dominated by
the uncertainty in the jet energy scale. The NLO QCD prediction gives
a good description of the data except for $\ptjetl<120$~\GeV, where in
the calculation of $A\cdot\alpha_{\mathrm{em}}\alpha_{\mathrm{s}}+B\cdot\alpha_{\mathrm{em}}\alpha_{\mathrm{s}}^2$~\cite{jhep:0205:028,pr:d73:094007} the Born term is zero,
i.e. $A=0$. The theoretical uncertainty grows from $<5\%$ at
$\ptjetl\sim 135$~\GeV\ to $\approx 25\%$ for $\ptjetl\sim 1.1$~\TeV\
and is dominated by the variation of $\mu_{\mathrm{R}}$ in the whole
measured range. The predictions from \sher\ and \pyt\ give an adequate
description of the data (see Figure~\ref{fig175b}(b)).

Figure~\ref{fig175}(c) shows ${\mathrm{d}}\sigma/{\mathrm{d}}\mgjn$; the
measured cross section decreases by four orders of magnitude as
$\mgjn$ increases from about $0.5$~\TeV\ to the highest measured
value, $\approx 2.45$~\TeV. The experimental uncertainty ranges from
$\approx 3\%$ to $\approx 22\%$ and is dominated by the jet energy
scale uncertainty in most of the measured range; for $\mgjn>1.5$~\TeV\
the statistical uncertainty dominates. The NLO QCD calculation gives a
good description of the data and no significant deviation from the
prediction from pQCD is observed. The theoretical uncertainty is
$\approx 10\%\ (15\%)$ at $\mgjn\approx 490\ (2450)$~\GeV; it is
dominated by the contribution arising from scale uncertainties, in
particular from the variation of $\mu_{\mathrm{R}}$ ($\approx 10\%$),
although for $\mgjn\gtrsim 2.15$~\TeV\ the uncertainty  from the PDF
grows to be of the same order and dominates  for higher $\mgjn$
values. The predictions from \pyt\ and \sher\ give a good description
of the data (see Figure~\ref{fig175b}(c)), except for
$\mgjn>1.8$~\TeV\ where, nevertheless, the differences are covered by
the theoretical uncertainties.

The measured cross-section ${\mathrm{d}}\sigma/{\mathrm{d}}|\ctgjn|$, shown in
Figure~\ref{fig175}(d), increases as $|\ctgjn|$ increases. The
experimental uncertainty is $\approx 3\%$; the only significant
contributions arise from the photon and jet energy scale uncertainties
and the model dependence. The NLO QCD prediction gives a good
description of the data.  The theoretical uncertainty is $\approx
10\%$, dominated by the contribution arising from scale uncertainties,
in particular from the variation of $\mu_{\mathrm{R}}$. The predictions
from \pyt\ and \sher\ give a good description of the data (see
Figure~\ref{fig175b}(d)).

\begin{figure}[p]
\setlength{\unitlength}{1.0cm}
\begin{picture} (18.0,14.0)
\put (0.0,0.0){\centerline{\includegraphics[width=15cm,height=15cm]{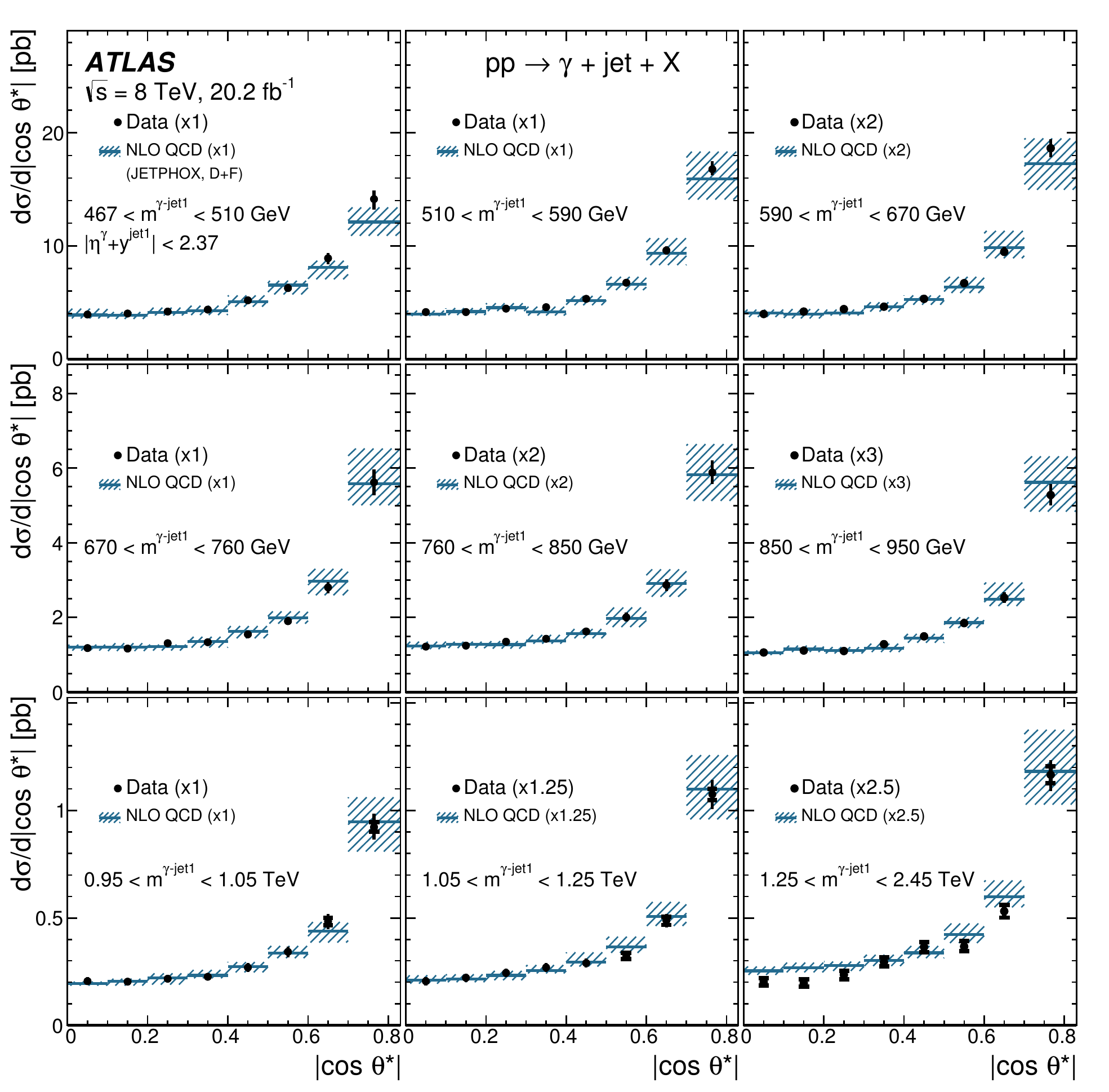}}}
\end{picture}
\caption
{
  Measured cross sections for isolated-photon plus one-jet production
  (dots) as functions of $|\ctgjn|$ in different regions of
  $\mgjn$. The NLO QCD predictions from \jetp\ corrected for
  hadronisation and underlying-event effects and using the CT10 PDF
  set (solid lines) are also shown. These predictions include direct
  and fragmentation contributions (D+F). The inner (outer) error bars
  represent the statistical uncertainties (the statistical and
  systematic uncertainties added in quadrature) and the shaded band
  represents the theoretical uncertainty. For most of the points, the
  inner error bars are smaller than the marker size and, thus, not
  visible. For visibility, the measured and predicted cross sections
  are scaled by the factors indicated in parentheses.
}
\label{fig176}
\end{figure}

\begin{figure}[p]
\setlength{\unitlength}{1.0cm}
\begin{picture} (18.0,14.0)
\put (0.0,0.0){\centerline{\includegraphics[width=15cm,height=15cm]{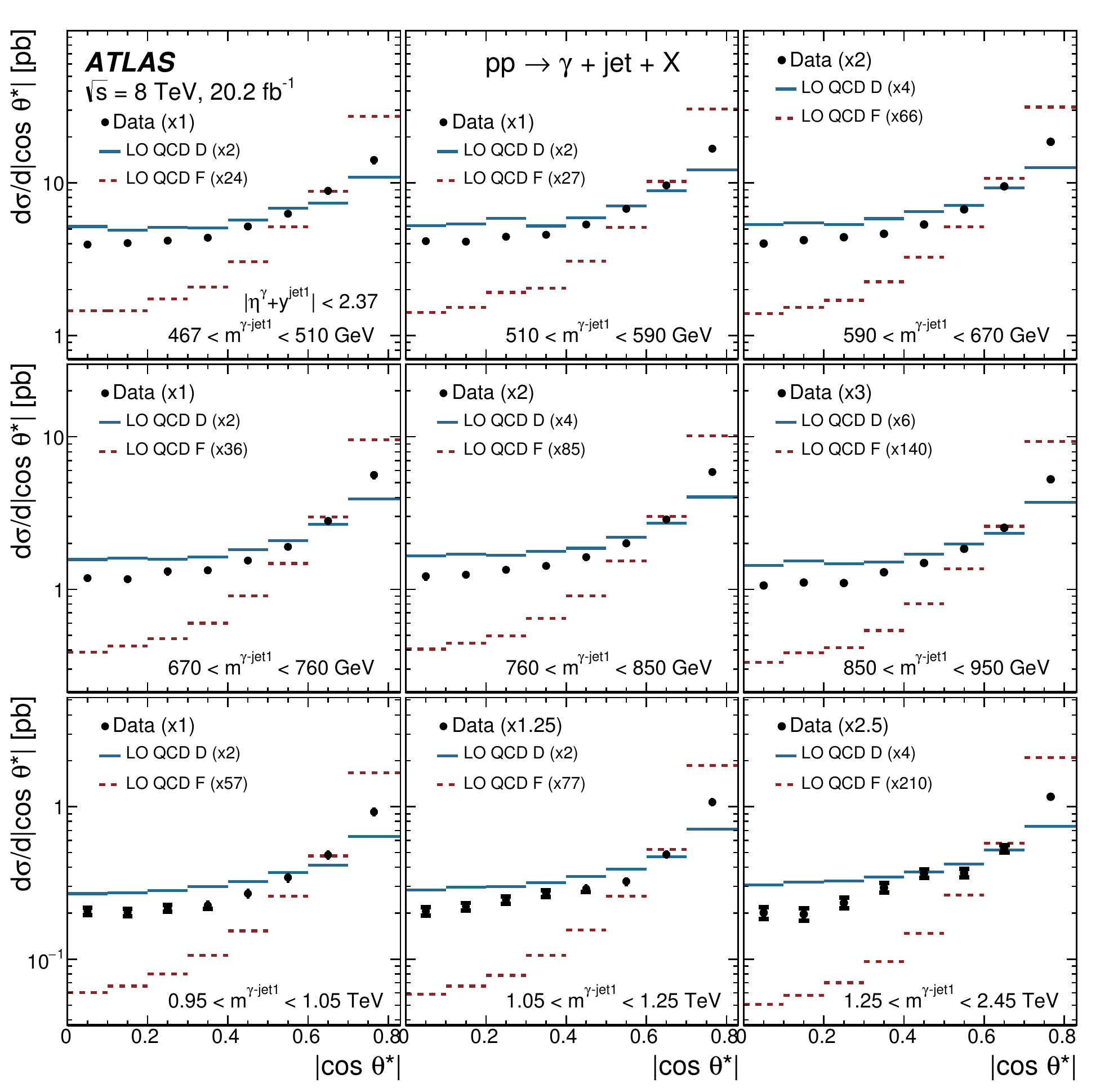}}}
\end{picture}
\caption
{
  Measured cross sections for isolated-photon plus one-jet production
  (dots) as functions of $|\ctgjn|$ in different regions of $\mgjn$,
  presented in Figure~\ref{fig176}. For visibility, the measured cross
  sections  are scaled by the factors indicated in parentheses. For
  comparison, the LO QCD predictions from \jetp\ corrected for
  hadronisation and underlying-event effects and using the CT10 PDF
  set for direct (solid lines) and fragmentation (dashed lines)
  processes are shown separately. In each region of $\mgjn$, the
  predictions are normalised to the integrated measured cross section
  by the factors shown in parentheses, which include the visibility
  factor. The inner (outer) error bars represent the statistical
  uncertainties (the statistical and systematic uncertainties added in
  quadrature). For most of the points, the inner error bars are
  smaller than the marker size and, thus, not visible.  
}
\label{fig176dp}
\end{figure}

\begin{figure}[p]
\setlength{\unitlength}{1.0cm}
\begin{picture} (18.0,14.0)
\put (0.0,0.0){\centerline{\includegraphics[width=15cm,height=15cm]{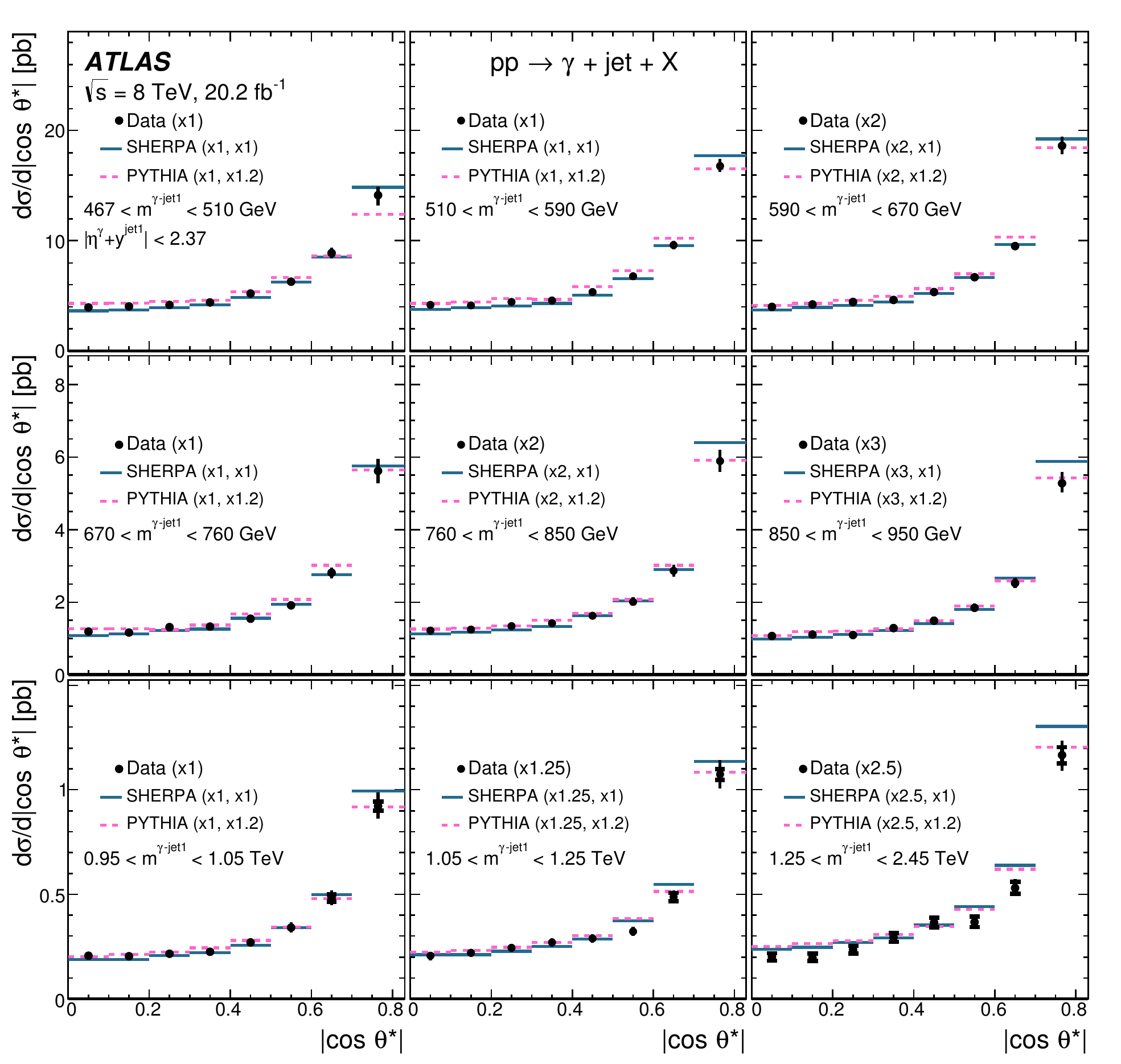}}}
\end{picture}
\caption
{
  Measured cross sections for isolated-photon plus one-jet production
  (dots) as functions of $|\ctgjn|$ in different regions of $\mgjn$,
  presented in Figure~\ref{fig176}. For comparison, the predictions
  from \sher\  (solid lines) and \pyt\ (dashed lines) are also shown;
  the predictions are normalised to the data by a global factor, which
  is shown as the second factor in parentheses. In addition, for
  visibility, the measured and predicted cross sections are scaled by
  the first factor indicated in parentheses. The inner (outer) error
  bars represent the statistical uncertainties (the statistical and
  systematic uncertainties added in quadrature). For most of the
  points, the inner error bars are smaller than the marker size and,
  thus, not visible.
}
\label{fig176b}
\end{figure}

To gain further insight into the dynamics of the photon--jet system,
cross sections are measured as functions of $|\ctgjn|$ in different
regions of $\mgjn$. Figure~\ref{fig176} shows the measured cross
sections and NLO QCD predictions in nine regions of $\mgjn$. The NLO
QCD predictions describe well the scale evolution of the measured
cross sections. The LO QCD predictions of the direct and fragmentation
contributions to the cross section are compared with the measurements
in Figure~\ref{fig176dp}. Even though at NLO the two components are no
longer distinguishable, the LO calculations are useful in illustrating
the basic differences in the dynamics of the two processes. The
contribution from fragmentation shows a steeper increase as
$|\ctgjn|\rightarrow 1$ than that from direct processes. This
different behaviour is due to the different spin of the exchanged
particle dominating each of the processes: a quark in the case of
direct processes and a gluon in the case of fragmentation
processes. The shape of the measured cross-section ${\mathrm{d}}\sigma/{\mathrm{d}}|\ctgjn|$ is much closer to that of the direct-photon processes
than that of fragmentation in all $\mgjn$ regions. This is consistent
with the dominance of processes in which the exchanged particle is a
quark. The predictions\footnote{The MC predictions for every region in
  $\mgjn$ are normalised using the same factors as for ${\mathrm{d}}\sigma/{\mathrm{d}}\mgjn$.} from \pyt\ and \sher\ are compared with
the data in Figure~\ref{fig176b} and also give an adequate description
of the measurements.

\FloatBarrier

\subsection{Cross sections for isolated-photon plus two-jet production}

\begin{figure}[p]
\setlength{\unitlength}{1.0cm}
\begin{picture} (18.0,15.0)
\put (0.0,8.0){\includegraphics[width=9cm,height=9cm]{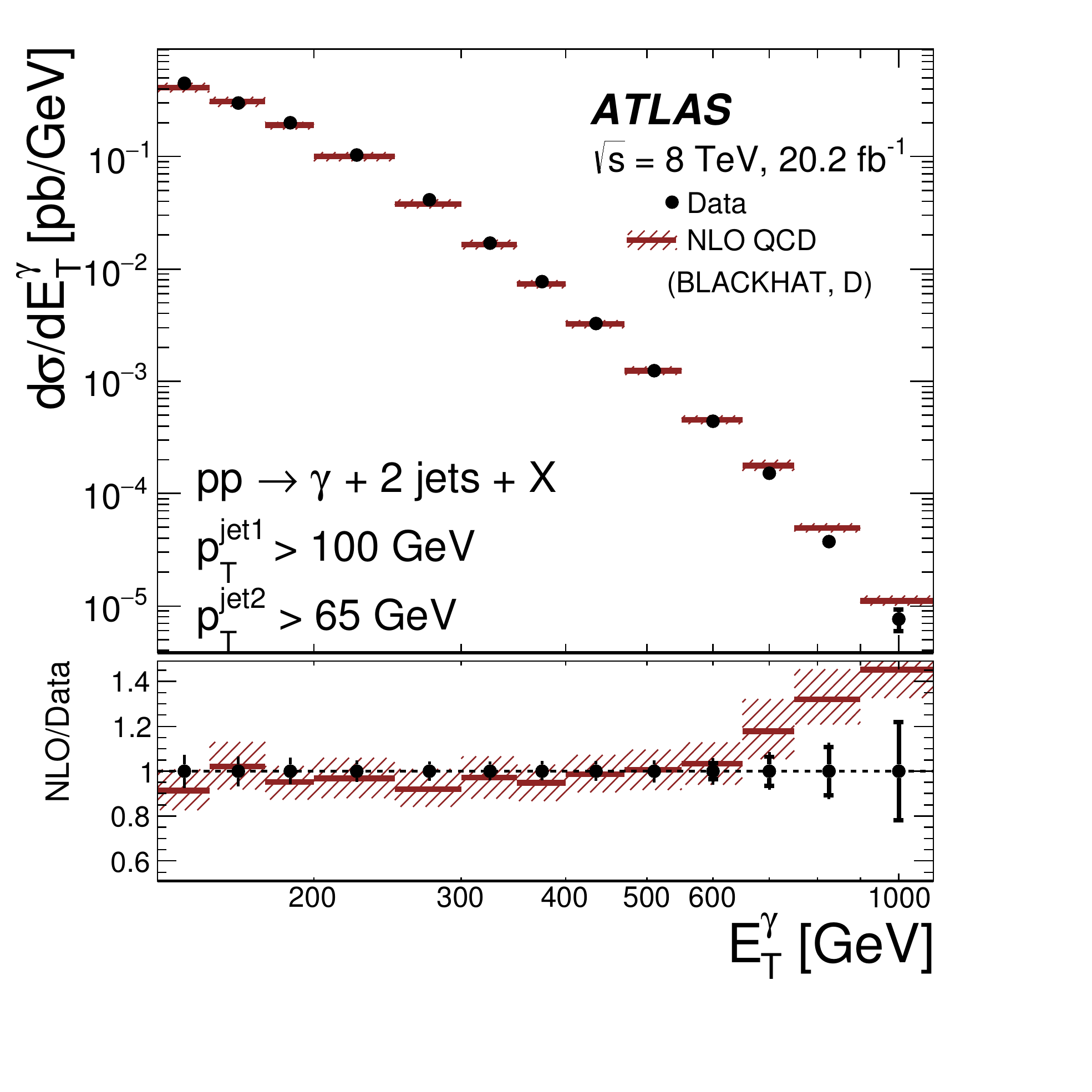}}
\put (8.0,8.0){\includegraphics[width=9cm,height=9cm]{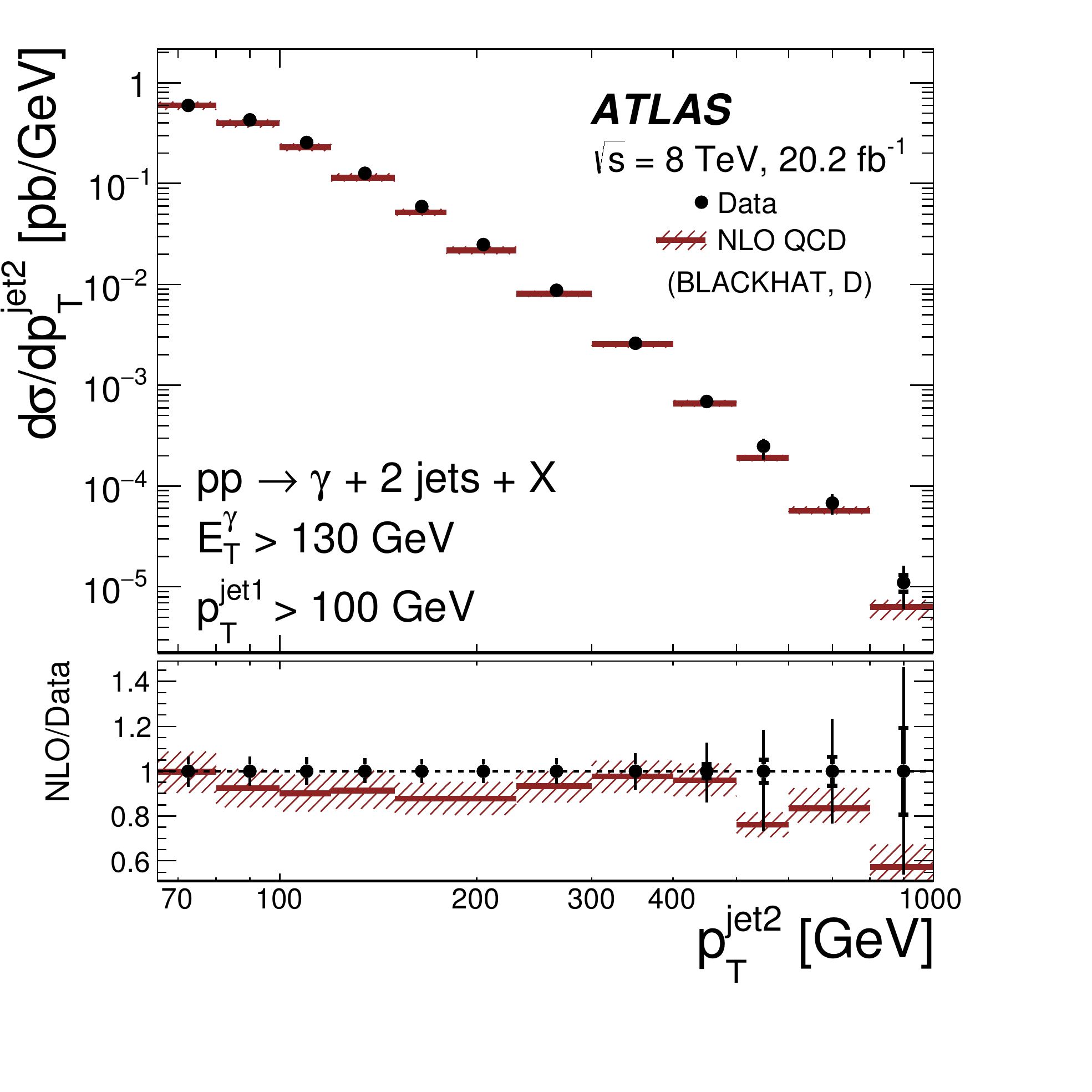}}
\put (0.0,0.0){\includegraphics[width=9cm,height=9cm]{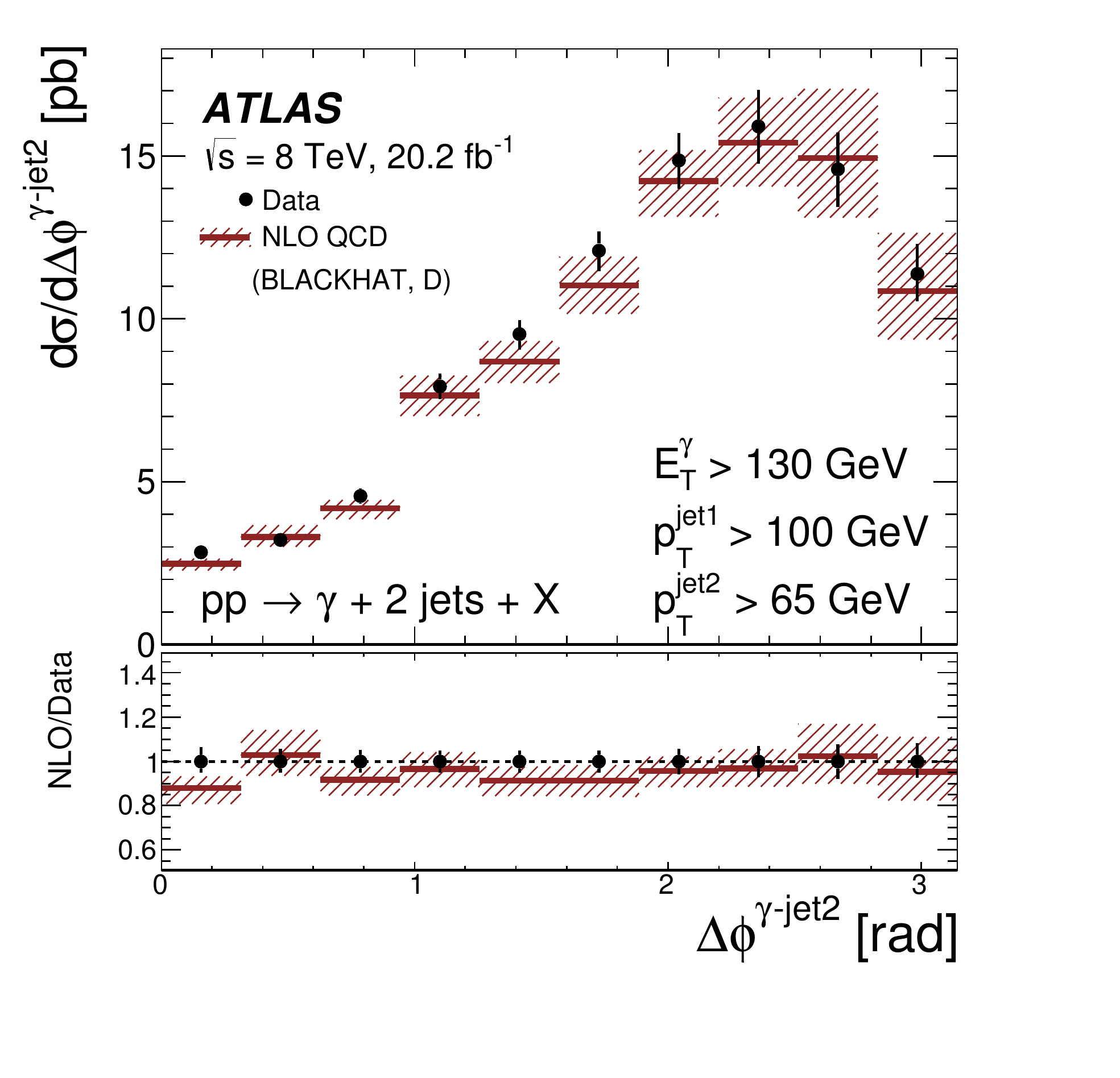}}
\put (8.0,0.0){\includegraphics[width=9cm,height=9cm]{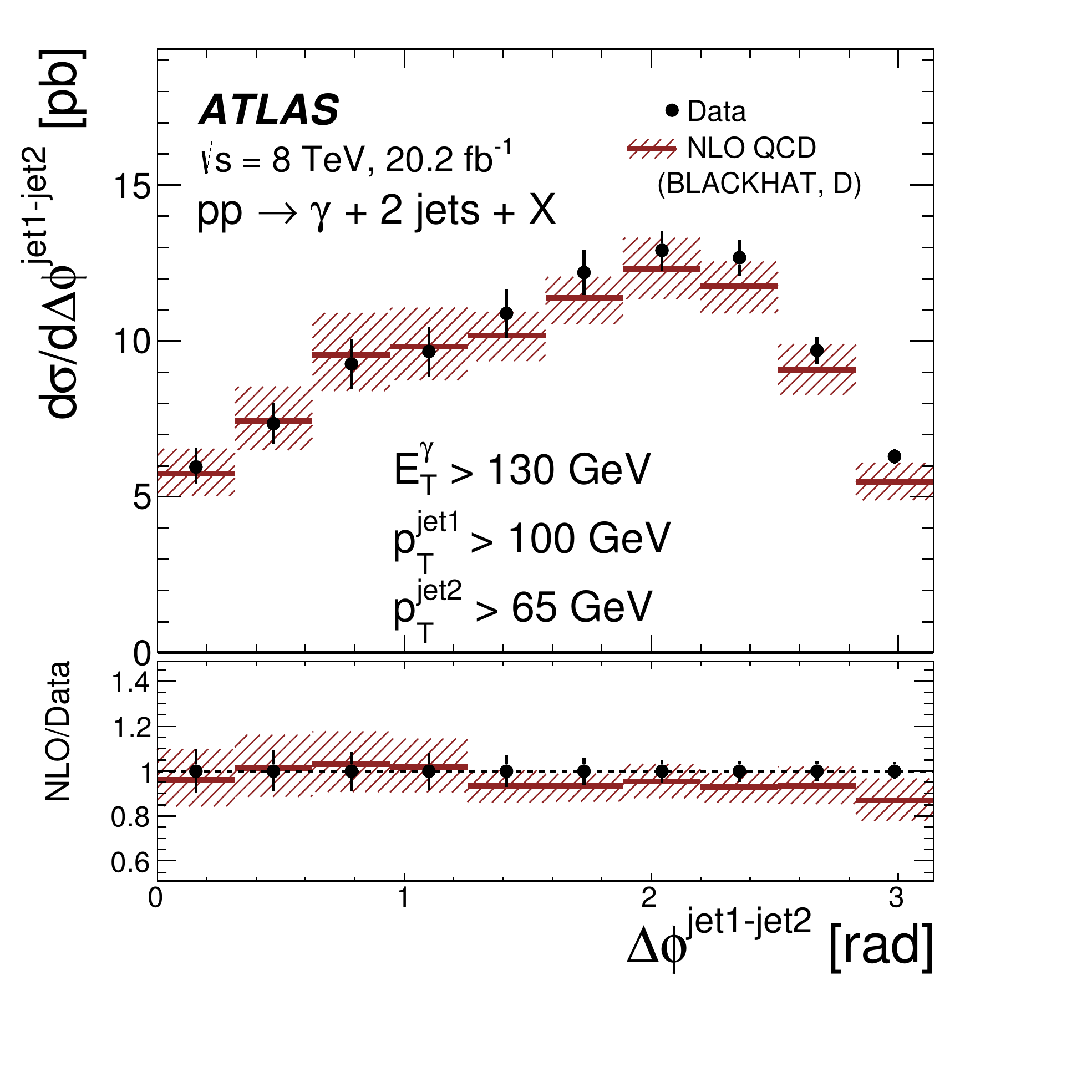}}
\put (3.7,9.0){{\textbf{\small (a)}}}
\put (11.7,9.0){{\textbf{\small (b)}}}
\put (3.7,1.0){{\textbf{\small (c)}}}
\put (11.7,1.0){{\textbf{\small (d)}}}
\end{picture}
\caption
{
  Measured cross sections for isolated-photon plus two-jet production
  (dots) as functions of (a) $\etg$, (b) $\ptjetsl$, (c) $\delphigsl$
  and (d) $\delphijjlsl$. The NLO QCD predictions from \blh\
  corrected for hadronisation and underlying-event effects and using
  the CT10 PDF set (solid lines) are also shown. These predictions
  include only the direct contribution (D). The bottom part of each
  figure shows the ratio of the NLO QCD prediction to the measured
  cross section. The inner (outer) error bars represent the
  statistical uncertainties (the statistical and systematic
  uncertainties added in quadrature) and the shaded band represents
  the theoretical uncertainty. For most of the points, the inner error
  bars are smaller than the marker size and, thus, not visible.
}
\label{fig177}
\end{figure}

The measured cross-section ${\mathrm{d}}\sigma/{\mathrm{d}}\etg$
(Figure~\ref{fig177}(a)) decreases by almost five orders of magnitude
as $\etg$ increases over the measured range. Values of $\etg$ up to
$1.1$~\TeV\ are measured. The experimental uncertainty ranges from
$7\%$ to $23\%$, dominated at low $\etg$ by the jet energy scale
uncertainty and at high $\etg$ by the statistical uncertainty. The NLO
QCD prediction from \blh\ is compared with the measurement in
Figure~\ref{fig177}(a). The NLO QCD prediction gives a good
description of the data within the experimental and theoretical
uncertainties for $\etg<750$~\GeV. The theoretical uncertainty amounts
to $\approx 10\%$; it is dominated by the contribution arising from
scale uncertainties, in particular from the variation of $\mu_{\mathrm{R}}$
for $\etg\lesssim 500$~\GeV, and by the uncertainty from the PDF for
higher $\etg$ values.

The measured cross-section ${\mathrm{d}}\sigma/{\mathrm{d}}\ptjetsl$
(Figure~\ref{fig177}(b)) decreases by almost five orders of magnitude
within the measured range. The experimental uncertainty varies from
$6\%$ to $46\%$ and is dominated by the jet energy scale uncertainty
at low $\ptjetsl$ and by the statistical uncertainty at high
$\ptjetsl$. The NLO QCD prediction gives a good description of the
data. No significant deviation from the prediction from NLO QCD is
observed up to the highest value measured of $\ptjetsl\approx
1$~\TeV. The theoretical uncertainty grows from $10\%$ at
$\ptjetsl\sim 70$~\GeV\ to $\approx 26\%$ for $\ptjetsl\sim 1$~\TeV\
and is dominated by the variation of $\mu_{\mathrm{R}}$ for
$\ptjetsl\lesssim 250$~\GeV\ and by the uncertainty from the PDF for
higher $\etg$ values.

The ${\mathrm{d}}\sigma/{\mathrm{d}}\delphigsl$ and ${\mathrm{d}}\sigma/{\mathrm{d}}\delphijjlsl$ cross sections are shown in Figures~\ref{fig177}(c)
and \ref{fig177}(d), respectively. The measured cross sections display
a maximum at $2$--$2.5$~radians. The NLO QCD predictions give a good
description of the data.

\begin{figure}[p]
\setlength{\unitlength}{1.0cm}
\begin{picture} (18.0,15.0)
\put (0.0,8.0){\includegraphics[width=9cm,height=9cm]{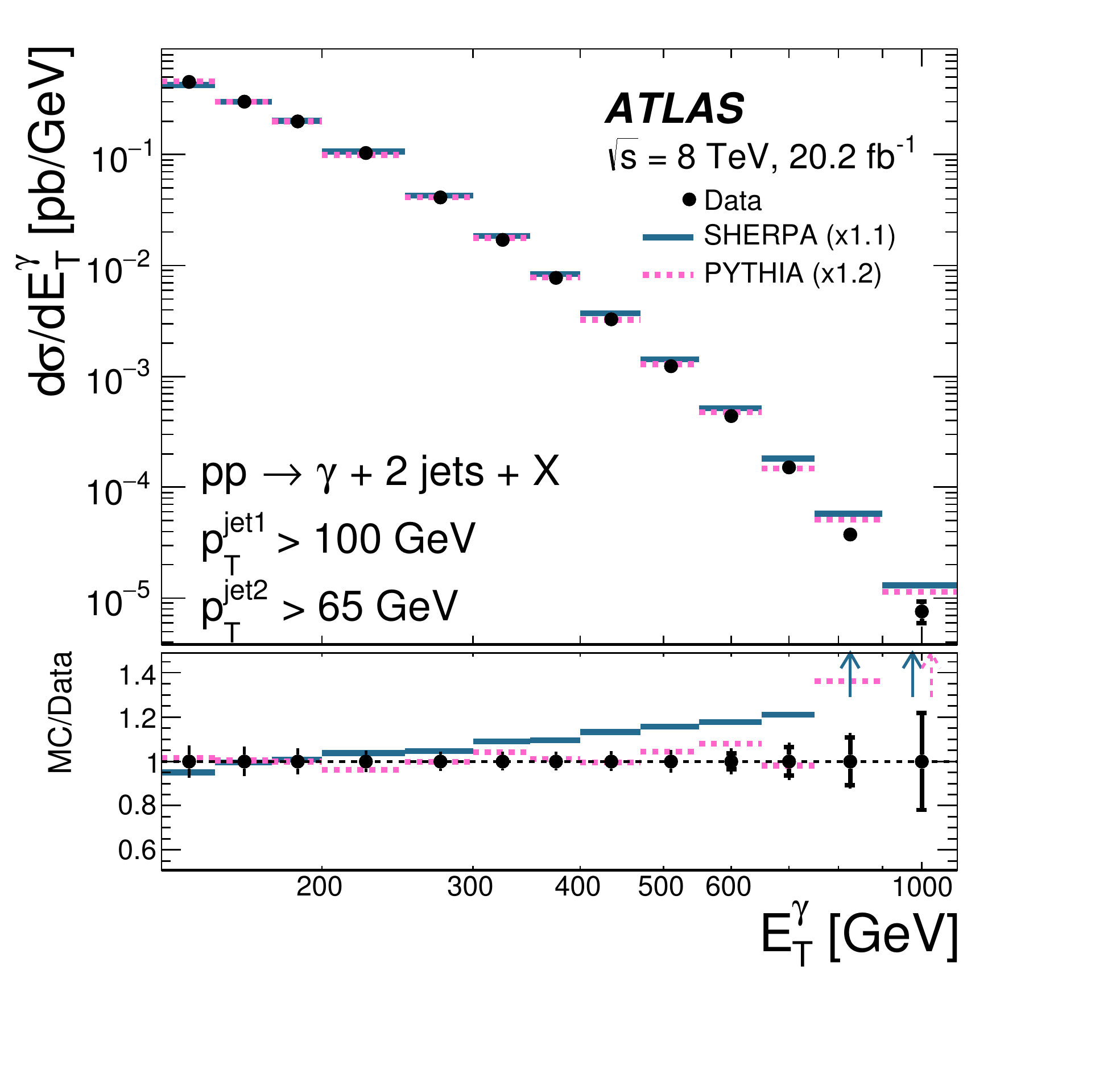}}
\put (8.0,8.0){\includegraphics[width=9cm,height=9cm]{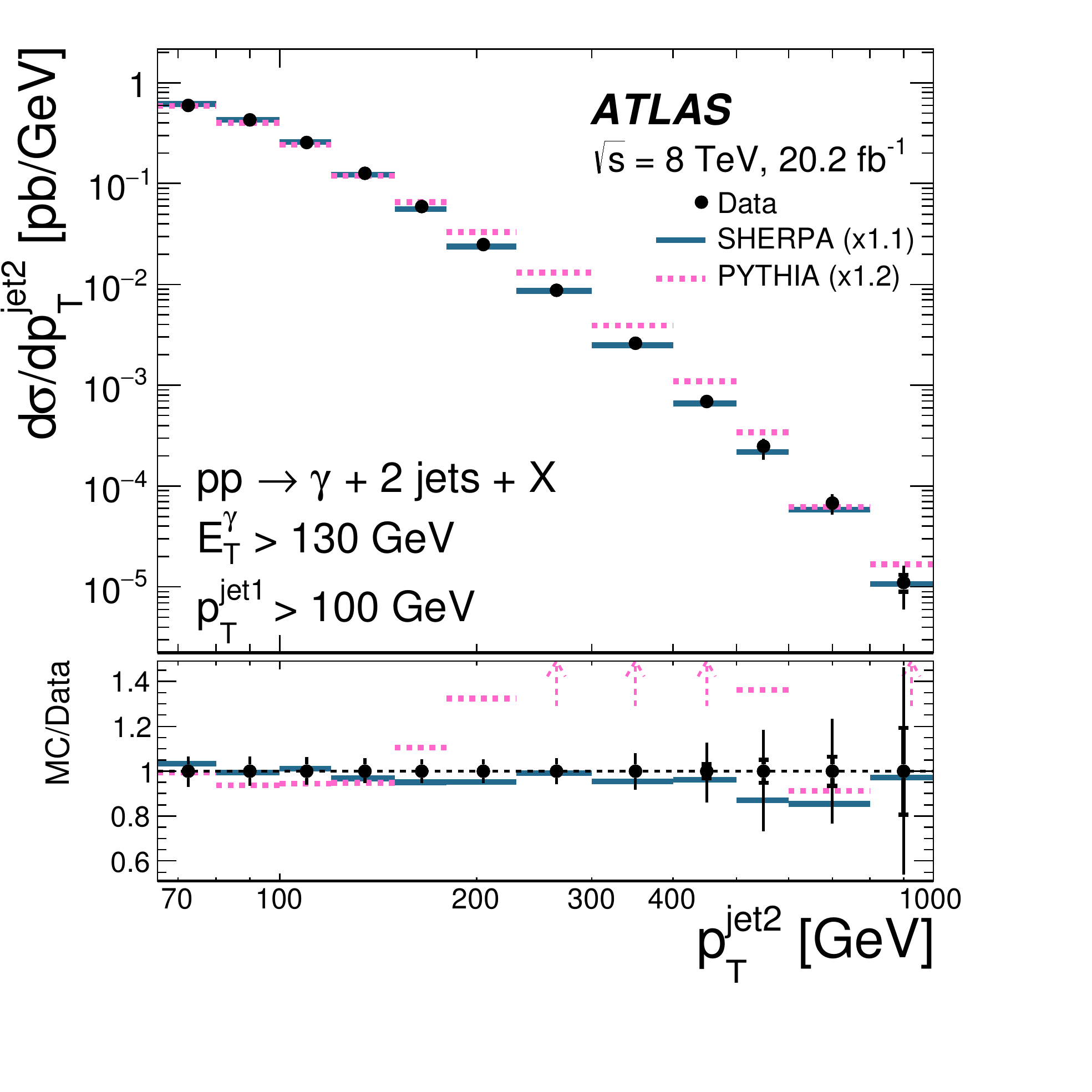}}
\put (0.0,0.0){\includegraphics[width=9cm,height=9cm]{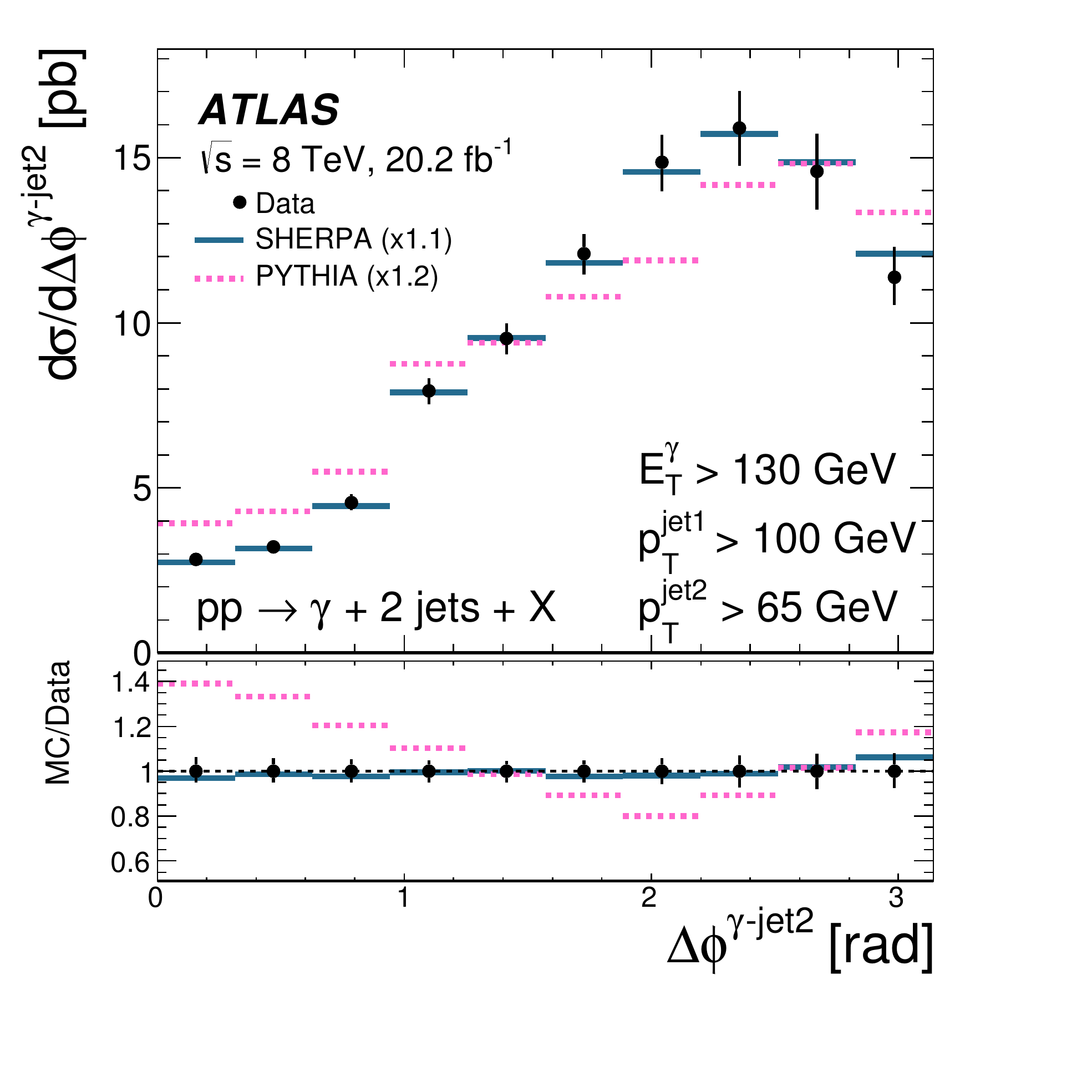}}
\put (8.0,0.0){\includegraphics[width=9cm,height=9cm]{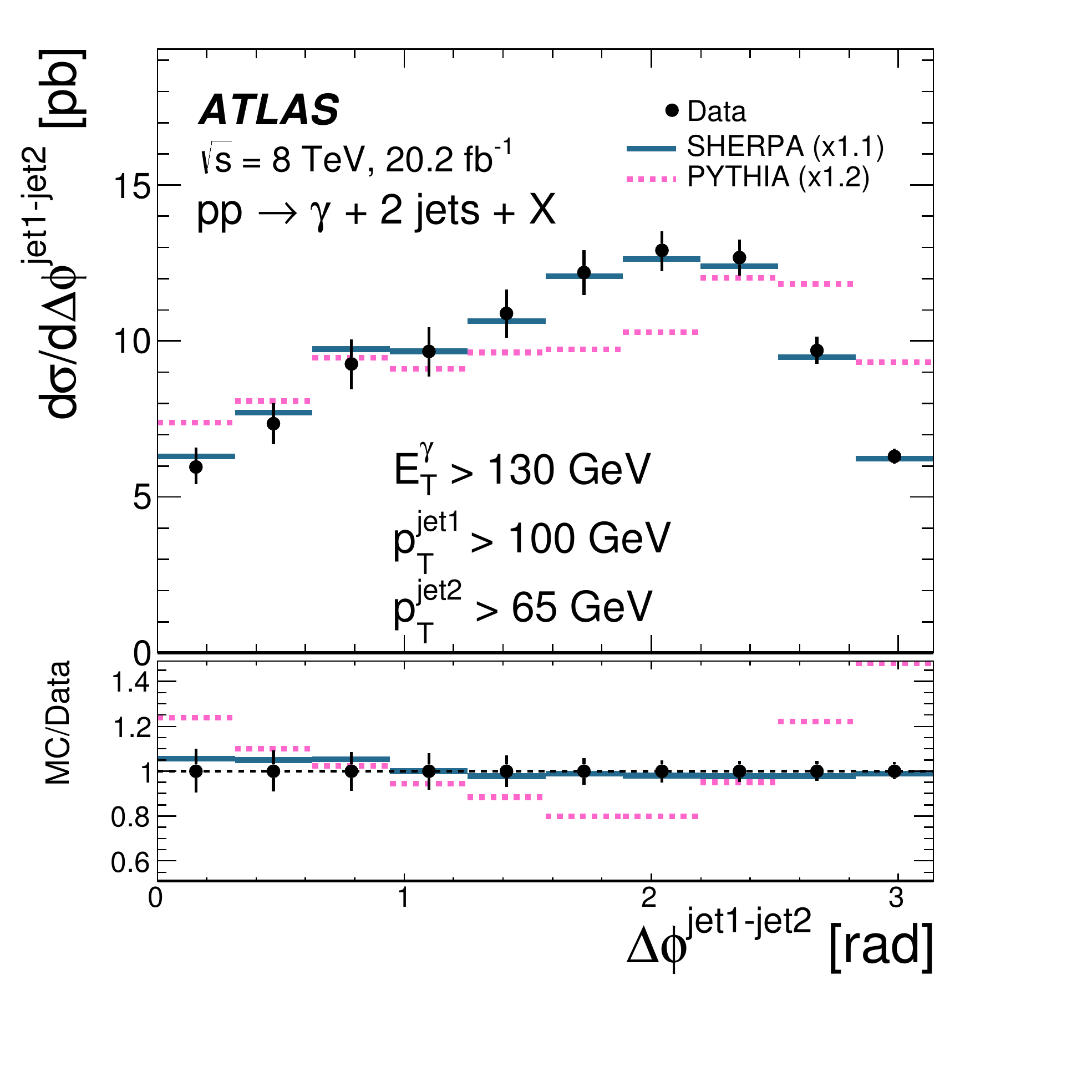}}
\put (3.7,9.0){{\textbf{\small (a)}}}
\put (11.7,9.0){{\textbf{\small (b)}}}
\put (3.7,1.0){{\textbf{\small (c)}}}
\put (11.7,1.0){{\textbf{\small (d)}}}
\end{picture}
\caption
{
  Measured cross sections for isolated-photon plus two-jet production
  (dots) as functions of (a) $\etg$, (b) $\ptjetsl$, (c) $\delphigsl$
  and (d) $\delphijjlsl$. For comparison, the predictions from \pyt\
  (dashed lines) and \sher\  (solid lines) normalised to the
  integrated measured cross sections (using the factors indicated in
  parentheses) are also shown. The bottom part of each figure shows
  the ratios of the MC predictions to the measured cross section. The
  inner (outer) error bars represent the statistical uncertainties
  (the statistical and systematic uncertainties added in
  quadrature). For most of the points, the inner error bars are
  smaller than the marker size and, thus, not visible.
}
\label{fig177b}
\end{figure}

The prediction from \pyt\  gives a good description of the measured
cross-section ${\mathrm{d}}\sigma/{\mathrm{d}}\etg$ up to $\etg \sim 750$~\GeV\
(see Figure~\ref{fig177b}(a)), whereas the prediction from \sher\
describes well the measured cross-section ${\mathrm{d}}\sigma/{\mathrm{d}}\ptjetsl$ (see Figure~\ref{fig177b}(b)).  The predictions from
\sher\  give a good description of the measured cross-section ${\mathrm{d}}\sigma/{\mathrm{d}}\delphigsl$ and ${\mathrm{d}}\sigma/{\mathrm{d}}\delphijjlsl$,
while the predictions from \pyt\ do not. In the predictions from \pyt\
a second jet can arise only from the parton shower, whereas in \sher,
$2\rightarrow n$ (with $n\geq 3$) matrix-element contributions are
included as well, with a higher probability of producing a second hard
jet.

\begin{figure}[p]
\setlength{\unitlength}{1.0cm}
\begin{picture} (18.0,12.0)
\put (0.0,5.2){\includegraphics[width=9cm,height=9cm]{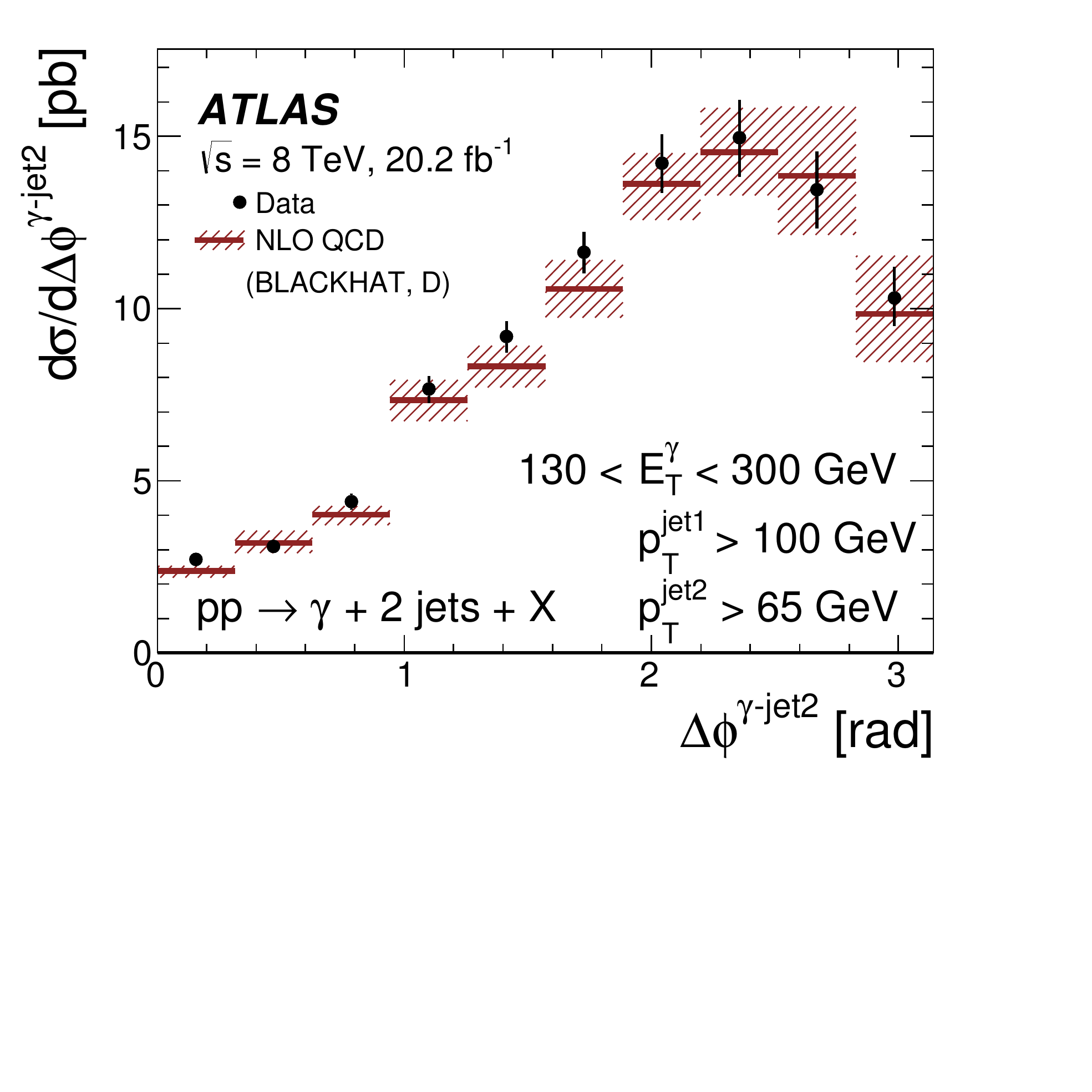}}
\put (8.0,5.2){\includegraphics[width=9cm,height=9cm]{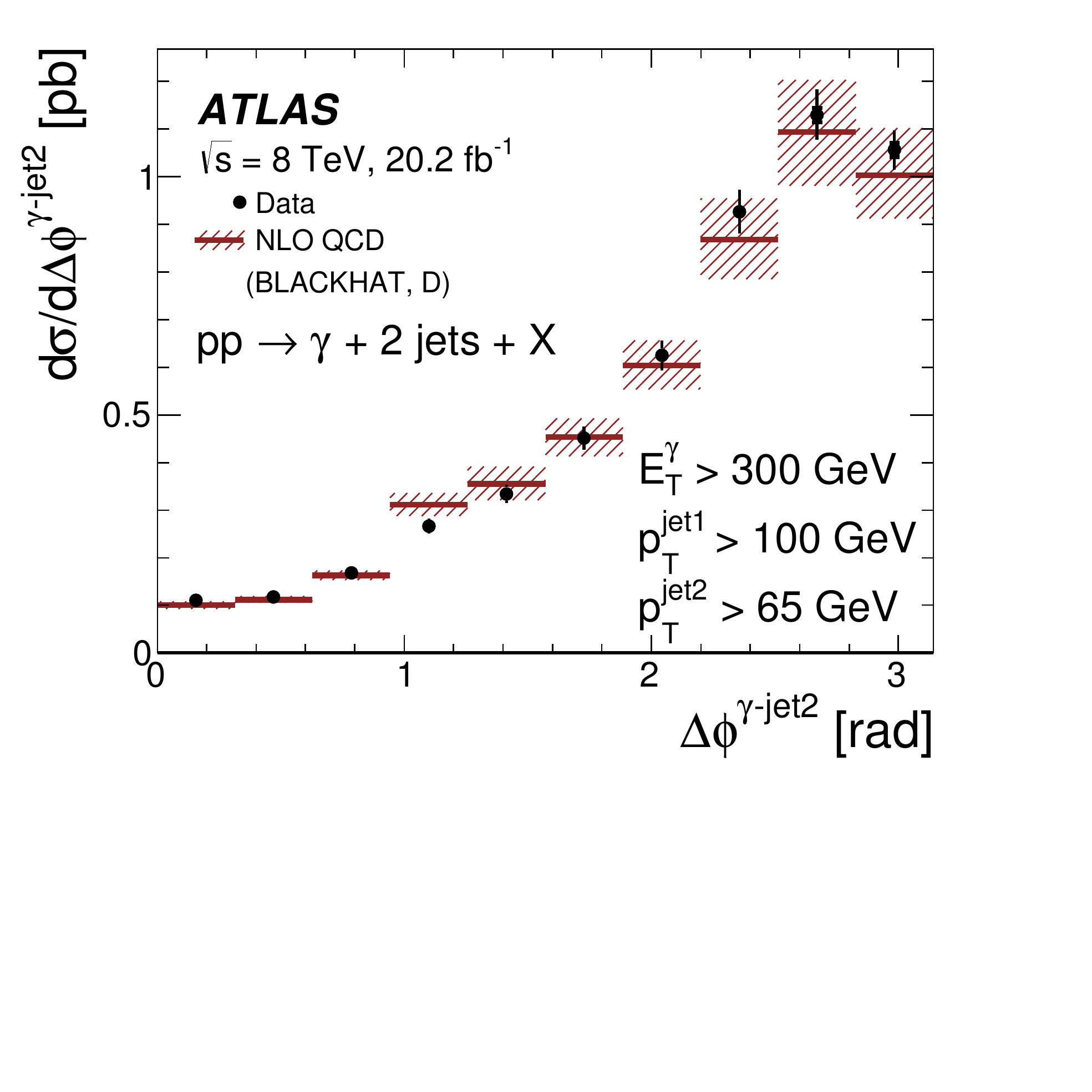}}
\put (0.0,-1.8){\includegraphics[width=9cm,height=9cm]{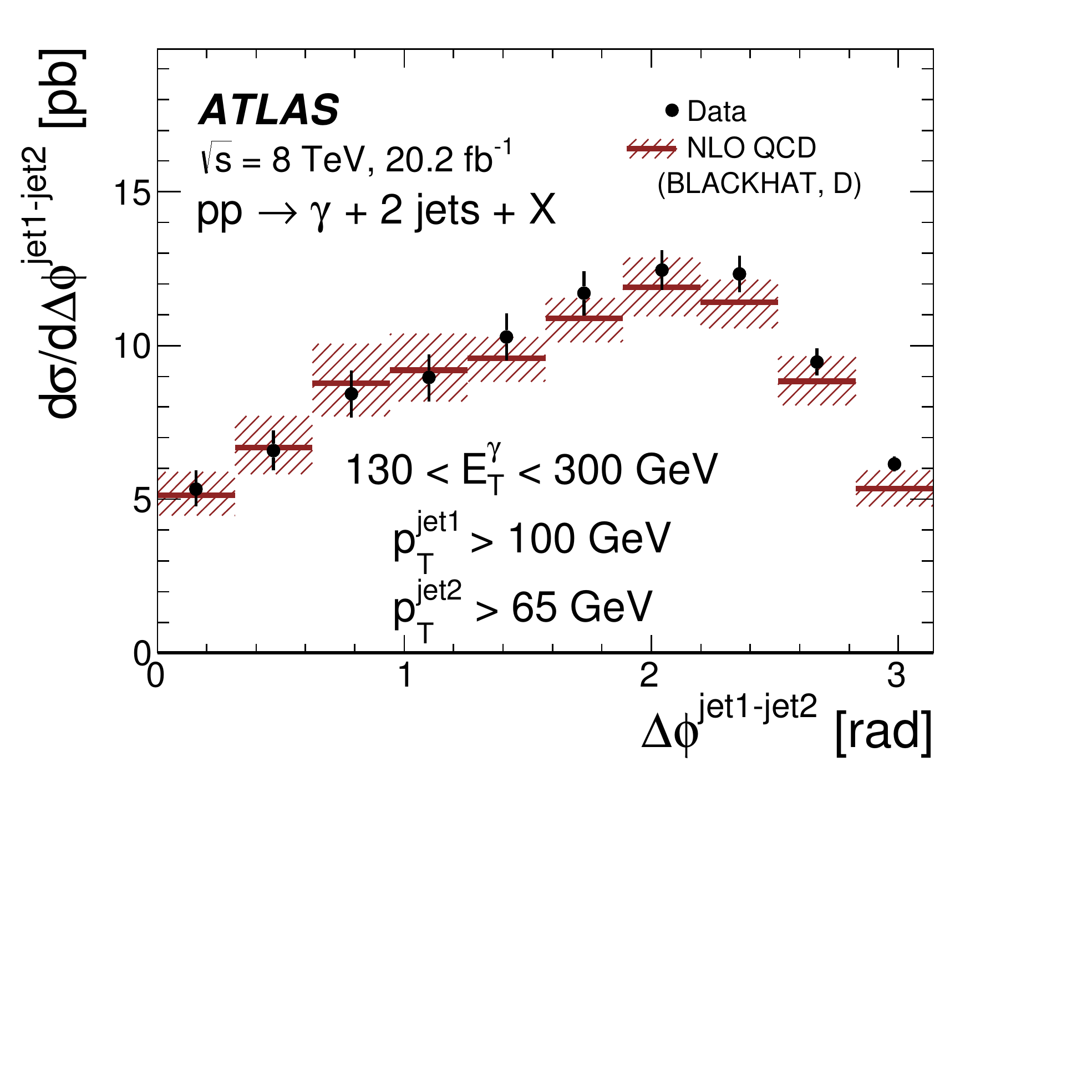}}
\put (8.0,-1.8){\includegraphics[width=9cm,height=9cm]{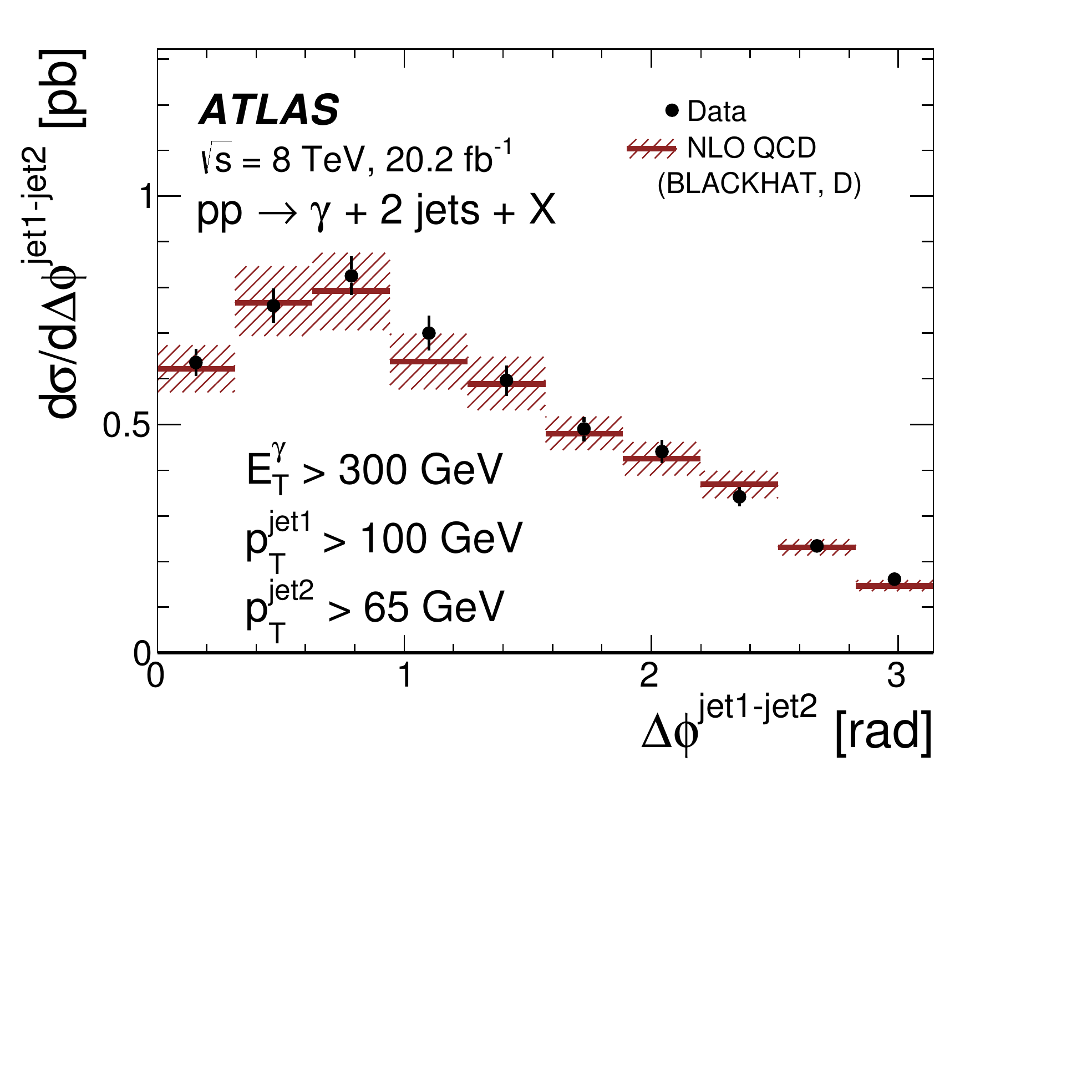}}
\put (3.7,8.0){{\textbf{\small (a)}}}
\put (11.7,8.0){{\textbf{\small (b)}}}
\put (3.7,1.0){{\textbf{\small (c)}}}
\put (11.7,1.0){{\textbf{\small (d)}}}
\end{picture}
\caption
{
  Measured cross sections for isolated-photon plus two-jet production
  (dots) as functions of (a,b) $\delphigsl$ and (c,d) $\delphijjlsl$
  for (a,c) $\etg<300$~\GeV\ and (b,d) $\etg>300$~\GeV. The NLO QCD
  predictions from \blh\ corrected for hadronisation and
  underlying-event effects and using the CT10 PDF set are also shown
  as solid lines. These predictions include only the direct
  contribution (D). The inner (outer) error bars represent the
  statistical uncertainties (the statistical and systematic
  uncertainties added in quadrature) and the shaded band represents
  the theoretical uncertainty. For most of the points, the inner error
  bars are smaller than the marker size and, thus, not visible.
}
\label{fig178}
\end{figure}

\begin{figure}[p]
\setlength{\unitlength}{1.0cm}
\begin{picture} (18.0,12.0)
\put (0.0,5.2){\includegraphics[width=9cm,height=9cm]{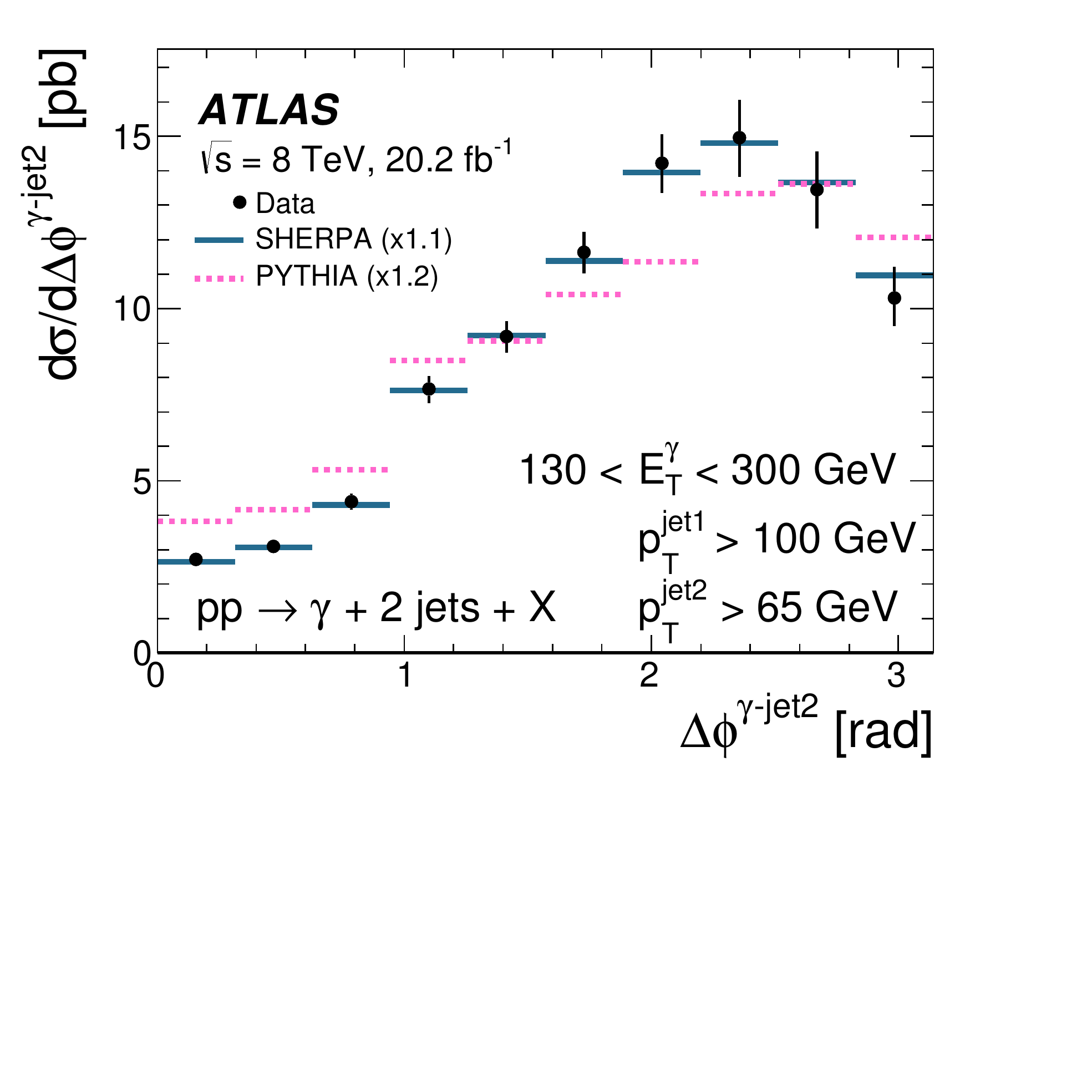}}
\put (8.0,5.2){\includegraphics[width=9cm,height=9cm]{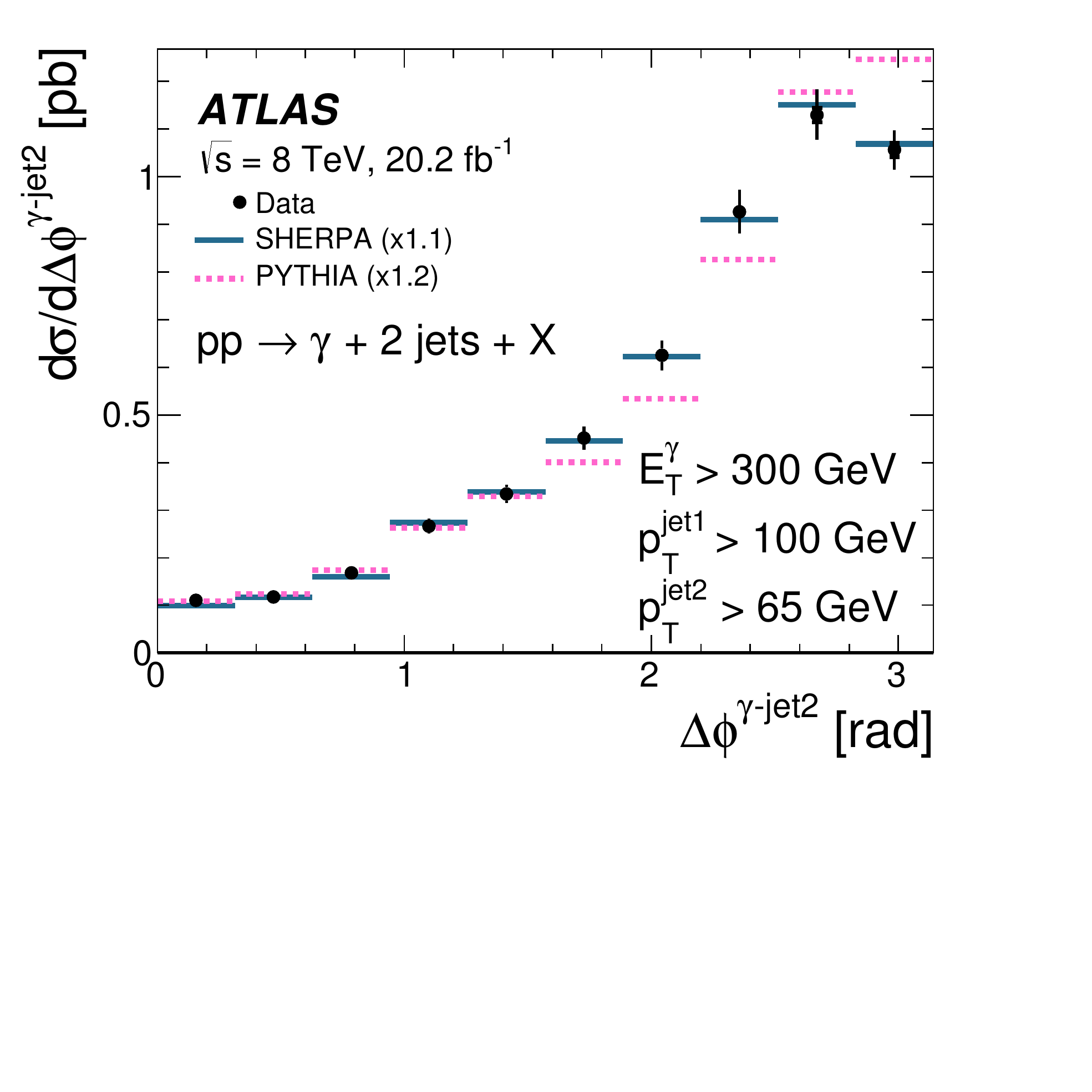}}
\put (0.0,-1.8){\includegraphics[width=9cm,height=9cm]{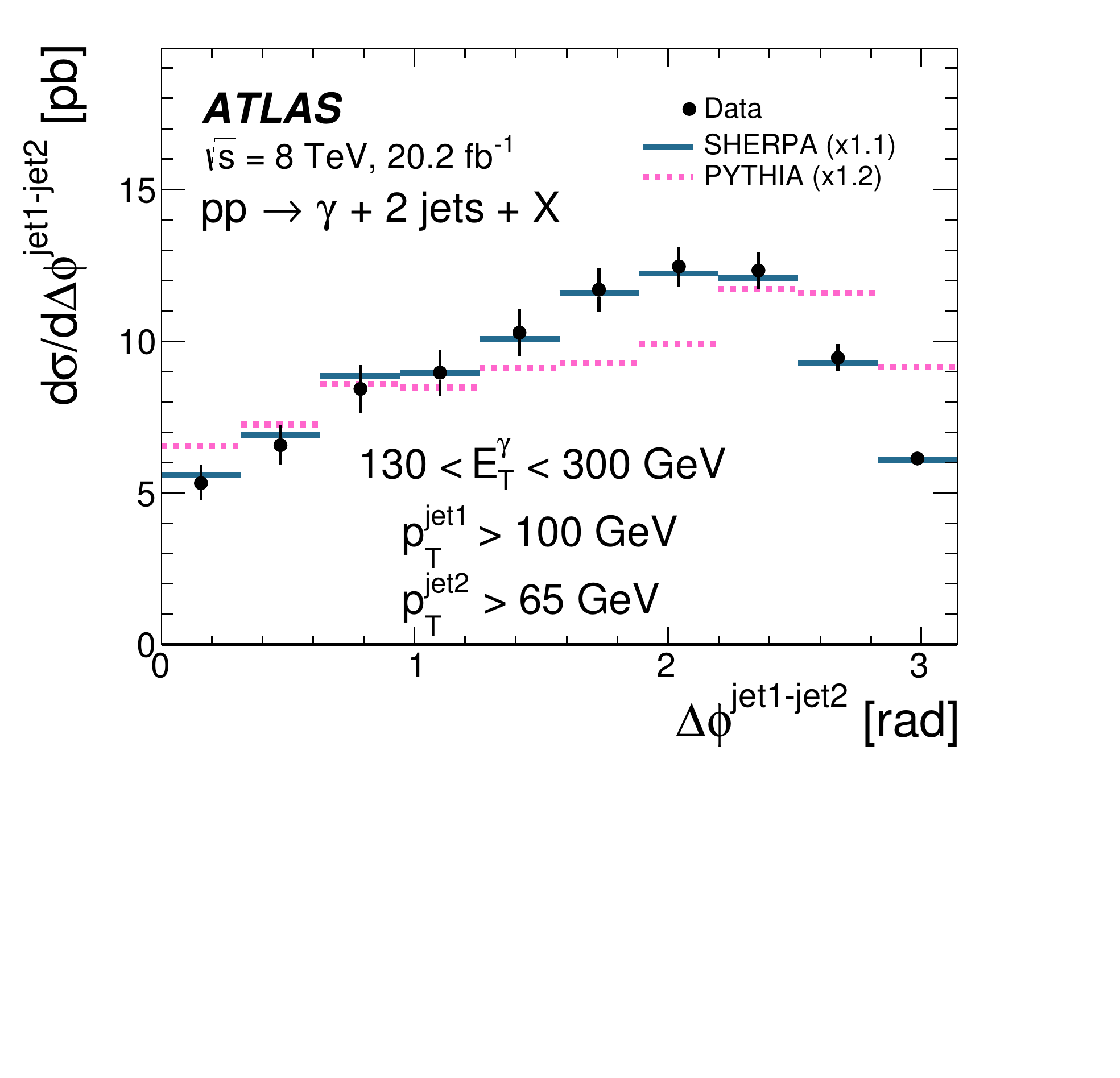}}
\put (8.0,-1.8){\includegraphics[width=9cm,height=9cm]{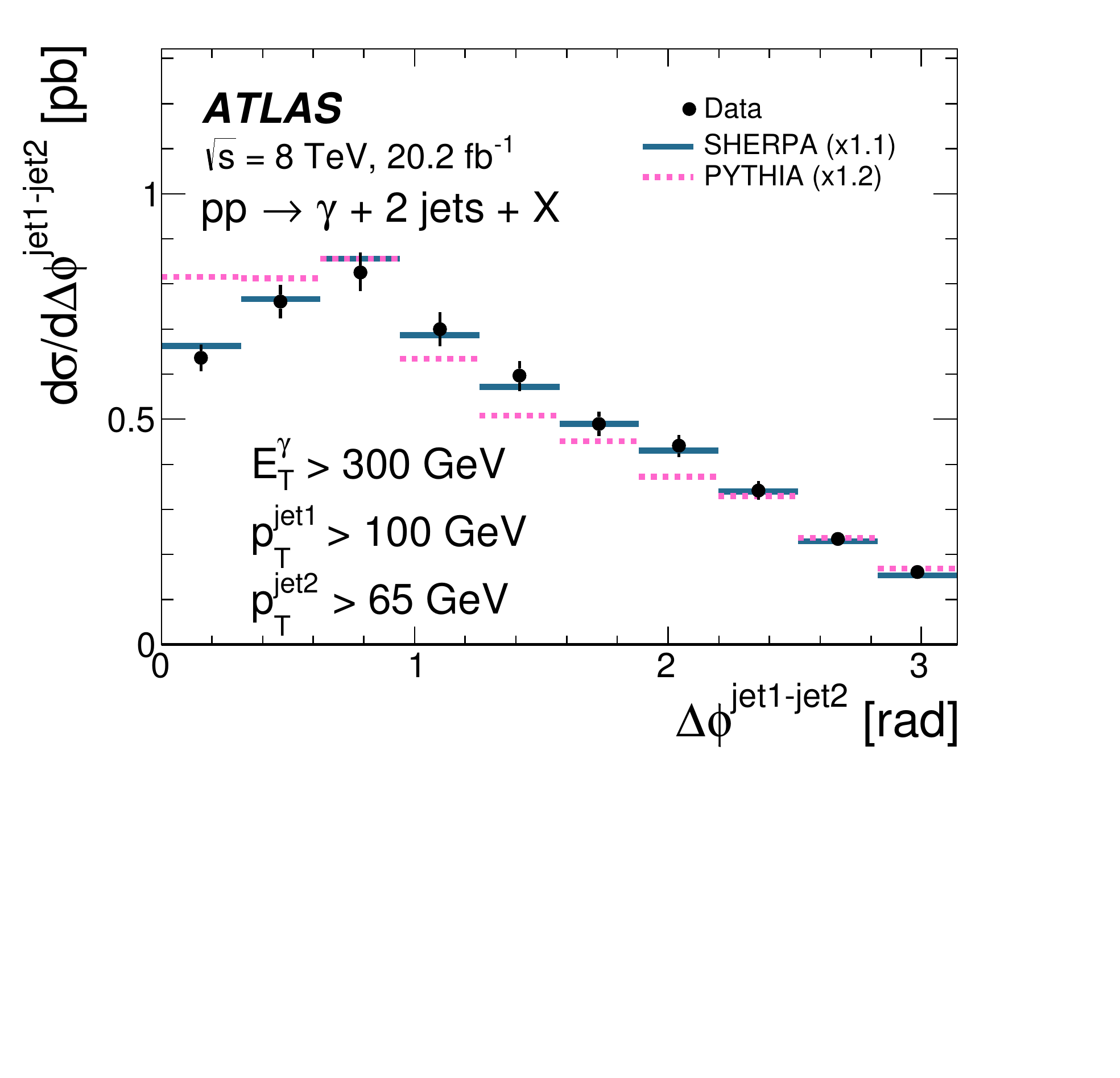}}
\put (3.7,8.0){{\textbf{\small (a)}}}
\put (11.7,8.0){{\textbf{\small (b)}}}
\put (3.7,1.0){{\textbf{\small (c)}}}
\put (11.7,1.0){{\textbf{\small (d)}}}
\end{picture}
\caption
{
  Measured cross sections for isolated-photon plus two-jet production
  (dots) as functions of (a,b) $\delphigsl$ and (c,d) $\delphijjlsl$
  for (a,c) $\etg<300$~\GeV\ and (b,d) $\etg>300$~\GeV. For
  comparison, the predictions from \pyt\ (dashed lines) and \sher\
  (solid lines) are also shown. The predictions are normalised to the
  data by a global factor, which is indicated in parentheses. The
  inner (outer) error bars represent the statistical uncertainties
  (the statistical and systematic uncertainties added in
  quadrature). For most of the points, the inner error bars are
  smaller than the marker size and, thus, not visible.
}
\label{fig178b}
\end{figure}

The scale evolution of photon plus two-jet production is tested by
measuring the azimuthal angle between jet2 and the photon or between
jet2 and jet1 for $\etg$ below/above $300$~\GeV. Figure~\ref{fig178}
shows the cross sections as functions of $\delphigsl$ and
$\delphijjlsl$ for the two $\etg$ ranges: both cross-section
distributions have different shapes for $\etg$ below/above
$300$~\GeV. The ${\mathrm{d}}\sigma/{\mathrm{d}}\delphigsl$ cross section is
more peaked towards large values of $\delphigsl$ for $\etg >
300$~\GeV\ than that for $\etg < 300$~\GeV; the ${\mathrm{d}}\sigma/{\mathrm{d}}\delphijjlsl$ cross section peaks at $\delphijjlsl\sim 0.8$~rad
($2.2$~rad) for $\etg > 300$~\GeV\ ($\etg < 300$~\GeV). The NLO QCD
predictions give a good description of the data and, in particular,
reproduce the scale evolution of the measured cross
sections. Figure~\ref{fig178b} shows the comparison of the data and
the predictions from \pyt\ and \sher. The predictions from \pyt\ fail
to describe the data whereas those from \sher\ describe well the shape
of the measured cross-section distributions and their evolution with
the scale.

\FloatBarrier

\subsection{Comparison of jet production around the photon and jet1}

\begin{figure}[h]
\setlength{\unitlength}{1.0cm}
\begin{picture} (18.0,11.0)
\put (0.0,0.1){\includegraphics[width=9cm,height=10.5cm]{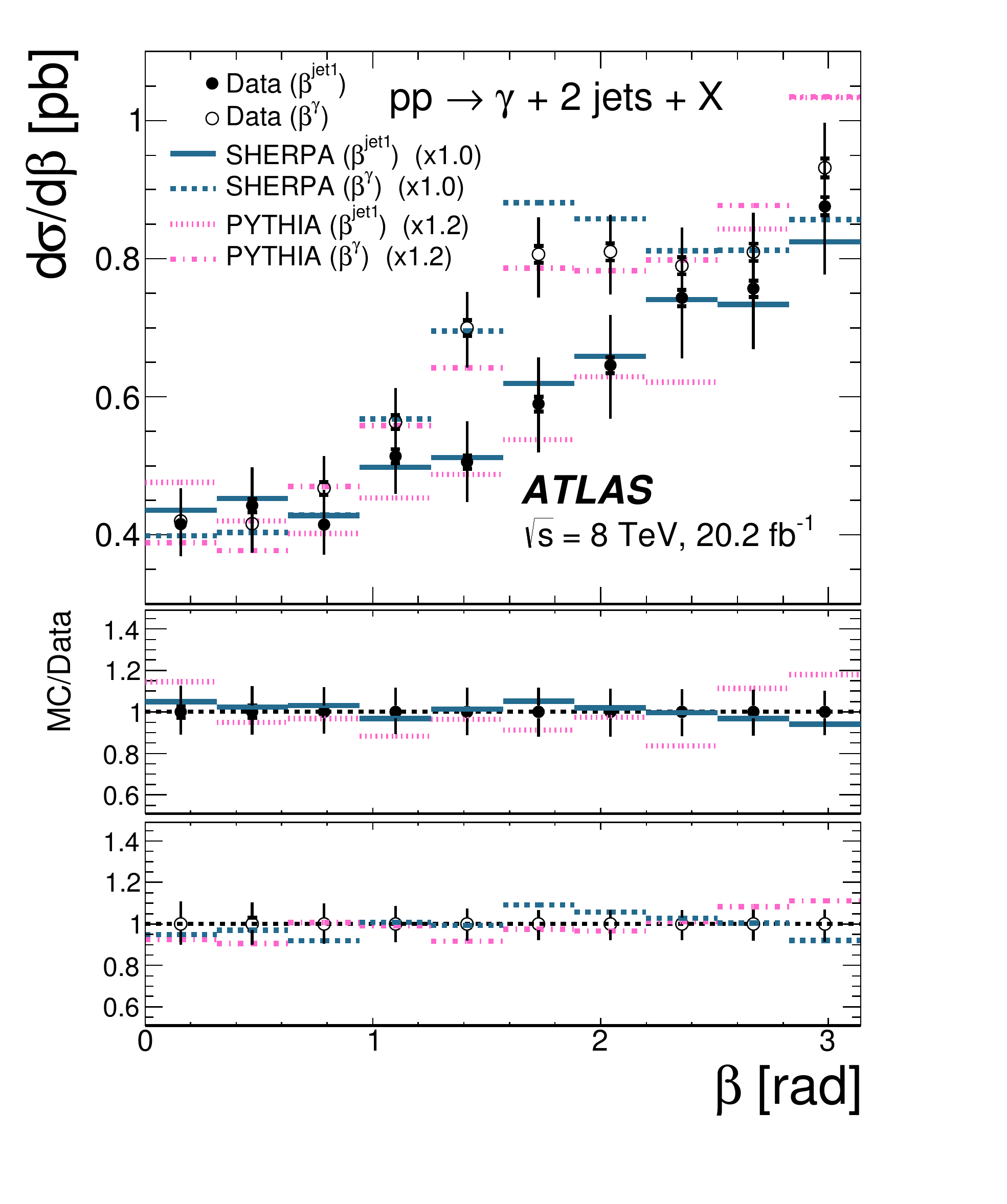}}
\put (8.0,1.5){\includegraphics[width=8cm,height=8cm]{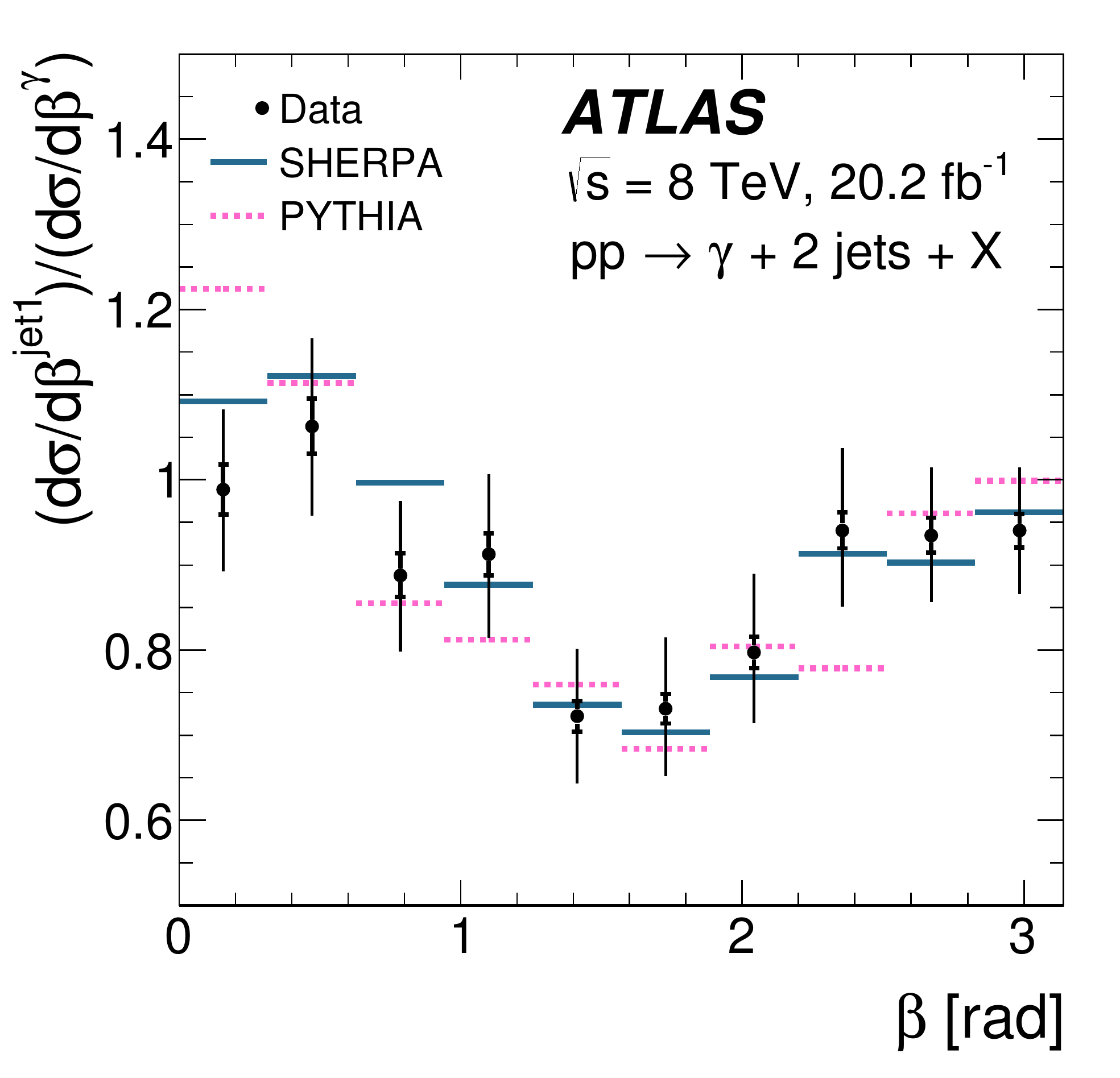}}
\put (3.5,0.85){{\textbf{\small (a)}}}
\put (11.5,0.85){{\textbf{\small (b)}}}
\end{picture}
\caption
{
  (a) Measured cross sections for isolated-photon plus two-jet
  production as functions of $\betaj$ (dots) and $\betag$ (open
  circles). For comparison, the predictions from \sher\  (blue solid
  and dashed lines) and \pyt\ (pink dash-dotted and dotted lines)
  normalised to the integrated measured cross sections (using the
  factors indicated in parentheses) are also shown. The bottom parts
  of the figure show the ratios of the MC predictions to the measured
  cross sections. (b) Ratio of the measured cross-section ${\mathrm{d}}\sigma/{\mathrm{d}}\betaj$ and ${\mathrm{d}}\sigma/{\mathrm{d}}\betag$ (dots);
  the ratios for the \sher\ and \pyt\ predictions are shown as solid
  and dashed lines, respectively. The inner (outer) error bars
  represent the statistical uncertainties (the statistical and
  systematic uncertainties added in quadrature). For some of the
  points, the inner error bars are smaller than the marker size and,
  thus, not visible. 
}
\label{fig177beta}
\end{figure}

Figure~\ref{fig177beta}(a) shows the cross sections for photon plus
two-jet production as functions of $\betaj$ and $\betag$.  The two
measured cross sections have different shapes: the measured
cross-section ${\mathrm{d}}\sigma/{\mathrm{d}}\betaj$ increases monotonically as
$\betaj$ increases, whereas the measured cross-section ${\mathrm{d}}\sigma/{\mathrm{d}}\betag$ increases up to $\betag\approx 1.8$~rad and
then remains approximately constant. The predictions from \sher\  give
a good description of the measured cross sections. To quantify the
differences in the patterns of jet production around the photon and
jet1, the ratio of the measured cross-sections ${\mathrm{d}}\sigma/{\mathrm{d}}\betaj$ and ${\mathrm{d}}\sigma/{\mathrm{d}}\betag$ is made. In the
estimation of the systematic uncertainties of the ratio of the cross
sections, the correlations between numerator and denominator are fully
taken into account leading to complete or partial cancellations
depending on the source of uncertainty. The ratio $({\mathrm{d}}\sigma/{\mathrm{d}}\betaj)/({\mathrm{d}}\sigma/{\mathrm{d}}\betag)$, shown in
Figure~\ref{fig177beta}(b), is enhanced at $\beta=0$ and $\pi$~rad
with respect to the value of the ratio at $\beta=\pi/2$~rad. The
measured ratio is tested against the hypothesis of being independent
of $\beta$ and the resulting $p$-value is $1.3\%$. Thus, it is
observed, for the first time, that the patterns of QCD radiation
around the photon and jet1 are different.

\FloatBarrier
\newpage
\subsection{Cross sections for isolated-photon plus
  three-jet production}

\begin{figure}[p]
\setlength{\unitlength}{1.0cm}
\begin{picture} (18.0,19.5)
\put (0.0,12.5){\includegraphics[width=7cm,height=7cm]{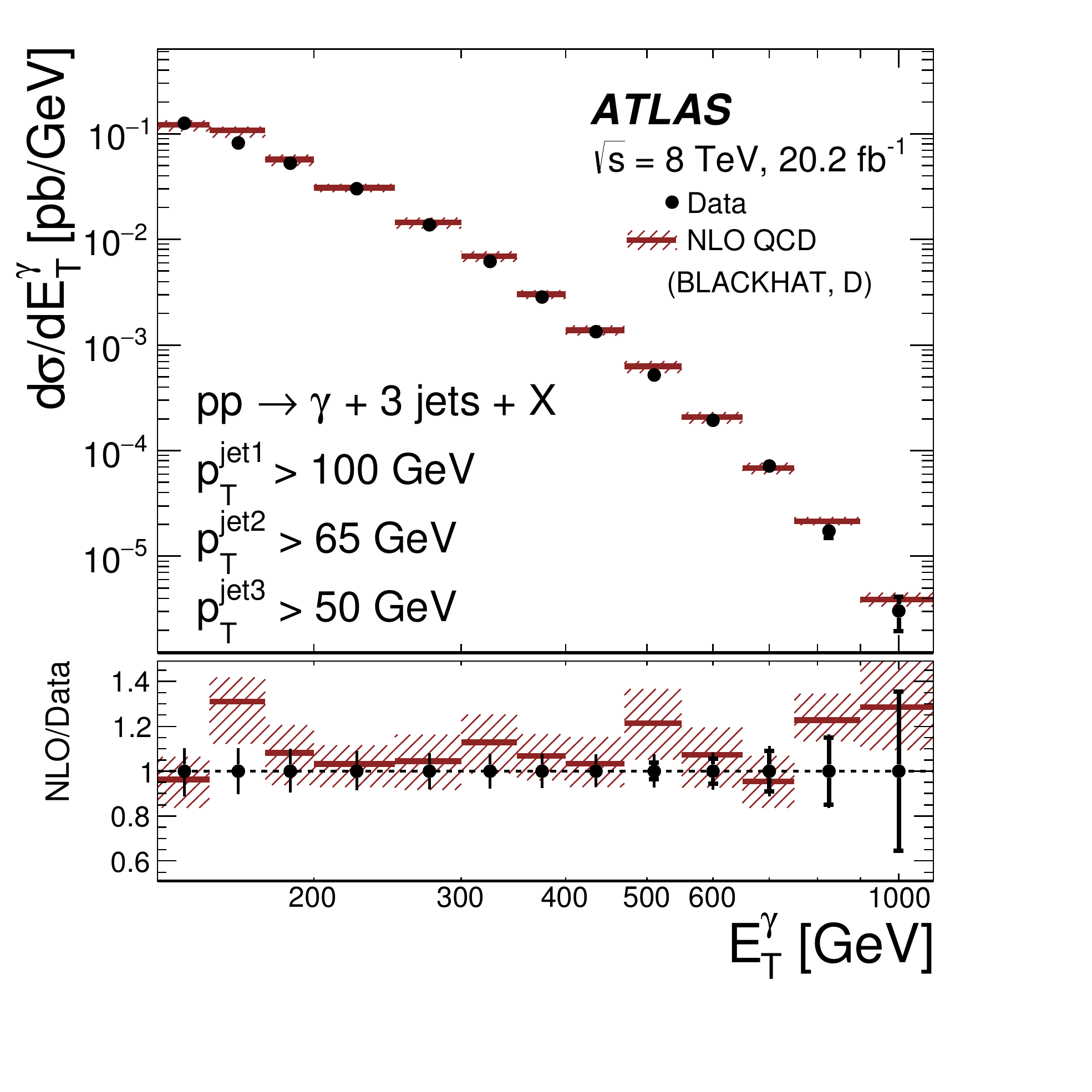}}
\put (8.0,12.5){\includegraphics[width=7cm,height=7cm]{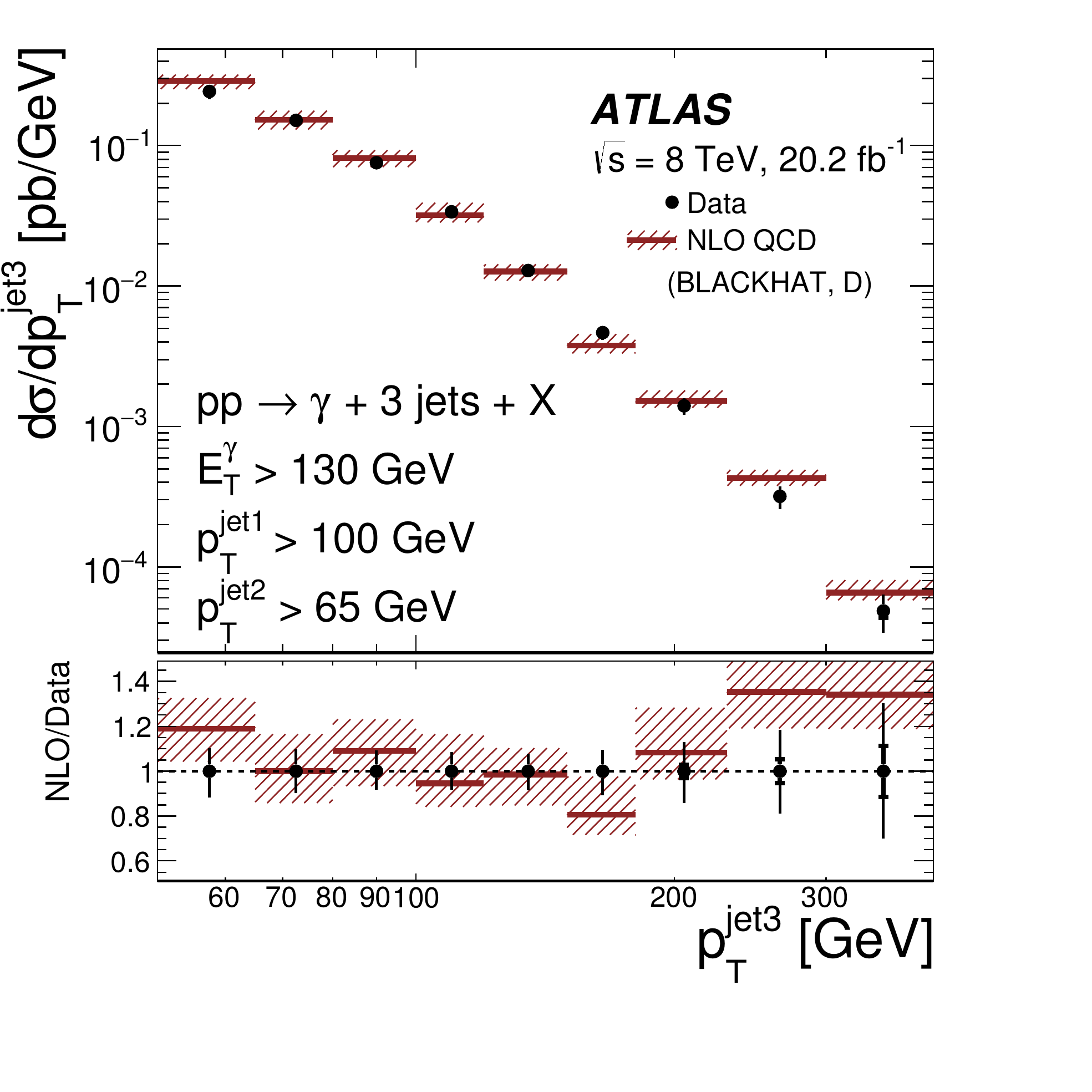}}
\put (0.0,6.0){\includegraphics[width=7cm,height=7cm]{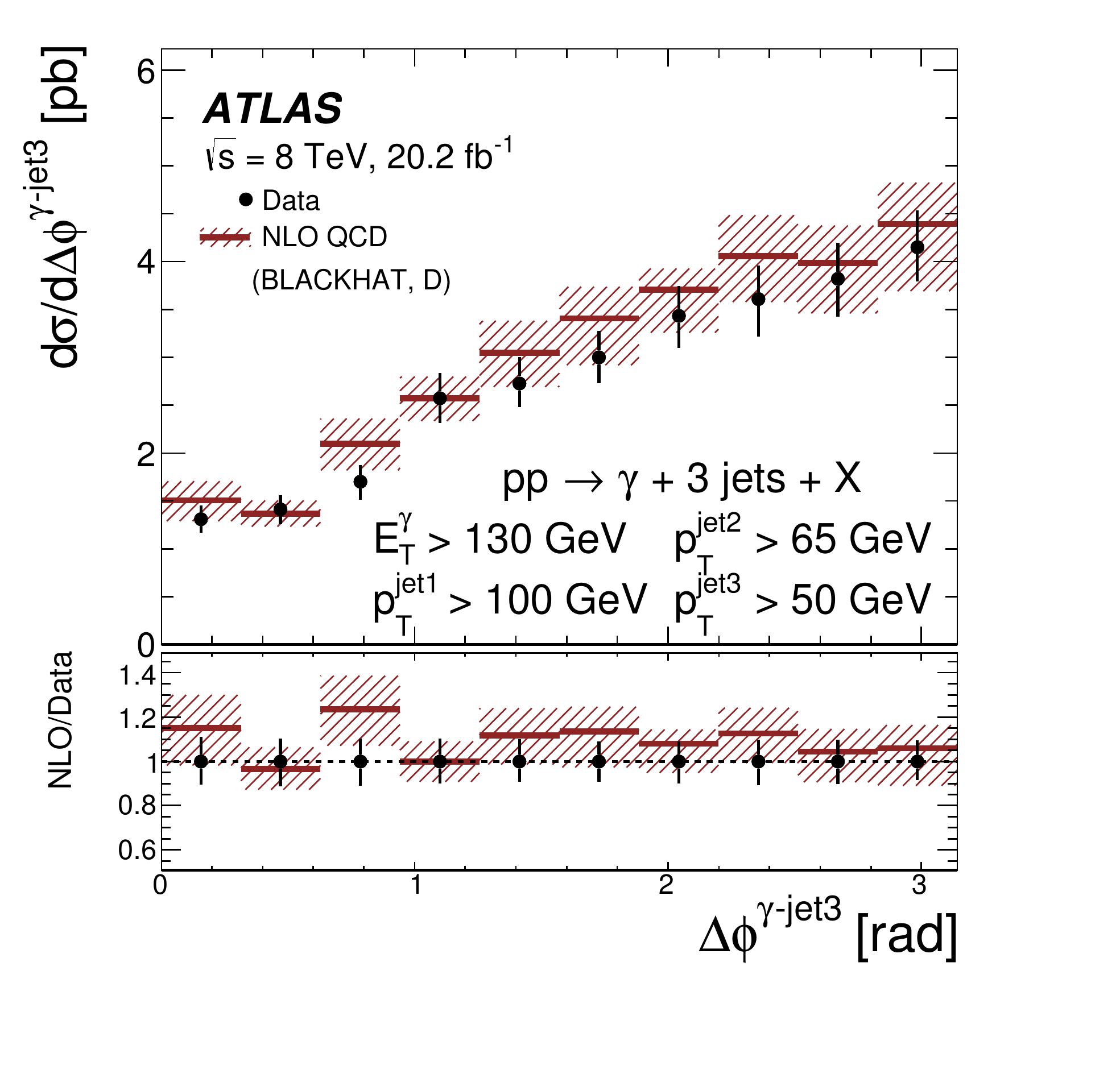}}
\put (8.0,6.0){\includegraphics[width=7cm,height=7cm]{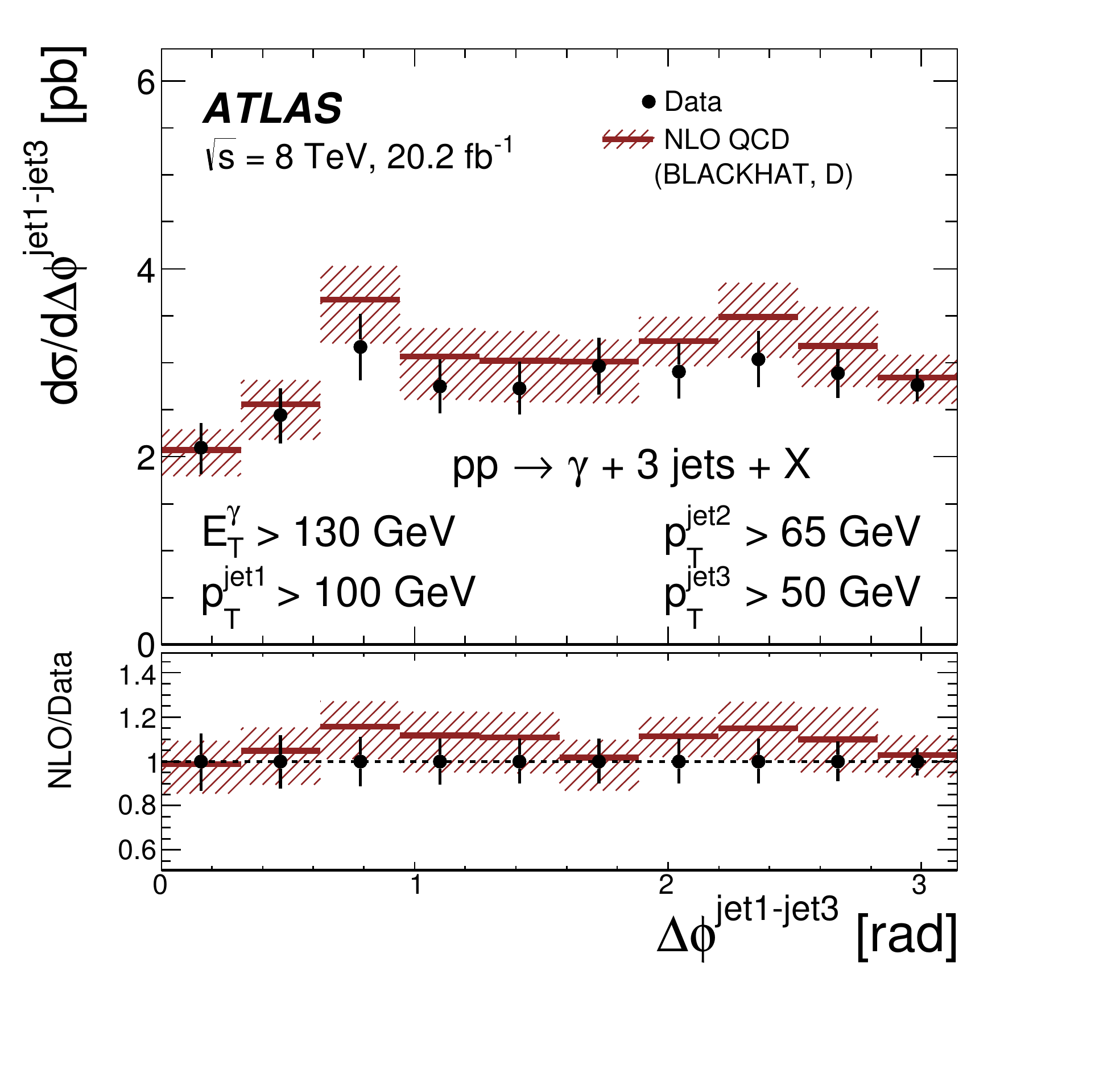}}
\put (0.0,-0.5){\includegraphics[width=7cm,height=7cm]{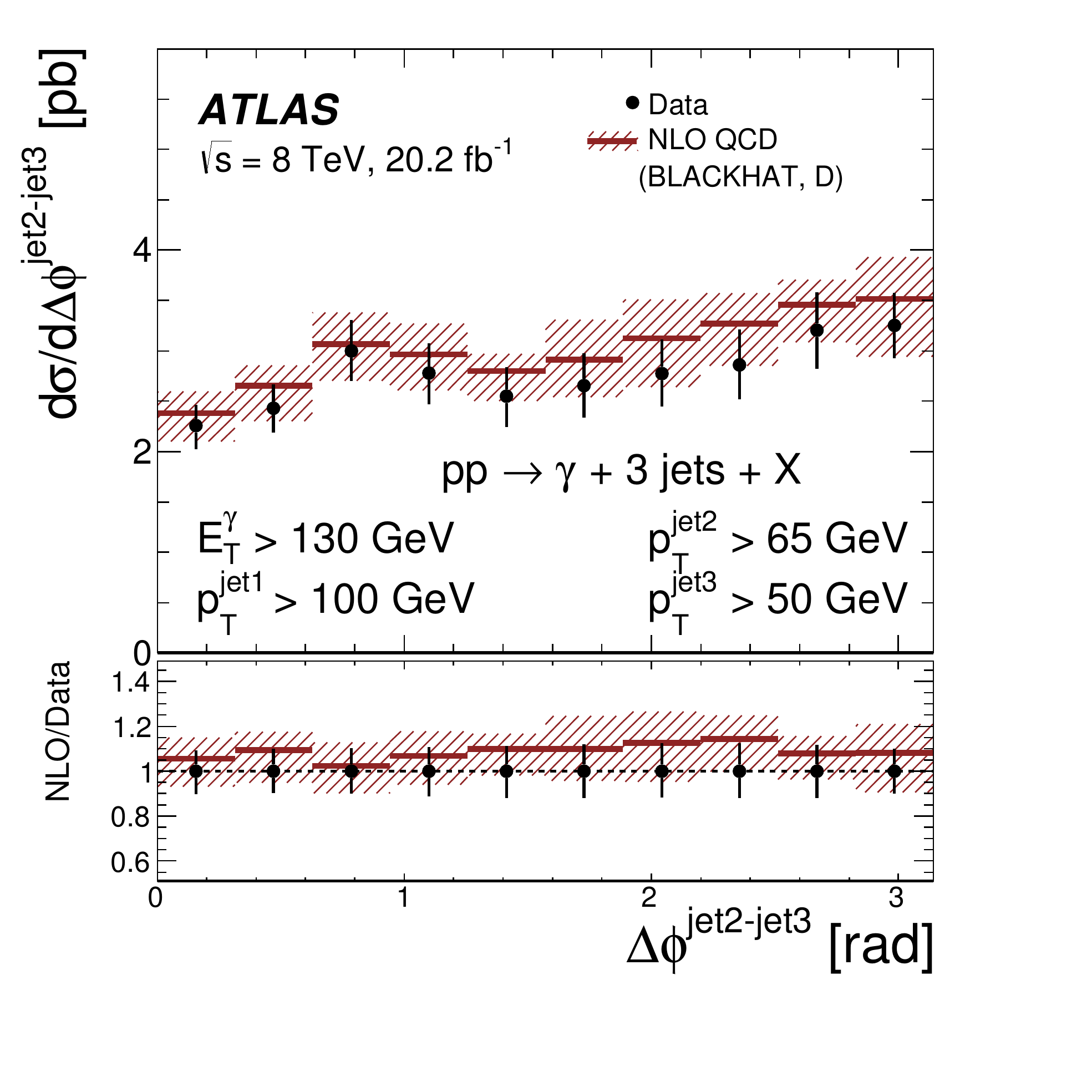}}
\put (2.5,13.2){{\textbf{\small (a)}}}
\put (10.5,13.2){{\textbf{\small (b)}}}
\put (2.5,6.7){{\textbf{\small (c)}}}
\put (10.5,6.7){{\textbf{\small (d)}}}
\put (2.5,0.2){{\textbf{\small (e)}}}
\end{picture}
\caption
{
  Measured cross sections for isolated-photon plus three-jet
  production (dots) as functions of (a) $\etg$, (b) $\ptjetssl$, (c)
  $\delphigssl$, (d) $\delphijjlssl$ and (e) $\delphijjslssl$. The NLO
  QCD predictions from \blh\ corrected for hadronisation and
  underlying-event effects and using the CT10 PDF set (solid lines)
  are also shown. These predictions include only the direct
  contribution (D). The bottom part of each figure shows the ratio of
  the NLO QCD prediction to the measured cross section. The inner
  (outer) error bars represent the statistical uncertainties (the
  statistical and systematic uncertainties added in quadrature) and
  the shaded band represents the theoretical uncertainty. For most of
  the points, the inner error bars are smaller than the marker size
  and, thus, not visible.
}
\label{fig179}
\end{figure}

The measured cross-section ${\mathrm{d}}\sigma/{\mathrm{d}}\etg$ (${\mathrm{d}}\sigma/{\mathrm{d}}\ptjetssl$), shown in Figure~\ref{fig179}(a)
(Figure~\ref{fig179}(b)), decreases by approximately five (three)
orders of magnitude within the measured range. The measured
cross-section ${\mathrm{d}}\sigma/{\mathrm{d}}\delphigssl$
(Figure~\ref{fig179}(c)) increases as $\delphigssl$ increases whereas
the measured cross sections as functions of $\delphijjlssl$
(Figure~\ref{fig179}(d)) and $\delphijjslssl$ (Figure~\ref{fig179}(e))
are approximately constant for $\delphijjlssl$, $\delphijjslssl
>1$~rad. The NLO QCD predictions from \blh\ give an adequate
description of the data within the experimental and theoretical
uncertainties; the predictions have a tendency to be systematically above the data.

\begin{figure}[p]
\setlength{\unitlength}{1.0cm}
\begin{picture} (18.0,19.5)
\put (0.0,12.5){\includegraphics[width=7cm,height=7cm]{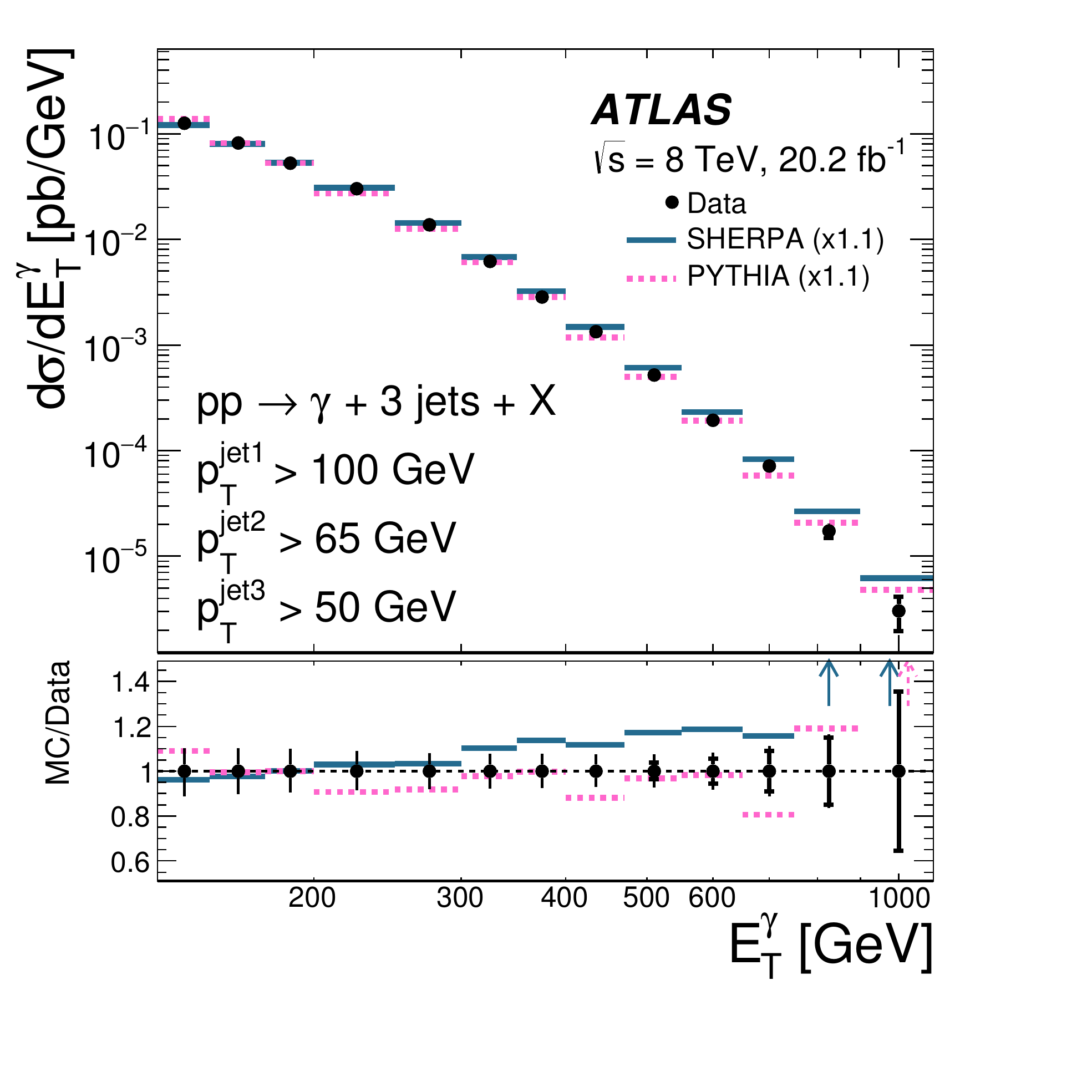}}
\put (8.0,12.5){\includegraphics[width=7cm,height=7cm]{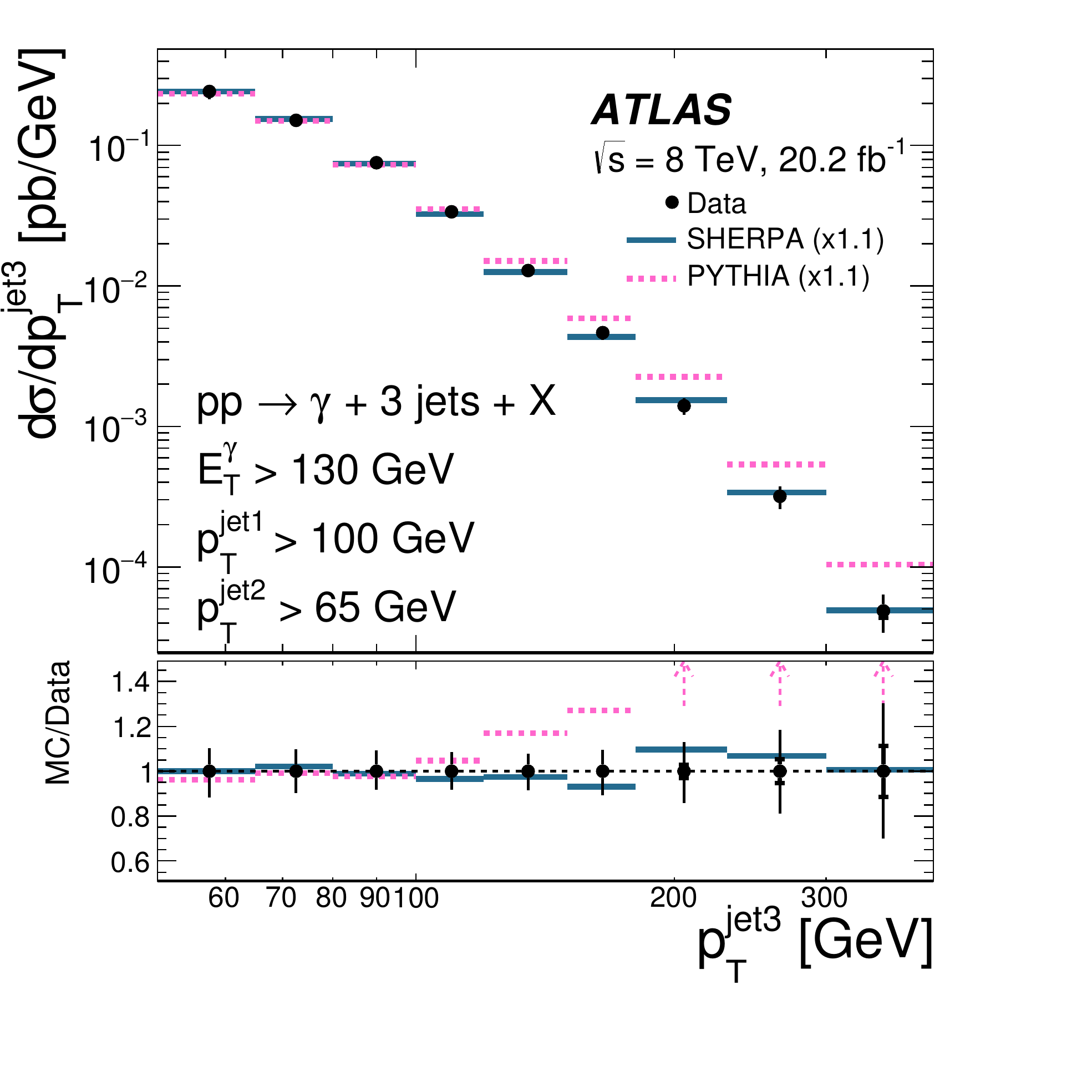}}
\put (0.0,6.0){\includegraphics[width=7cm,height=7cm]{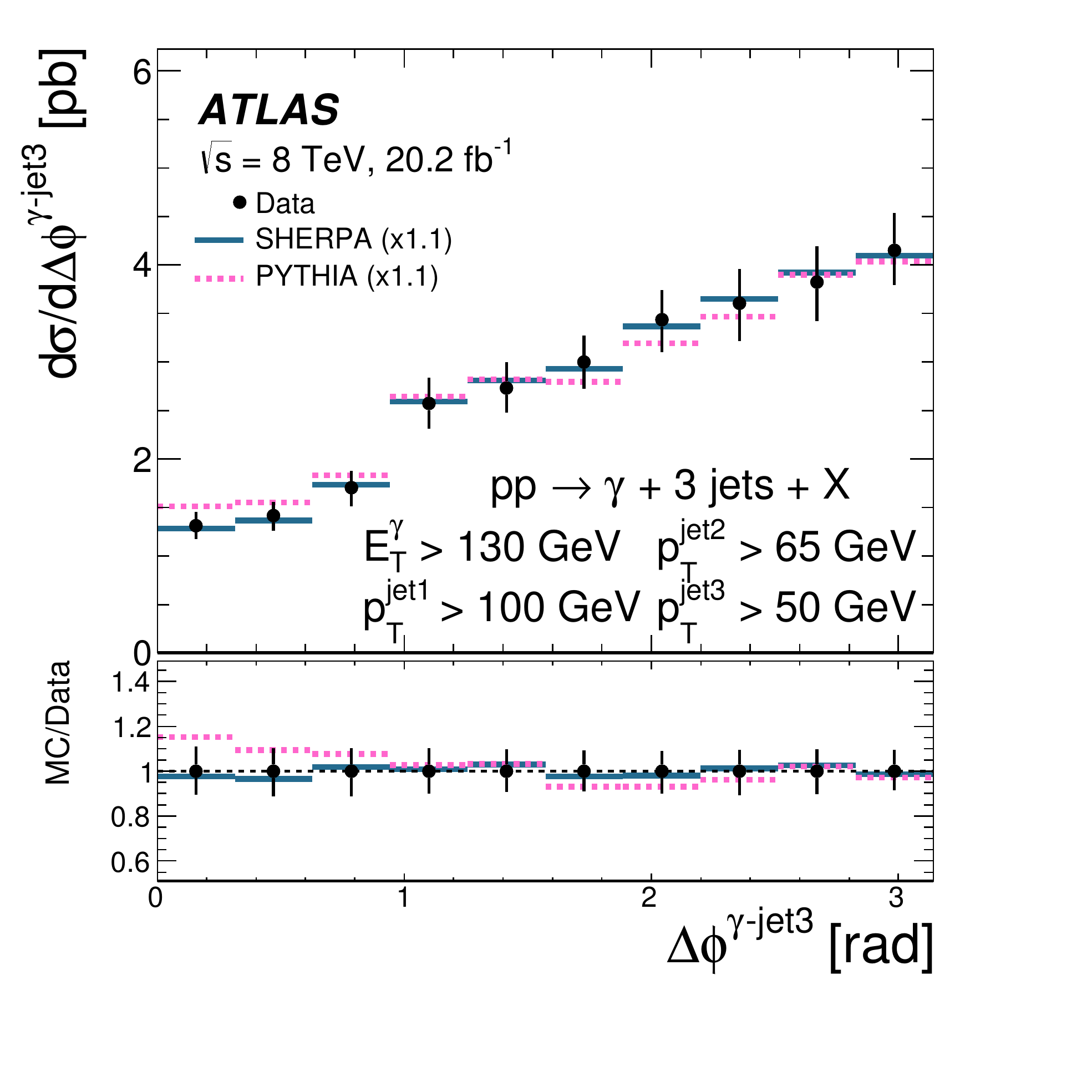}}
\put (8.0,6.0){\includegraphics[width=7cm,height=7cm]{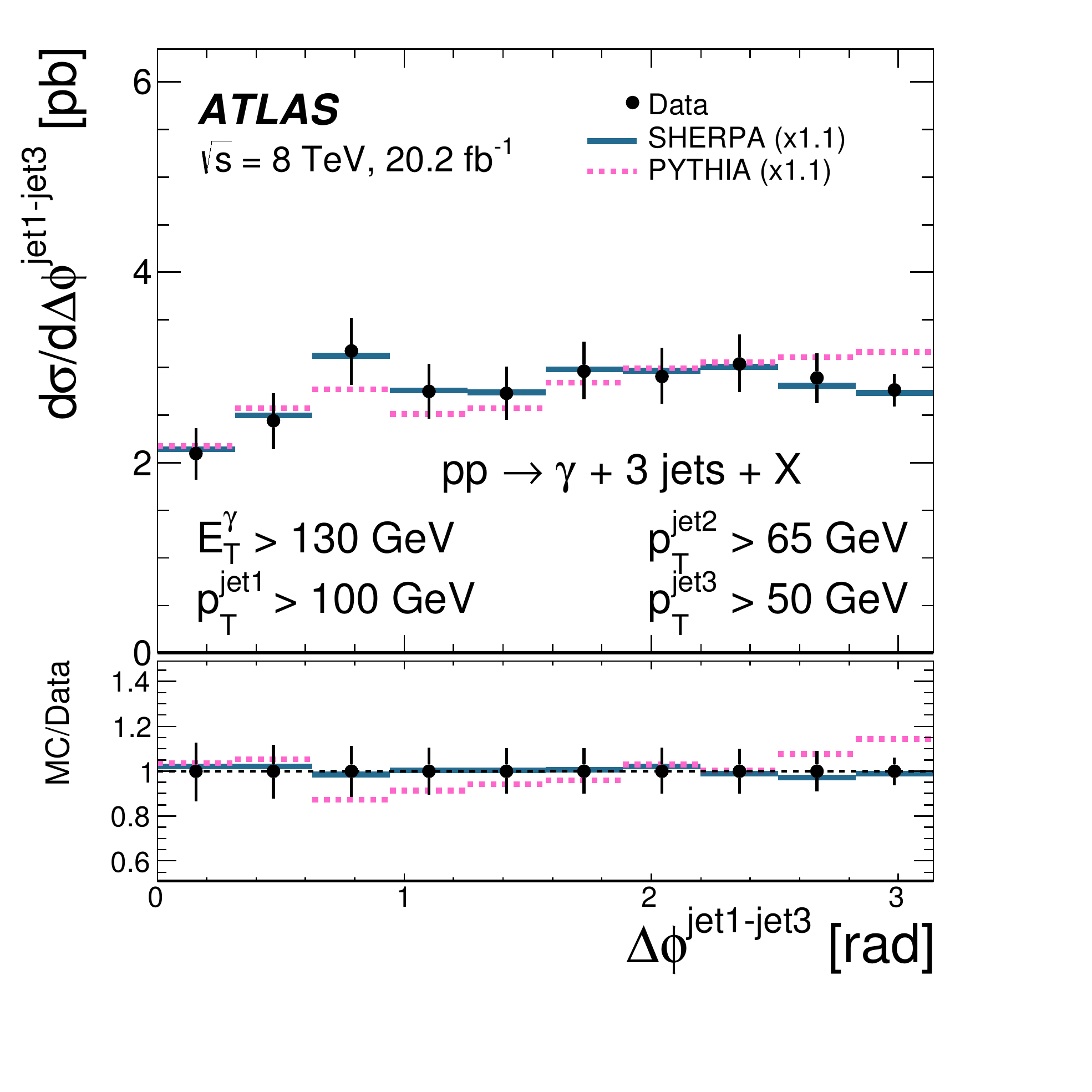}}
\put (0.0,-0.5){\includegraphics[width=7cm,height=7cm]{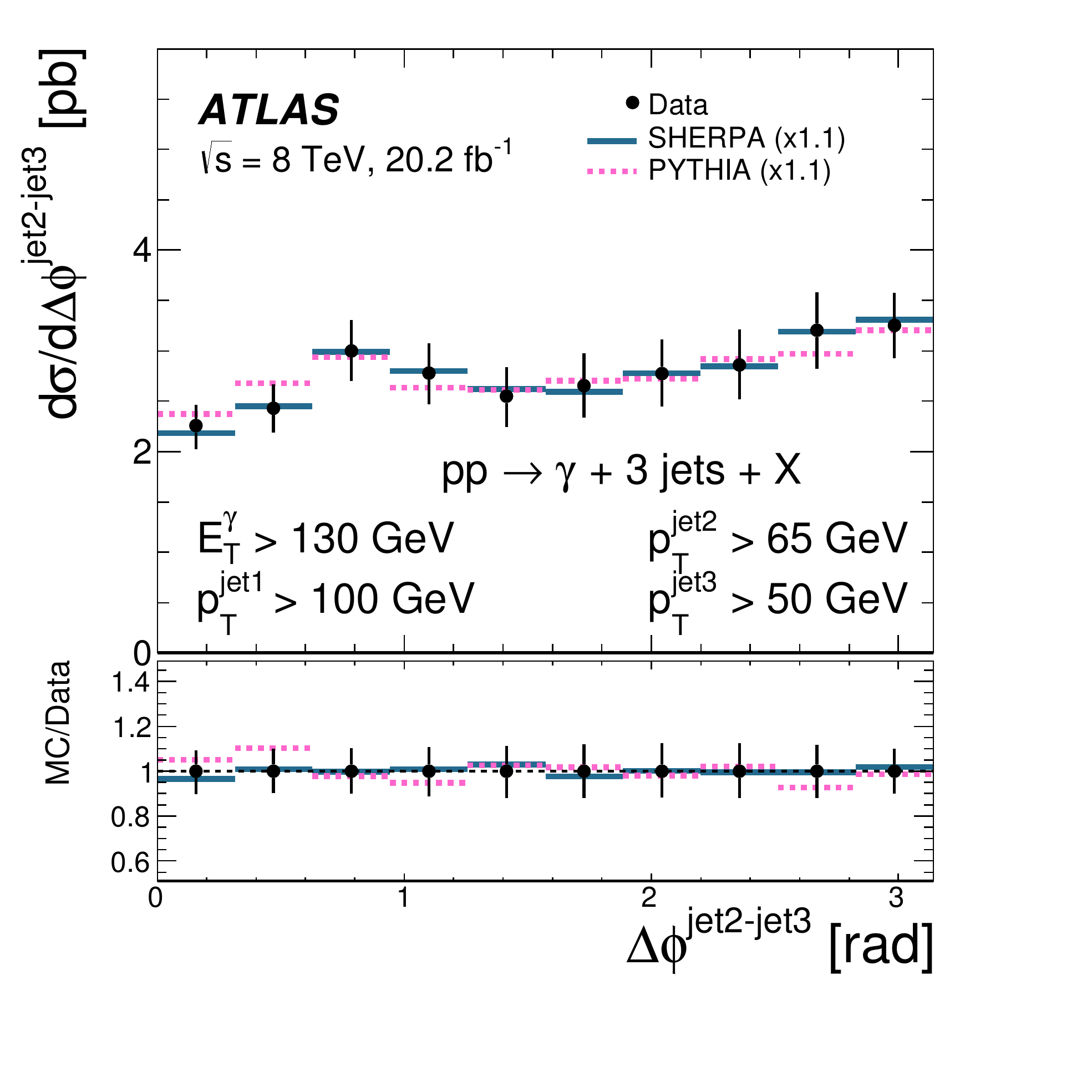}}
\put (2.5,13.2){{\textbf{\small (a)}}}
\put (10.5,13.2){{\textbf{\small (b)}}}
\put (2.5,6.7){{\textbf{\small (c)}}}
\put (10.5,6.7){{\textbf{\small (d)}}}
\put (2.5,0.2){{\textbf{\small (e)}}}
\end{picture}
\caption
{
  Measured cross sections for isolated-photon plus three-jet
  production (dots) as functions of (a) $\etg$, (b) $\ptjetssl$, (c)
  $\delphigssl$, (d) $\delphijjlssl$ and (e) $\delphijjslssl$. For
  comparison, the predictions from \sher\ (solid lines) and \pyt\
  (dashed lines)  normalised to the integrated measured cross sections
  (using the factors indicated in parentheses) are also shown. The
  bottom part of each figure shows the ratios of the MC predictions to
  the measured cross section. The inner (outer) error bars represent
  the statistical uncertainties (the statistical and systematic
  uncertainties added in quadrature). For most of the points, the
  inner error bars are smaller than the marker size and, thus, not
  visible.
}
\label{fig179b}
\end{figure}

The prediction from \pyt\  gives a good description of the measured
cross-section ${\mathrm{d}}\sigma/{\mathrm{d}}\etg$, whereas the prediction from
\sher\ describes well the measured cross-section ${\mathrm{d}}\sigma/{\mathrm{d}}\ptjetssl$ (see Figure~\ref{fig179b}). The predictions from \sher\
and \pyt\ give a good description of the measured cross sections as
functions of $\delphigssl$, $\delphijjlssl$ and $\delphijjslssl$.

\begin{figure}[p]
\setlength{\unitlength}{1.0cm}
\begin{picture} (18.0,19.0)
\put (0.0,10.2){\includegraphics[width=9cm,height=9cm]{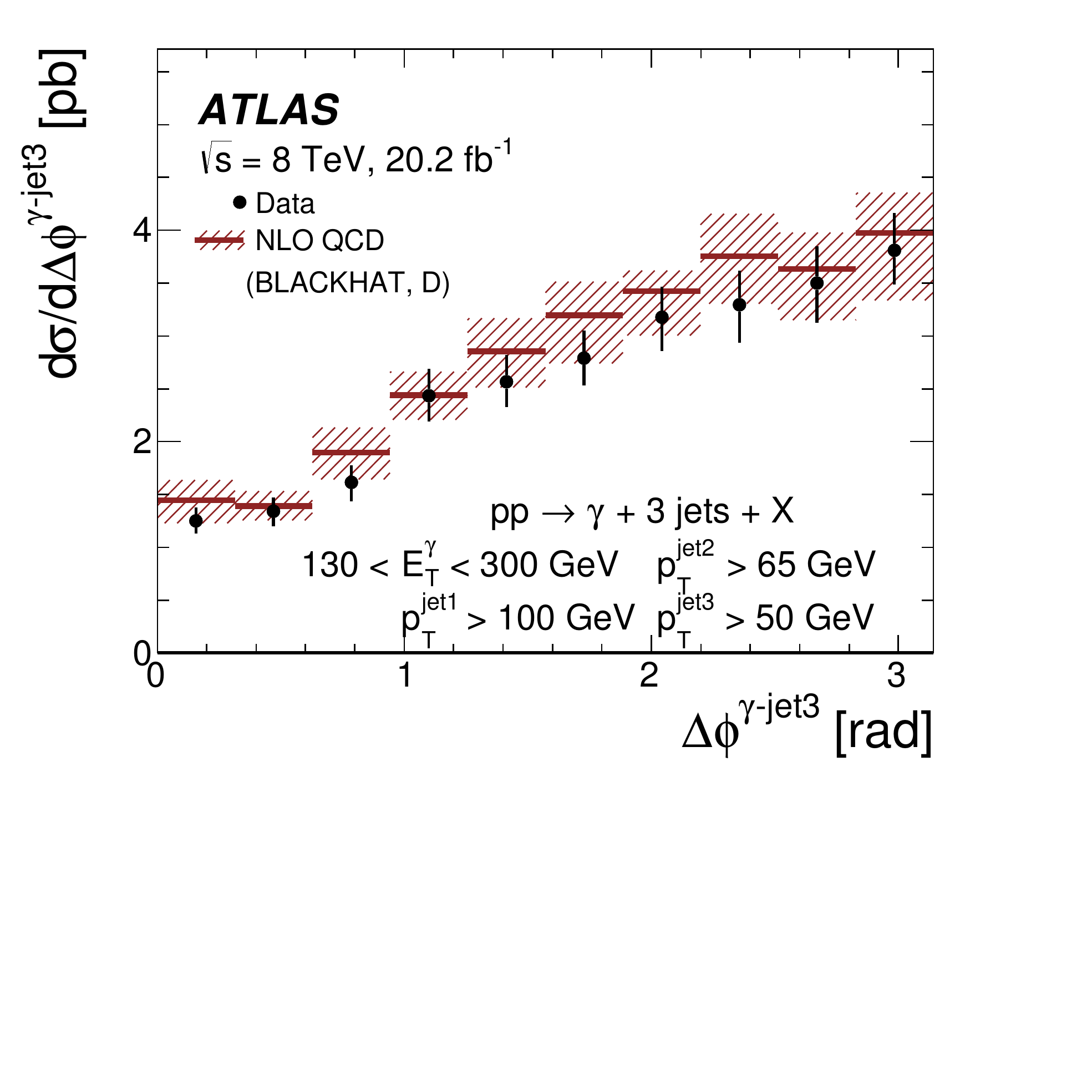}}
\put (8.0,10.2){\includegraphics[width=9cm,height=9cm]{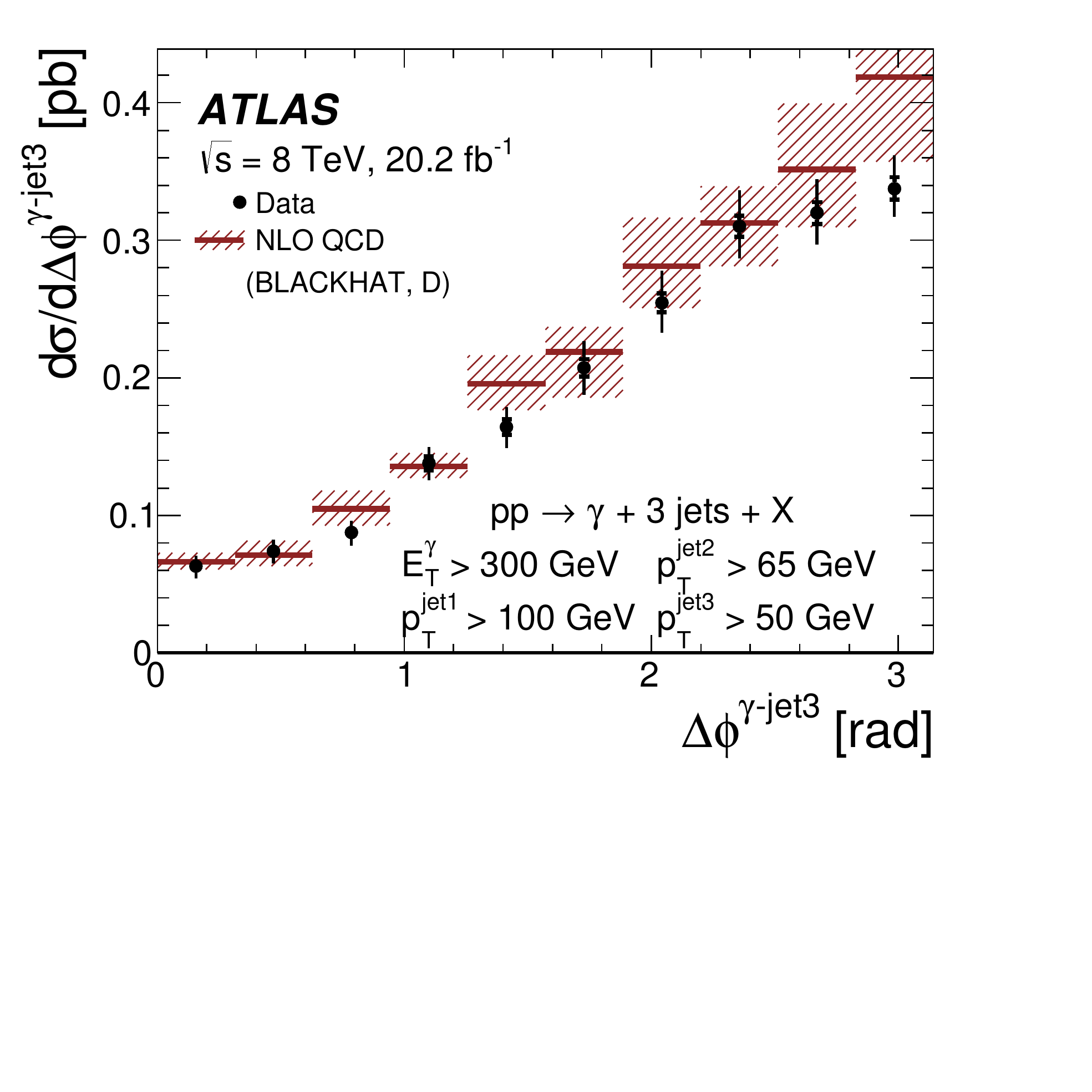}}
\put (0.0,3.7){\includegraphics[width=9cm,height=9cm]{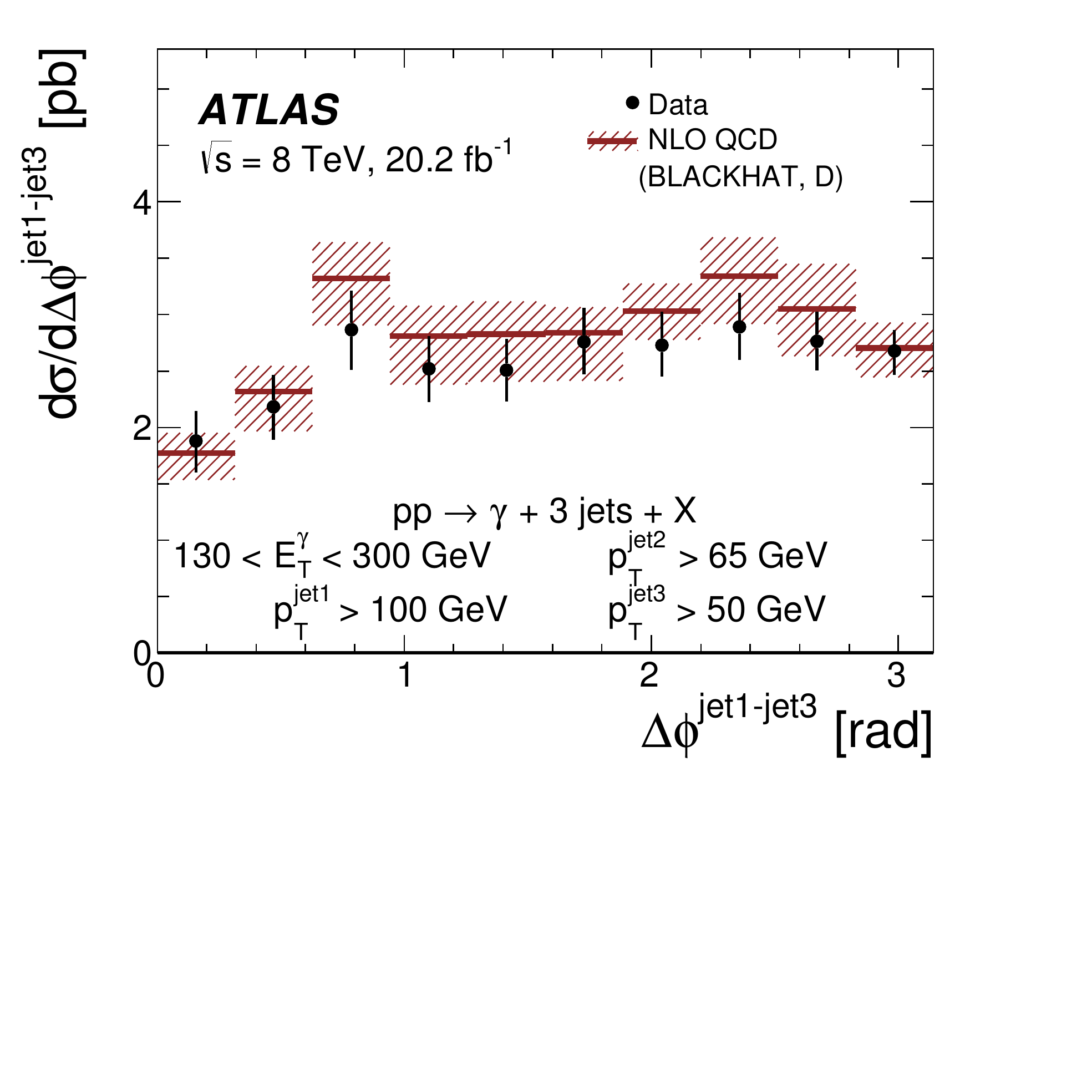}}
\put (8.0,3.7){\includegraphics[width=9cm,height=9cm]{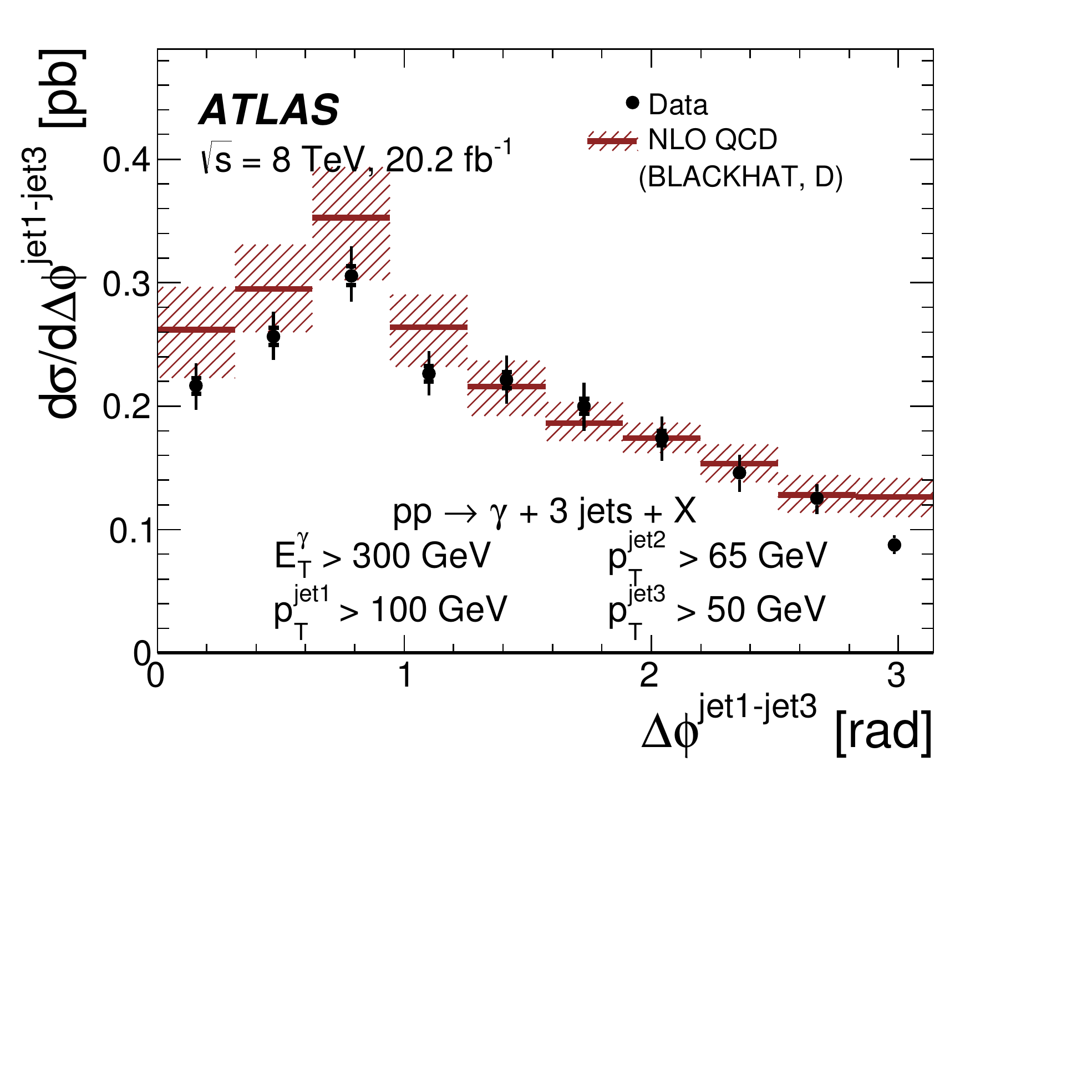}}
\put (0.0,-2.8){\includegraphics[width=9cm,height=9cm]{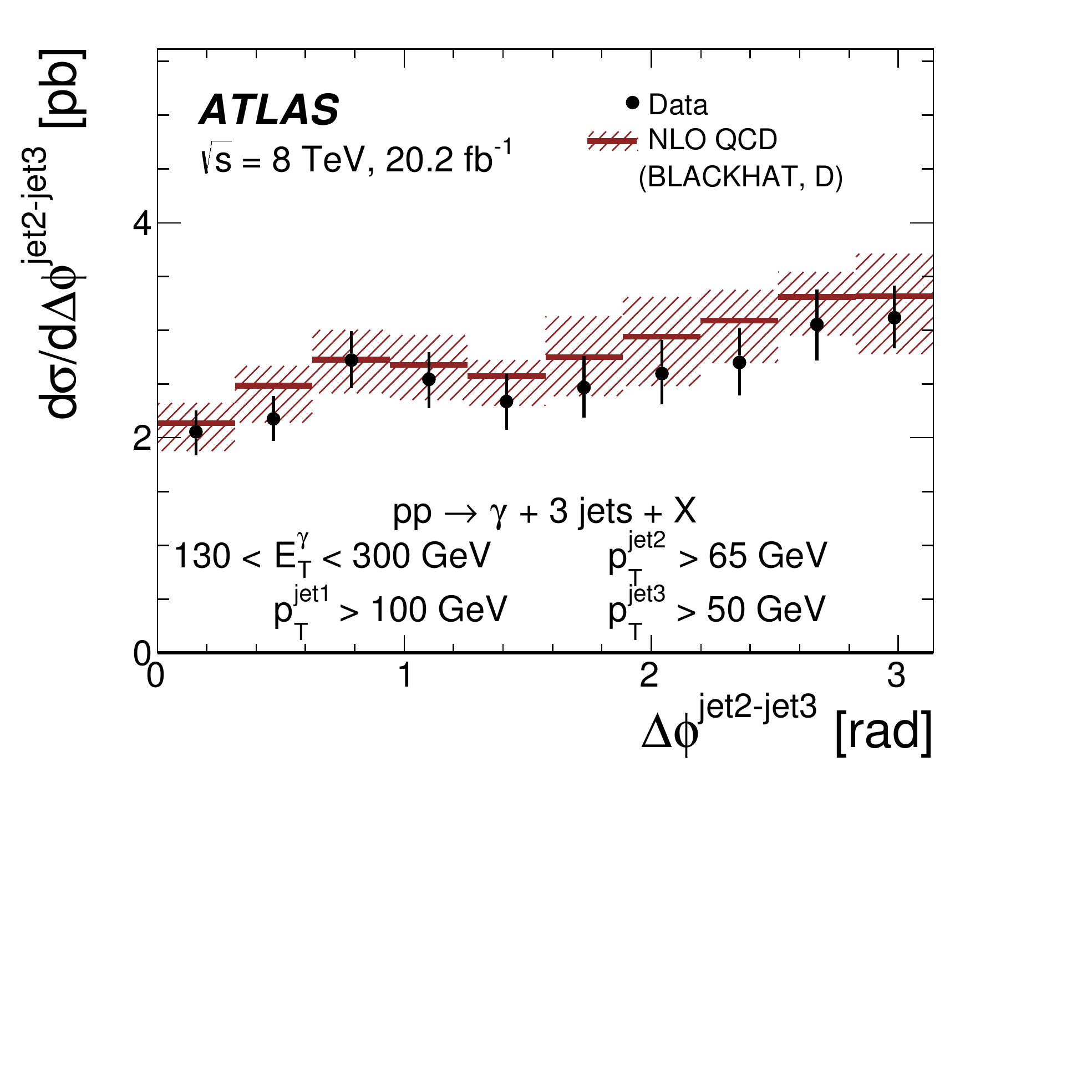}}
\put (8.0,-2.8){\includegraphics[width=9cm,height=9cm]{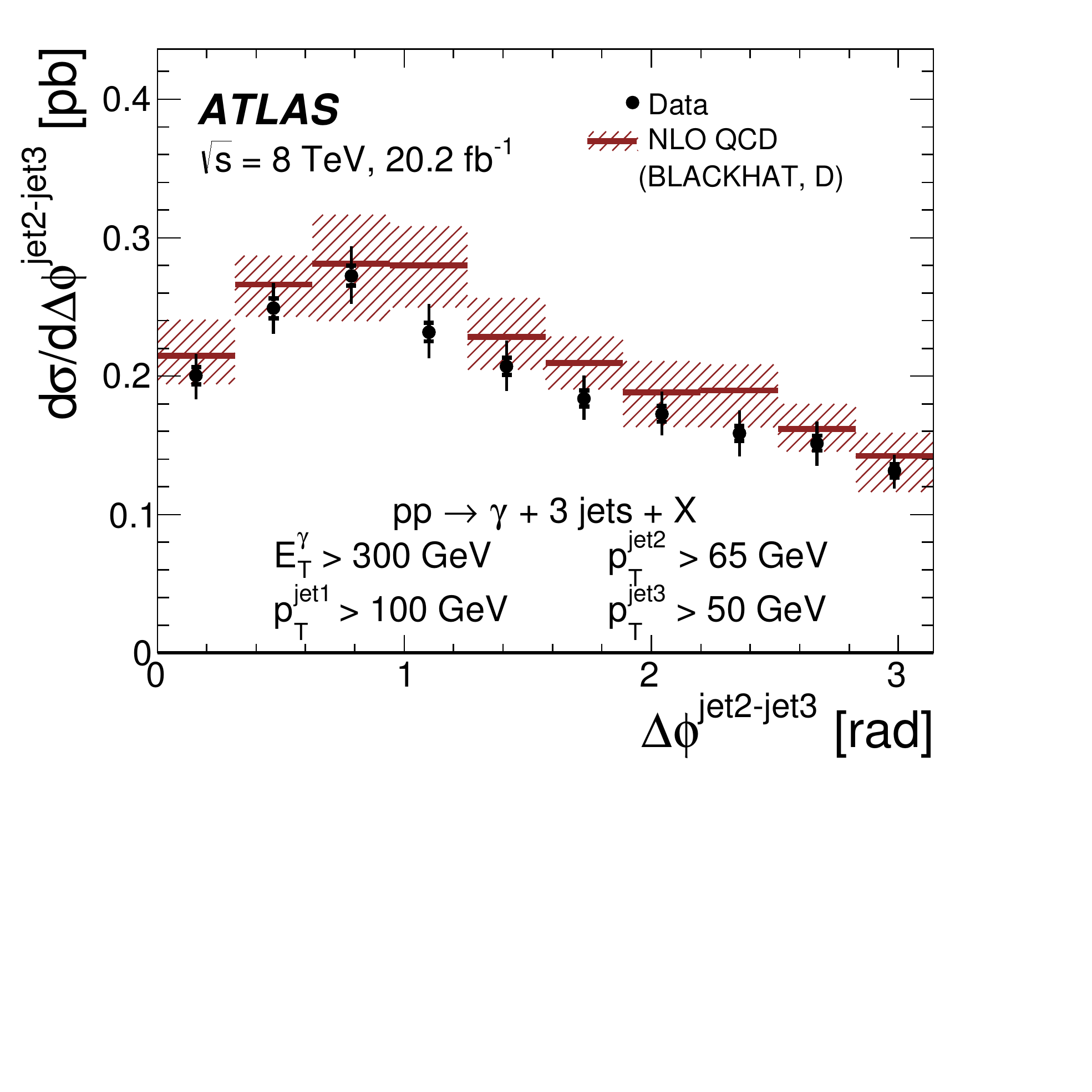}}
\put (3.7,13.0){{\textbf{\small (a)}}}
\put (11.7,13.0){{\textbf{\small (b)}}}
\put (3.7,6.5){{\textbf{\small (c)}}}
\put (11.7,6.5){{\textbf{\small (d)}}}
\put (3.7,0.0){{\textbf{\small (e)}}}
\put (11.7,0.0){{\textbf{\small (f)}}}
\end{picture}
\caption
{
  Measured cross sections for isolated-photon plus three-jet
  production (dots) as functions of (a,b) $\delphigssl$, (c,d)
  $\delphijjlssl$ and (e,f) $\delphijjslssl$ for (a,c,e)
  $\etg<300$~\GeV\ and (b,d,f) $\etg>300$~\GeV. The NLO QCD
  predictions from \blh\ corrected for hadronisation and
  underlying-event effects and using the CT10 PDF set are also shown
  as solid lines. These predictions include only the direct
  contribution (D). The inner (outer) error bars represent the
  statistical uncertainties (the statistical and systematic
  uncertainties added in quadrature) and the shaded band represents
  the theoretical uncertainty. For most of the points, the inner error
  bars are smaller than the marker size and, thus, not visible.
}
\label{fig180}
\end{figure}

\begin{figure}[p]
\setlength{\unitlength}{1.0cm}
\begin{picture} (18.0,19.0)
\put (0.0,10.2){\includegraphics[width=9cm,height=9cm]{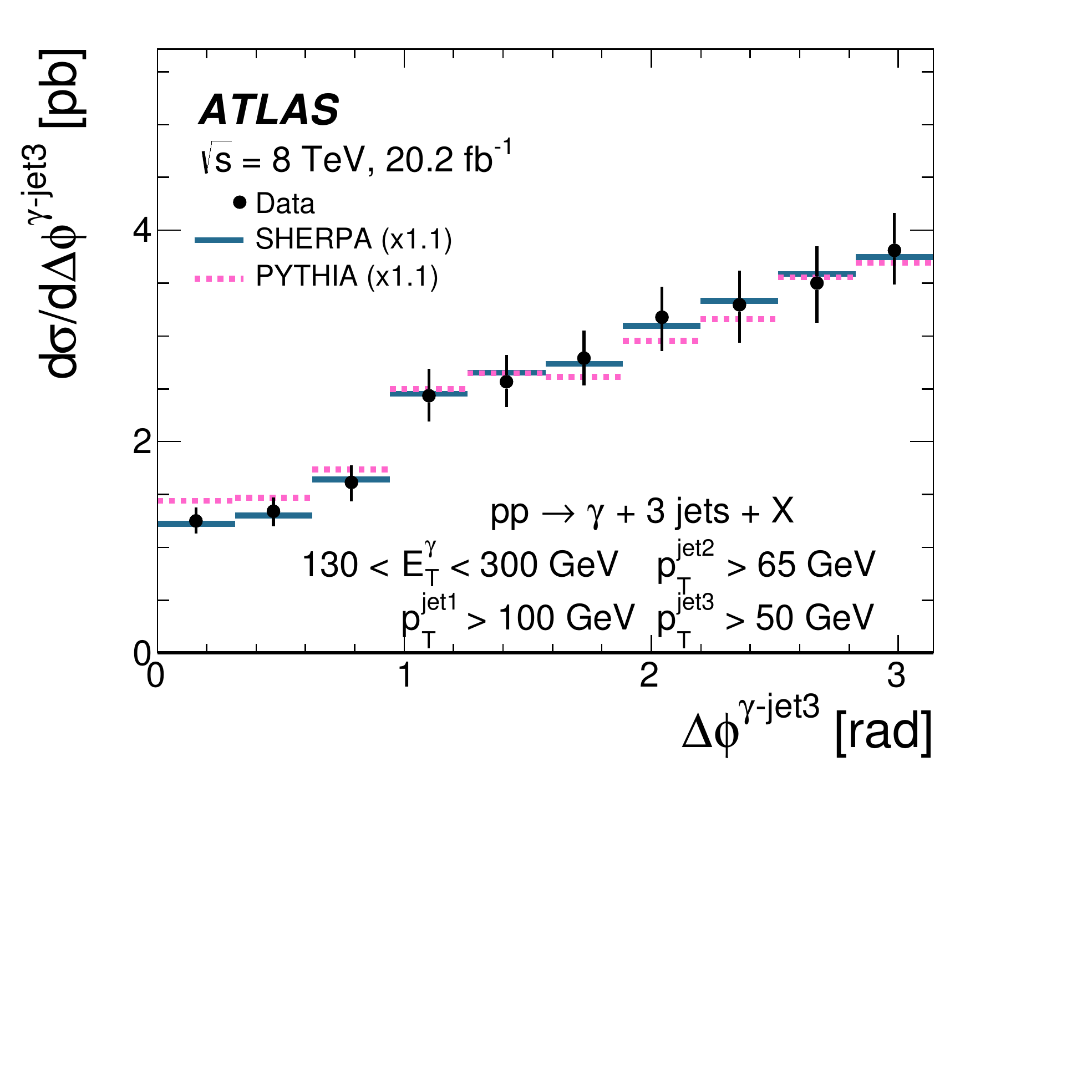}}
\put (8.0,10.2){\includegraphics[width=9cm,height=9cm]{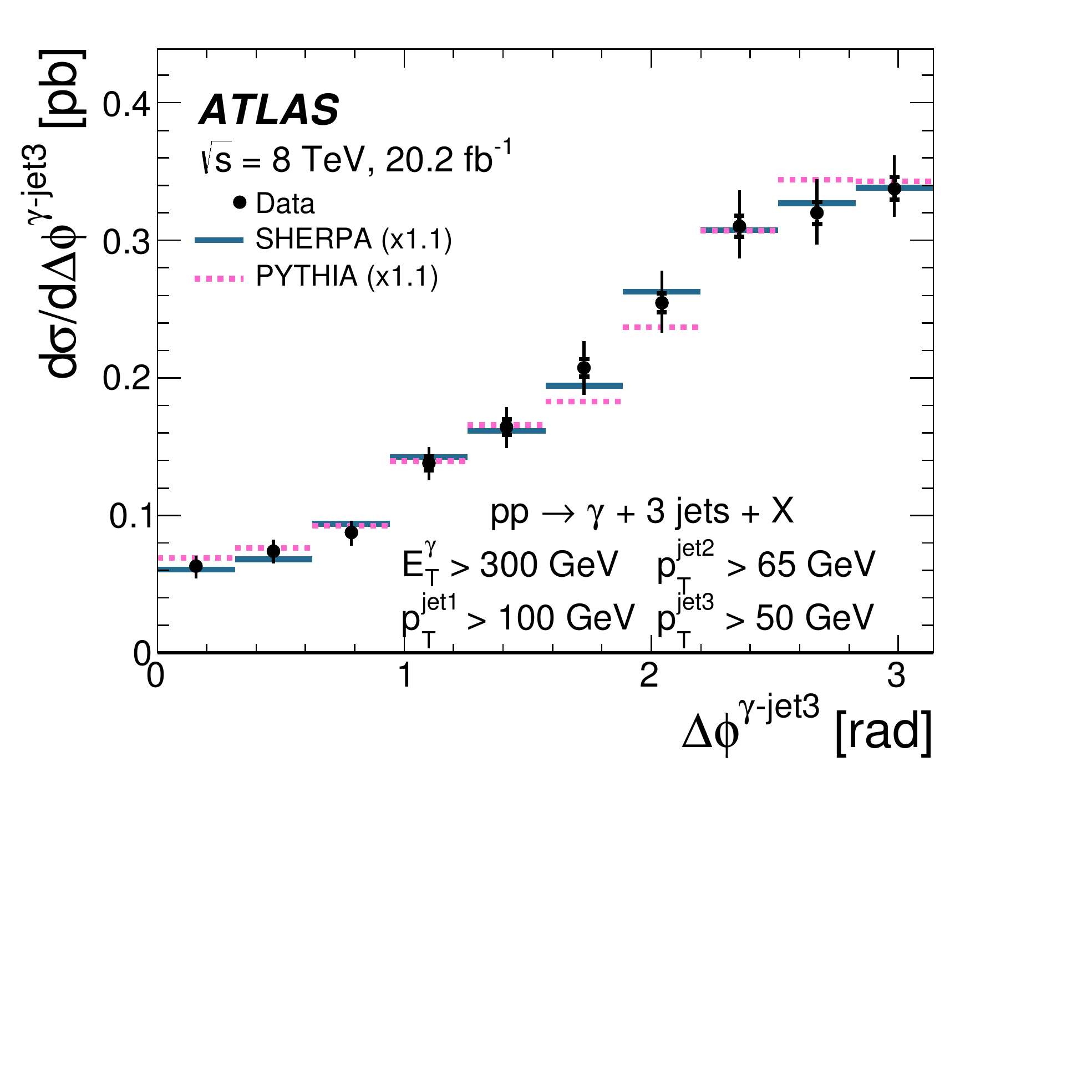}}
\put (0.0,3.7){\includegraphics[width=9cm,height=9cm]{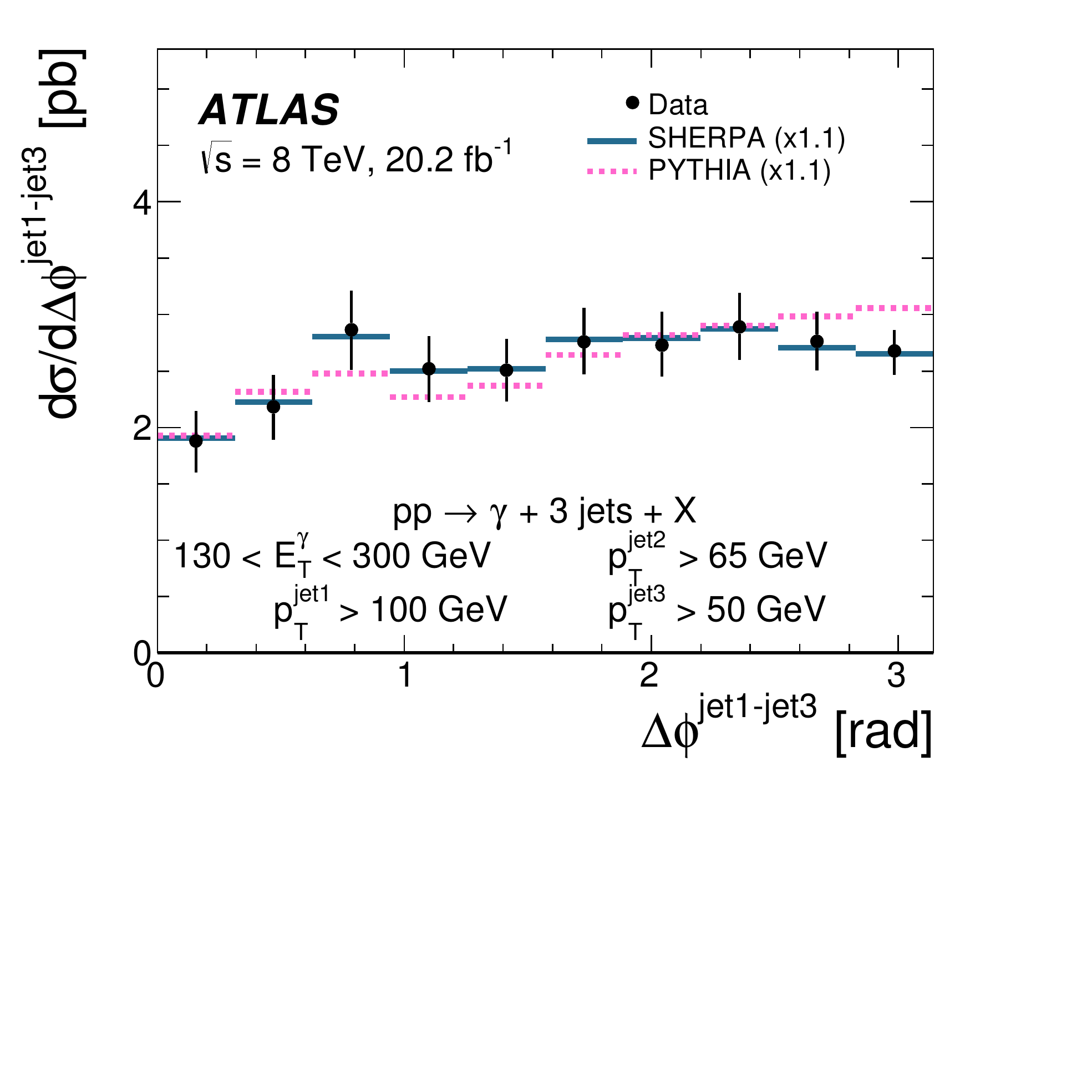}}
\put (8.0,3.7){\includegraphics[width=9cm,height=9cm]{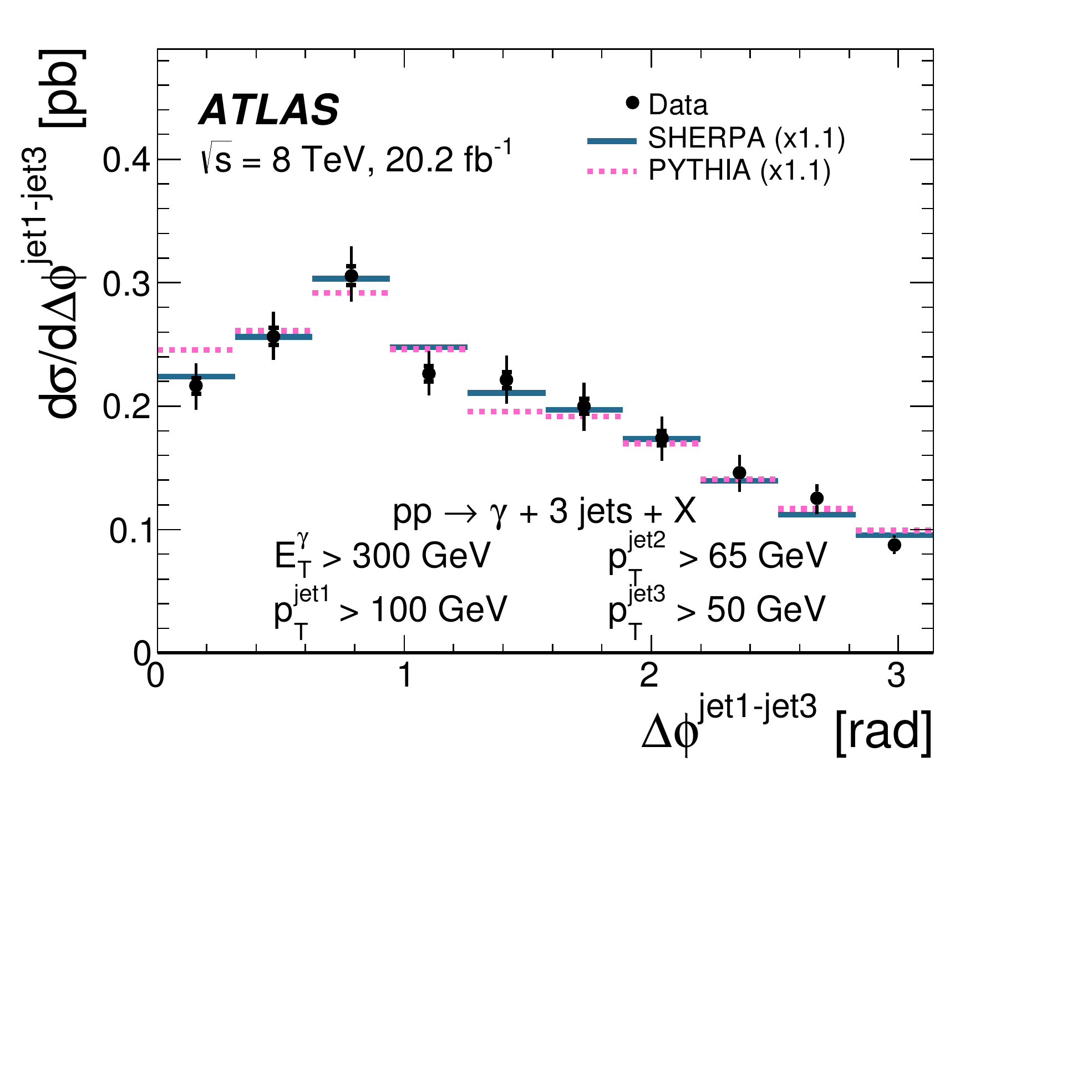}}
\put (0.0,-2.8){\includegraphics[width=9cm,height=9cm]{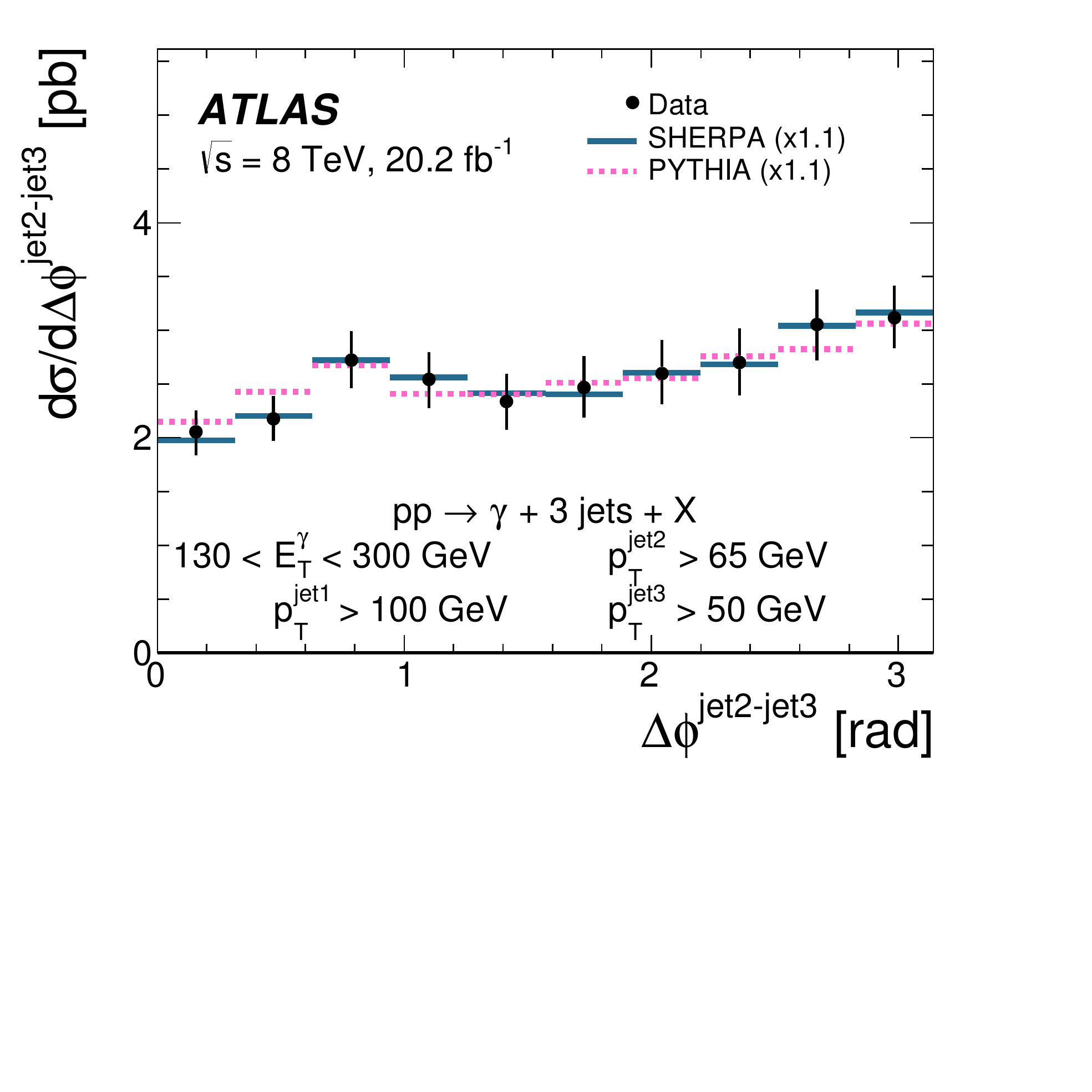}}
\put (8.0,-2.8){\includegraphics[width=9cm,height=9cm]{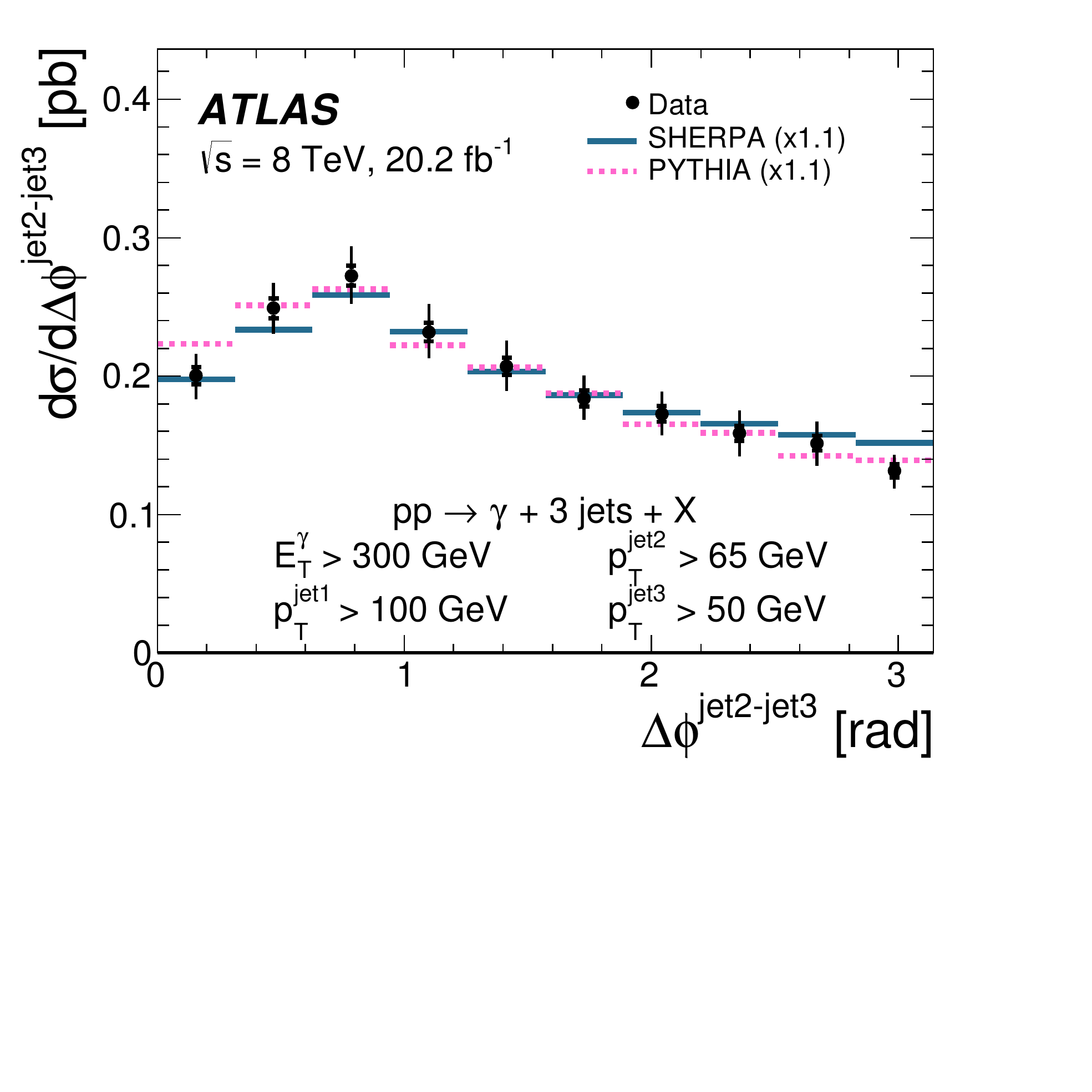}}
\put (3.7,13.0){{\textbf{\small (a)}}}
\put (11.7,13.0){{\textbf{\small (b)}}}
\put (3.7,6.5){{\textbf{\small (c)}}}
\put (11.7,6.5){{\textbf{\small (d)}}}
\put (3.7,0.0){{\textbf{\small (e)}}}
\put (11.7,0.0){{\textbf{\small (f)}}}
\end{picture}
\caption
{
  Measured cross sections for isolated-photon plus three-jet
  production (dots) as functions of (a,b) $\delphigssl$, (c,d)
  $\delphijjlssl$ and (e,f) $\delphijjslssl$ for (a,c,e)
  $\etg<300$~\GeV\ and (b,d,f) $\etg>300$~\GeV. For comparison, the
  predictions from \pyt\  (dashed lines) and \sher\  (solid lines) are
  also shown.  The predictions are normalised to the data by a global
  factor, which is  indicated in parentheses. The inner (outer) error
  bars represent the statistical uncertainties (the statistical and
  systematic uncertainties added in quadrature). For most of the
  points, the inner error bars are smaller than the marker size and,
  thus, not visible.
}
\label{fig180b}
\end{figure}

The scale evolution of the photon plus three-jet production is tested
by measuring the distributions of the azimuthal angle between jet3 and
the photon, jet1 or jet2 for $\etg$ below/above
$300$~\GeV. Figure~\ref{fig180} shows the cross sections as functions
of $\delphigssl$, $\delphijjlssl$ and $\delphijjslssl$ for the two
$\etg$ ranges: the shape of the cross-section distributions is
different for $\etg$ below/above $300$~\GeV. The ${\mathrm{d}}\sigma/{\mathrm{d}}\delphigssl$ cross section is more peaked towards large values of
$\delphigssl$ for $\etg > 300$~\GeV\ than that for $\etg < 300$~\GeV;
the ${\mathrm{d}}\sigma/{\mathrm{d}}\delphijjlssl$ (${\mathrm{d}}\sigma/{\mathrm{d}}\delphijjslssl$) cross section decreases beyond the peak as
$\delphijjlssl$ ($\delphijjslssl$) increases for $\etg > 300$~\GeV\
whereas it stays approximately constant for $\etg < 300$~\GeV. The NLO
QCD predictions provide an adequate description of the measured cross
sections and, thereby, of the observed scale evolution of the angular
correlations. The predictions from \pyt\ and \sher, shown in
Figure~\ref{fig180b}, give an adequate description of the shape of the
measured cross sections.

\FloatBarrier

\section{Summary}
\label{conc}
Measurements of the cross sections for the production of an isolated
photon in association with one, two or three jets in proton--proton
collisions at $\sqrt s=8$~\TeV, $pp\rightarrow\gamma+{\mathrm{jet(s)}}+{\mathrm{X}}$, using a data set with an integrated luminosity of $20.2$~\fb1\
recorded by the ATLAS detector at the LHC are presented. The photon is
required to have $\etg>130$~\GeV\ and $|\etag|<2.37$, excluding the
region $1.37<|\etag|<1.56$, and to be isolated with
$\etisop<10$~\GeV. The jets are reconstructed using the anti-$\kt$
algorithm with radius parameter $R=0.6$.

The cross sections for photon plus one-jet are measured as functions
of $\etg$ and $\ptjetl$ with $\ptjetl>100$~\GeV; the measurements
extend up to values of $\etg\ (\ptjetl)$ of $1.1$~\TeV\
($1.2$~\TeV). The dependence on $\mgjn$ and $|\ctgjn|$ is also
measured for $\mgjn>467$~\GeV\ and extends up to $\mgjn$ of
$2.45$~\TeV. The NLO QCD predictions from \jetp, corrected for
hadronisation and underlying-event effects, give a good description of
the measured cross-section distributions in both shape and
normalisation. In particular, the measured dependence on $|\ctgjn|$
and its scale dependence is consistent with the dominance of processes
in which a quark is being exchanged; the experimental (theoretical)
uncertainty in ${\mathrm{d}}\sigma/{\mathrm{d}}|\ctgjn|$ amounts to $\approx
3\%\ (10\%)$.

Photon plus two-jet production is investigated by measuring cross
sections as functions of $\etg$ and $\ptjetsl$ and angular
correlations between the final-state objects for $\ptjetl>100$~\GeV\
and $\ptjetsl>65$~\GeV. The NLO QCD predictions from \blh\
provide a good description of the measurements except for
$\etg>750$~\GeV. The predictions from \sher, which include
higher-order tree-level matrix elements, are found to be superior to
those from \pyt, based on $2\rightarrow 2$ processes, in describing
the distributions in $\ptjetsl$ and the angular correlations.

The patterns of QCD radiation around the photon and the leading jet
are compared by measuring the production of the subleading jet in an
annular region centred on the given final-state object. The cross
sections as functions of $\betag$ and $\betaj$ are observed to be
different. The ratio of the cross sections shows enhancements in the
directions towards the beams, $\beta=0$ and $\pi$~rad.

Photon plus three-jet production is characterised by measurements of
cross sections as functions of $\etg$, $\ptjetssl$ and angular
correlations for $\ptjetl>100$~\GeV, $\ptjetsl>65$~\GeV\ and
$\ptjetssl>50$~\GeV. The NLO QCD predictions from \blh\
provide an adequate description of the measurements. Whereas the
prediction from \sher\ for $\ptjetssl$ is superior to that from \pyt,
both give adequate descriptions of the angular correlations.

All these studies provide stringent tests of pQCD and scrutinise the
description of the dynamics of isolated-photon plus jets production in
$pp$ collisions up to ${\cal O}(\alpha_{\mathrm{em}}\alpha_{\mathrm{s}}^4)$.

\section*{Acknowledgements}


We thank CERN for the very successful operation of the LHC, as well as
the support staff from our institutions without whom ATLAS could not
be operated efficiently.

We acknowledge the support of ANPCyT, Argentina; YerPhI, Armenia; ARC,
Australia; BMWFW and FWF, Austria; ANAS, Azerbaijan; SSTC, Belarus;
CNPq and FAPESP, Brazil; NSERC, NRC and CFI, Canada; CERN; CONICYT,
Chile; CAS, MOST and NSFC, China; COLCIENCIAS, Colombia; MSMT CR, MPO
CR and VSC CR, Czech Republic; DNRF and DNSRC, Denmark; IN2P3-CNRS,
CEA-DSM/IRFU, France; SRNSF, Georgia; BMBF, HGF, and MPG, Germany;
GSRT, Greece; RGC, Hong Kong SAR, China; ISF, I-CORE and Benoziyo
Center, Israel; INFN, Italy; MEXT and JSPS, Japan; CNRST, Morocco; FOM
and NWO, Netherlands; RCN, Norway; MNiSW and NCN, Poland; FCT,
Portugal; MNE/IFA, Romania; MES of Russia and NRC KI, Russian
Federation; JINR; MESTD, Serbia; MSSR, Slovakia; ARRS and MIZ\v{S},
Slovenia; DST/NRF, South Africa; MINECO, Spain; SRC and Wallenberg
Foundation, Sweden; SERI, SNSF and Cantons of Bern and Geneva,
Switzerland; MOST, Taiwan; TAEK, Turkey; STFC, United Kingdom; DOE and
NSF, United States of America. In addition, individual groups and
members have received support from BCKDF, the Canada Council, CANARIE,
CRC, Compute Canada, FQRNT, and the Ontario Innovation Trust, Canada;
EPLANET, ERC, ERDF, FP7, Horizon 2020 and Marie Sk{\l}odowska-Curie
Actions, European Union; Investissements d'Avenir Labex and Idex, ANR,
R{\'e}gion Auvergne and Fondation Partager le Savoir, France; DFG and
AvH Foundation, Germany; Herakleitos, Thales and Aristeia programmes
co-financed by EU-ESF and the Greek NSRF; BSF, GIF and Minerva,
Israel; BRF, Norway; CERCA Programme Generalitat de Catalunya,
Generalitat Valenciana, Spain; the Royal Society and Leverhulme Trust,
United Kingdom.

The crucial computing support from all WLCG partners is acknowledged
gratefully, in particular from CERN, the ATLAS Tier-1 facilities at
TRIUMF (Canada), NDGF (Denmark, Norway, Sweden), CC-IN2P3 (France),
KIT/GridKA (Germany), INFN-CNAF (Italy), NL-T1 (Netherlands), PIC
(Spain), ASGC (Taiwan), RAL (UK) and BNL (USA), the Tier-2 facilities
worldwide and large non-WLCG resource providers. Major contributors of
computing resources are listed in Ref.~\cite{ATL-GEN-PUB-2016-002}.

\printbibliography

\newpage
\begin{flushleft}
{\Large The ATLAS Collaboration}

\bigskip

M.~Aaboud$^\textrm{\scriptsize 137d}$,
G.~Aad$^\textrm{\scriptsize 88}$,
B.~Abbott$^\textrm{\scriptsize 115}$,
J.~Abdallah$^\textrm{\scriptsize 8}$,
O.~Abdinov$^\textrm{\scriptsize 12}$,
B.~Abeloos$^\textrm{\scriptsize 119}$,
O.S.~AbouZeid$^\textrm{\scriptsize 139}$,
N.L.~Abraham$^\textrm{\scriptsize 151}$,
H.~Abramowicz$^\textrm{\scriptsize 155}$,
H.~Abreu$^\textrm{\scriptsize 154}$
% ...
\bigskip
\\
% ...
$^{8}$ Department of Physics, The University of Texas at Arlington, Arlington TX, United States of America\\
% ...
$^{12}$ Institute of Physics, Azerbaijan Academy of Sciences, Baku, Azerbaijan\\
% ...
$^{88}$ CPPM, Aix-Marseille Universit{\'e} and CNRS/IN2P3, Marseille, France\\
% ...
$^{115}$ Homer L. Dodge Department of Physics and Astronomy, University of Oklahoma, Norman OK, United States of America\\
% ...
$^{119}$ LAL, Univ. Paris-Sud, CNRS/IN2P3, Universit{\'e} Paris-Saclay, Orsay, France\\
% ...
$^{137}$ $^{(a)}$ Facult{\'e} des Sciences Ain Chock, R{\'e}seau Universitaire de Physique des Hautes Energies - Universit{\'e} Hassan II, Casablanca; $^{(b)}$ Centre National de l'Energie des Sciences Techniques Nucleaires, Rabat; $^{(c)}$ Facult{\'e} des Sciences Semlalia, Universit{\'e} Cadi Ayyad, LPHEA-Marrakech; $^{(d)}$ Facult{\'e} des Sciences, Universit{\'e} Mohamed Premier and LPTPM, Oujda; $^{(e)}$ Facult{\'e} des sciences, Universit{\'e} Mohammed V, Rabat, Morocco\\
% ...
$^{139}$ Santa Cruz Institute for Particle Physics, University of California Santa Cruz, Santa Cruz CA, United States of America\\
% ...
$^{151}$ Department of Physics and Astronomy, University of Sussex, Brighton, United Kingdom\\
% ...
$^{154}$ Department of Physics, Technion: Israel Institute of Technology, Haifa, Israel\\
$^{155}$ Raymond and Beverly Sackler School of Physics and Astronomy, Tel Aviv University, Tel Aviv, Israel\\
% ...
$^{a}$ Also at Department of Physics, King's College London, London, United Kingdom\\
$^{b}$ Also at Institute of Physics, Azerbaijan Academy of Sciences, Baku, Azerbaijan\\
$^{c}$ Also at Novosibirsk State University, Novosibirsk, Russia\\
$^{d}$ Also at TRIUMF, Vancouver BC, Canada\\
$^{e}$ Also at Department of Physics {\&} Astronomy, University of Louisville, Louisville, KY, United States of America\\
% ...
\end{flushleft}

\end{document}